\documentclass[aps,prd,nofootinbib,twocolumn,superscriptaddress,letterpaper,preprintnumbers, longbibliography,notitlepage]{revtex4-1}
\usepackage{braket}
\usepackage{amssymb}
\usepackage{amsmath}
\usepackage{epsfig}
\usepackage{tabularx}
\usepackage{float}
\usepackage{subfigure}
\usepackage[colorlinks=true
,urlcolor=blue
,anchorcolor=blue
,citecolor=blue
,filecolor=blue
,linkcolor=red
,menucolor=blue
,hyperfootnotes=false,
,linktocpage=true
,pdfproducer=medialab
,pdfa=true
]{hyperref}
\usepackage{array}
\usepackage{booktabs}
\usepackage{comment}
\usepackage{cleveref}
\usepackage{multirow}

\newcolumntype{L}[1]{>{\raggedright\let\newline\\\arraybackslash\hspace{0pt}}m{#1}}
\newcolumntype{C}[1]{>{\centering\let\newline\\\arraybackslash\hspace{0pt}}m{#1}}
\newcolumntype{R}[1]{>{\raggedleft\let\newline\\\arraybackslash\hspace{0pt}}m{#1}}

\usepackage{longtable}
\usepackage{capt-of}
\usepackage{graphicx}

\usepackage{siunitx}
\DeclareSIUnit\electronvolt{e\kern-.05em V}
\DeclareSIUnit\parsec{pc}
\DeclareSIUnit\erg{erg}

\newcommand\beq{\begin{alignat}{1}}
\newcommand\eeq{\end{alignat}}

\usepackage{listings}
\begin{document}
\preprint{MIT-CTP/5390}

\title{Characterizing the Expected Behavior of Non-Poissonian Template Fitting}

\author{Luis Gabriel C. Bariuan}
\email{lbariuan@mit.edu}
\affiliation{Center for Theoretical Physics, Massachusetts Institute of Technology, Cambridge, MA 02139, U.S.A.}

\author{Tracy R. Slatyer}
\email{tslatyer@mit.edu}
\affiliation{Center for Theoretical Physics, Massachusetts Institute of Technology, Cambridge, MA 02139, U.S.A.}
\affiliation{The NSF AI Institute for Artificial Intelligence and Fundamental Interactions}

\begin{abstract} 
We have performed a systematic study of the statistical behavior of non-Poissonian template fitting (NPTF), a method designed to analyze and characterize unresolved point sources in general counts datasets. In this paper, we focus on the properties and characteristics of the \textit{Fermi}-LAT gamma-ray data set. In particular, we have simulated and analyzed gamma-ray sky maps under varying conditions of exposure, angular resolution, pixel size, energy window, event selection, and source brightness. We describe how these conditions affect the sensitivity of NPTF to the presence of point sources, for inner-galaxy studies of point sources within the Galactic Center excess, and for the simplified case of isotropic emission. We do not find opportunities for major gains in sensitivity from varying these choices, within the range available with current \textit{Fermi-LAT} data. We provide an analytic estimate of the NPTF sensitivity to point sources for the case of isotropic emission and perfect angular resolution, and find good agreement with our numerical results for that case.
\end{abstract}
\maketitle

\section{Introduction} \label{sec:Introduction}

Recent years have seen a number of efforts to apply photon pixel count statistics to gamma-ray data, in order to characterize populations of point sources (PSs) too faint to be individually detected at high significance (e.g. \cite{2011ApJ...738..181M, Lee:2014mza,Lee:2015fea, Linden:2016rcf, Lisanti:2016jub, Zechlin:2015wdz, Zechlin:2017wsy,Zechlin:2017uzo, Daylan:2016tia, Portillo_2017, Collin:2021ufc}). The general idea of these methods is to exploit the fact that an unmodeled PS population gives rise to non-Poissonian fluctuations in the number of photons per pixel, with ``hot spots'' corresponding to the locations of sources. Even if no individual hot spot is significant enough to be established as a PS with high probability, the distribution of fluctuations can be used to infer the properties of the population. These methods have been applied to characterize contributions to the extragalactic gamma-ray background (e.g.~\cite{Lisanti:2016jub, Zechlin:2017wsy, Zechlin:2017uzo}) and to study inner Galaxy PS populations (e.g.~\cite{Lee:2015fea, Linden:2016rcf, Calore:2021jvg}); they have also been applied to other datasets, e.g. crowded optical fields \cite{Portillo_2017} and high-energy neutrinos \cite{IceCube:2019xiu}. 

Initially these methods focused on the case of isotropic PS populations, which is likely to be a good approximation for all-sky background radiation generated by a large ensemble of faint extragalactic sources. However, subsequent studies \cite{Lee:2015fea, Daylan:2016tia, Collin:2021ufc}  extended this approach to the case of source populations with an arbitrary spatial distribution.

In this work we focus on one such method, {\it Non-Poissonian Template Fitting} (NPTF) \cite{Lee:2014mza, Lee:2015fea,Mishra-Sharma:2016gis}, which has been applied in a range of contexts but particularly to study the Galactic Center Excess (GCE) in public data from the {\it Fermi} Gamma-Ray Space Telescope (hereafter {\it Fermi}). The GCE is an extended and roughly spherical (not disk-like) source of GeV-scale gamma rays filling the region within $1.5 \text{ kpc}$ of the Galactic Center (GC) \cite{Goodenough:2009gk,Hooper:2010mq, Hooper:2011ti,Hooper:2013rwa,Daylan:2014rsa,Calore:2014xka,TheFermi-LAT:2015kwa}. 

The origin of the GCE has been the subject of active controversy for the past decade, with two explanations receiving the most attention. One possibility is that the GCE originates from diffuse particle dark matter (DM) undergoing annihilation (e.g. \cite{Goodenough:2009gk,Daylan:2014rsa,Karwin:2016tsw}), as the flux, energy spectrum, and spatial morphology of the GCE appear broadly consistent with a DM origin. If this hypothesis were confirmed, it would be a discovery of profound importance, representing the first evidence of non-gravitational interactions between DM and visible particles. However, the energy spectrum of the GCE also closely resembles that of gamma-ray pulsars observed by {\it Fermi}, and a number of studies have found that the spatial morphology of the GCE is a closer match to the stellar bulge than to a DM annihilation signal \cite{Macias:2016nev, Bartels:2017vsx, Macias:2019omb, Pohl:2022nnd}.\footnote{However, other recent studies \cite{DiMauro:2021raz,Cholis:2021rpp} have found the opposite preference; the result appears to be sensitive to how the Galactic background emission is modeled.} For these reasons, it seems plausible that the GCE represents the detection of a pulsar population in the Galactic bulge (e.g.  \cite{Abazajian:2012pn,Abazajian:2014fta,Hooper:2013nhl,Mirabal:2013rba,Calore:2014oga,Cholis:2014lta,Yuan:2014yda,OLeary:2015qpx,Ploeg:2017vai,Hooper:2018fih,Bartels:2018xom,Bartels:2018eyb}). If this population includes sources with brightness approaching the {\it Fermi} sensitivity threshold, then NPTF methods have the potential to characterize at least the bright end of this new population, and provide strong evidence against the DM hypothesis.

Previous NPTF studies have claimed evidence for a GCE source population comprised of relatively bright and rare PSs \cite{Lee:2015fea}, but recent studies have found that those claims may have been premature due to unaccounted-for systematic errors \cite{Leane:2019xiy,Chang:2019ars,Buschmann:2020adf,Leane:2020nmi,Leane:2020pfc}. Other analyses have found a preference for a significant diffuse emission component \cite{List:2020mzd, Calore:2021jvg}, although this does not exclude the pulsar hypothesis, since the sources might simply be too faint to be detected with current methods. At the same time, work on modeling the  pulsar population in the bulge has suggested that plausible pulsar luminosity functions could generate very few {\it Fermi} detected sources, while yielding an appreciable number of sources in the flux range potentially detectable by NPTF methods \cite{Ploeg:2020jeh, Gautam:2021wqn, Dinsmore:2021nip} or related approaches using machine learning \cite{List:2021aer,Mishra-Sharma:2021oxe}.

Given this uncertain situation, it is timely to understand how well NPTF can be expected to perform in detecting faint PS populations, and how this performance can be optimized by analysis choices. For example, many previous studies have chosen {\it Fermi} event selections to optimize angular resolution, at the cost of exposure. While several studies have explored the effect on their results of varying the event selection (e.g. \cite{Lisanti:2016jub, Leane:2020nmi, Leane:2020pfc}), this has not yet been done in a systematic way.

In this work, we systematically explore the ability of the public \texttt{NPTFit} algorithm (as described in Ref.~\cite{Mishra-Sharma:2016gis}) to reconstruct faint sources in simulated data, as a function of the instrument capabilities and analysis choices. We focus primarily on the analysis of the inner Milky Way, as relevant for the GCE, but also provide results for the simpler case where signal and background are both isotropic.

We begin in Sec.~\ref{sec:analyticforms} by discussing how we expect the likelihood ratio in favor of a point-source population to behave, in a simplified approximate context that can be treated analytically, by approximating some or all of the relevant Poisson distributions as Gaussian. This approximation is not expected to hold in detail in the cases of greatest interest to us, but it is helpful for building intuition.

In Sec.~\ref{sec:Methodology}, we then move on to our numerical study, starting by discussing the procedure by which we perform fits to the real {\it Fermi} data to derive reasonable baseline estimates for the properties of the background model and PSs. We use these results to generate simulated data that is similar to the true gamma-ray sky as observed by {\it Fermi}, using the public code \texttt{NPTFit-Sim}, a package designed to simulate populations of unresolved PSs.\footnote{\url{https://github.com/nickrodd/NPTFit-Sim}} In this section we also discuss our methodology for fitting to simulated data, and the test statistic we will use to describe the sensitivity of NPTF methods to faint sources.

In Sec.~\ref{sec:varyingdiffparams} we lay out the parameters we will vary in our simulations: exposure, angular resolution, energy window, pixel size, and source brightness. We describe the procedure for varying each of these parameters using \texttt{NPTFit} and \texttt{NPTFit-Sim}, including any associated modifications to the prior ranges.

In Sec.~\ref{sec:isotropic} we perform an initial analysis and comparison between simulated data and our analytic approximations, in the simplified scenario where the PS and smooth contributions to the gamma-ray sky are both isotropic.

We then move on to a full realistic inner Galaxy analysis; conduct variations of the various analysis parameters, singly and in combination; and present the (numerical) results in Sec.~\ref{sec:results}. In particular, we explore the individual effects of varying the exposure level and the point spread function (PSF), and map out the tradeoff when exposure level is increased (reduced) with the effect of worsening (improving) angular resolution, using the specific examples of {\it Fermi} event classes sorted by angular resolution. Modifying the energy window varies the effective exposure, the PSF, and also (in real data) the relative amplitude of the various background and signal components; we explore these effects independently. We then demonstrate the effect of varying the brightness of the PSs while keeping the total flux of the population constant (as appropriate for hypothetical source populations that explain the bulk of the GCE). Finally, we examine the question of the optimal pixel size for NPTF analyses, exploring both the sensitivity to faint PSs and accuracy of the parameter reconstruction. 

In Sec.~\ref{sec:conclusion} we summarize our results and discuss some implications for NPTF analyses of {\it Fermi} gamma-ray data in the inner Galaxy.

In Appendix~\ref{app:detailedmethodology} we present further details of our simulation parameters and fitting methodology; in Appendix~\ref{app:sharpSCF} we discuss the degree to which our source count functions model a single-brightness PS population;
in Appendix~\ref{app:isotropic} we show additional results for the simpler case where both signal and background are isotropic; and in Appendix~\ref{app:modela} we show the results of using an alternative Galactic diffuse model as the basis for our simulations.

\section{Analytic approximations for non-Poissonian template fitting}
\label{sec:analyticforms}

Let us begin by building some intuition for how the detectability of PSs is likely to scale in a NPTF-like setup. We will initially follow the approach of Ref.~\cite{Leane:2020pfc}, essentially replacing the Poisson distributions with Gaussians; this will be a good approximation when the number of sources per pixel and number of counts/source are both large, and can more generally provide qualitative insights into how various inputs affect the PS sensitivity.

Here we will compute likelihoods and likelihood ratios as a measure of sensitivity, whereas in the numerical analysis of later sections we will perform a Bayesian analysis and evaluate Bayes factors. The Bayesian evidence is an integral of the likelihood weighted by the priors, and so loosely speaking we expect them to have qualitatively similar properties under variations of the source brightness, exposure, etc. However, our expressions for the likelihood ratios should not a priori be expected to accurately approximate the Bayes factors, since Bayes factors incorporate information from the priors (including the number of free parameters in the model), and the likelihood ratios do not.\footnote{However, in practice, we will find that for our default choice of priors, the differences between the likelihood ratios and Bayes factors are small compared with other differences between the analytic and numerical results.} We summarize the key results within this section in Table~\ref{tab:likelihoodratio}.

\subsection{Pixel likelihood to observe $N$ photons} \label{subsubsec:settingupanalytic}

Let us first review some relevant results from Ref.~\cite{Leane:2020pfc}. Consider a simplified scenario where our PS population model predicts $n_0$ sources per pixel, and all sources are identical, with an expected number of photons per source of $s$. For the moment, we will ignore leakage out of the pixel due to the non-trivial angular resolution, but as a first approximation the effect of such leakage would be to reduce $s$. We are interested in calculating the probability to observe $N$ photons in a pixel. 

If we fix the number of observed sources (in a given pixel) to be $n$, then the total number of photons in the pixel will follow a Poisson distribution with mean $n s$. For $n s \gg 1$, we can approximate this distribution as a Gaussian with a mean and variance of $n s$, via the Central Limit Theorem. Then the probability to observe $N$ photons is given approximately by:
\begin{equation}
    P(N|\{n, s \}) = \frac{1}{\sqrt{2 \pi n s}} e^{-(N-n s)^{2}/(2 n s)}
    \label{eqn:p(n|nchis)}
\end{equation}
As a note, this expression can be thought of as a continuous probability density function (PDF), but also as a measure of the finite probability to observe $N$ photons by integrating the PDF over a bin of width $dN=1$ (i.e. the difference between adjacent values of $N$). Provided the PDF does not vary rapidly over the bin, this integral can simply be approximated by the value of the PDF at the center of the bin. We will use both interpretations of $P(N|\{n,s\})$ and similar quantities in the following calculations.

This distribution function is convolved with $P(n|n_{0})$, a distribution that describes the probability of drawing $n$ sources given that the expected number of sources is $n_{0}$. The resulting function, which we denote $P(N|{n_0, s})$, describes the likelihood of obtaining $N$ photons given that the number of sources is described by a Poisson distribution with an expectation value of $n_{0}$. If the number of sources is large, we can also approximate the distribution that describes the number of sources with a Gaussian with mean and variance $n_{0}$. Furthermore, the integrand is dominated by the region where $n s \approx N$, so we can set $n s \approx N$ except where $N- n s$ appears in an exponent. 

These approximations yield the following equation for the probability to observe $N$ photons given $n_0$ and $s$:

\begin{align}
    P(N|\{n_{0},s \}) &= \int dn P(N |\{n, s\}) P(n|n_{0}) \nonumber \\
    &\approx \int dn \frac{1}{\sqrt{2 \pi N}} e^{-(n s -N)^{2}/(2N)} \nonumber \\ & \times \frac{1}{\sqrt{2 \pi  n_{0}}}e^{-(n_{0}-n)^2/(2 n_{0})}
    \label{eqn:convolution}
\end{align}

The integral over $n$ can be performed analytically and takes a simple form, if we assume the peak in the integrand is sufficiently far away from the limits of integration that we can take those limits to $\pm \infty$ without affecting the result. Furthermore, around the peak of the probability distribution we have $N \approx n_{0} s$, so we can approximate $N\approx n_0 s$ except when $N - n_0 s$ appears in an exponent. These approximations yield:
\begin{equation}
\begin{split}
    P(N|\{n_{0},s \}) &\approx \frac{1}{\sqrt{2 \pi (N+n_{0} s^{2})}} e^{\frac{-(N-n_{0} s)^{2}}{2(N+n_{0} s^{2})}}\\
    &\approx \frac{1}{\sqrt{2 \pi n_{0}  s(1+ s)}} e^{\frac{-(N-n_{0}  s)^{2}}{2 n_{0} s (1+ s)}}
\end{split}
\label{eqn:integraresult}
\end{equation}

That is, under these approximations the probability of observing $N$ photons takes a Gaussian form (at least near the peak of the distribution), but with an inflated variance of $n_0 s (1 + s)$, a factor of $(1+s)$ greater than the expectation value of $n_0 s$.

If $s \ll 1$, our model corresponds to a very faint source population that should be indistinguishable from diffuse emission. In this case, we recover the standard Gaussian approximation to the Poisson distribution, with equal mean and variance of $n_0 s$, 

\begin{equation}
    P(N|\{n_{0}, s \}) \approx \frac{1}{\sqrt{2 \pi n_{0}  s}} e^{\frac{-(N-n_{0} s)^{2}}{2 n_{0}  s }}
\label{eqn:poissondistribution}
\end{equation}

Thus the characteristic feature of a PS population (within these approximations) is an enhanced variance, by a factor of $1+s$. 

In the event that the number of counts per source satisfies $s \gg 1$ but the number of sources per pixel $n_0 \lesssim 1$, a more refined approximation for the distribution may be useful (going beyond the results of Ref.~\cite{Leane:2020pfc}). If $i$ sources are drawn in a given pixel ($i$ being an integer), the number of counts from those sources will be Poisson-distributed with expectation $s i$; for $s\gg 1$ and $i > 0$, we can approximate each of these individual distributions as a Gaussian with mean and variance $s i$, so overall we have:
\begin{equation}
    P(N|\{n_{0}, s \}) \approx p_0(n_0) \delta_{N,0} + \sum_{i=1}^\infty p_i(n_0) \frac{1}{\sqrt{2 \pi s i}} e^{\frac{-(N- s i)^{2}}{2 s i}},
\label{eqn:doublepoissondistribution}
\end{equation}
where $p_i(n_0)$ is the Poisson probability of drawing $i$ sources when $n_0$ are expected.

\subsection{Likelihood ratio between models (Gaussian approximation)}
Now suppose the true underlying model for a given pixel yields a Gaussian distribution for $N$ with mean $X$ and variance $\sigma^2$. We wish to evaluate the expected log likelihood ratio between the correct model and an alternative model that predicts mean $Y$ and variance $\tau^2$. We will denote these models respectively as $(X,\sigma^2)$ and $(Y,\tau^2)$. This result has been computed previously in Ref.~\cite{Leane:2020pfc}; we review it here.

For context, the correct model might represent a linear combination of a PS population and a diffuse signal, while the alternative model allows only for a diffuse signal; the expected log likelihood ratio in this case then gives a measure of how well we will be able to exclude the all-diffuse model and thus detect the PS population. We will work out the case for general $(X,Y,\sigma^2,\tau^2)$ first, under the approximation where all the relevant probability distributions are Gaussian, and then apply this general result to several scenarios in which we might wish to detect PS populations.

For a single pixel, the probability of finding $N$ photons predicted by the model $(Y,\tau^2)$) is:
\begin{equation}
    \mathcal{L} = P(N|\{Y,\tau^2 \}) = e^{[-(N-Y)^{2}]/(2 \tau^{2})}/\sqrt{2 \pi \tau^{2}}
    \label{eqn:likelihoodeqn}
\end{equation}
corresponding to a log likelihood of $\ln{\mathcal{L}} = -\frac{(N-Y)^2}{2\tau^2} - \frac{1}{2} \ln{2\pi \tau^2}$. To get the expected value of the log likelihood with respect to the true model, we can integrate against the true distribution of $N$, i.e. $P(N|\{X,\sigma^2\})$. This yields \cite{Leane:2020pfc}:
\begin{equation}
    \left \langle \ln{\mathcal{L}}(Y,\tau^2) \right \rangle = -\frac{\left[(X-Y)^2 + \sigma^2\right]}{2\tau^2} - \frac{1}{2} \ln(2\pi\tau^2)
 \end{equation}
Note we use the $\langle \rangle$ notation generally to denote expected values with respect to the true model.

Now we diverge from Ref.~\cite{Leane:2020pfc}, which focused on determining the best-fit choice for $\tau^2$ given a discrepancy between $X$ and $Y$. Let us instead simply examine the expected $\Delta \ln\mathcal{L}$ between the fitted model $(Y,\tau^2)$ and the best-fit model $(X,\sigma^2)$, which is given by:
\begin{align}
    \left \langle \Delta \ln{\mathcal{L}} \right \rangle & \equiv
     \left \langle \ln{\mathcal{L}}(X,\sigma^2) - \ln{\mathcal{L}}(Y,\tau^2)  \right \rangle \nonumber \\
     & =\frac{\left[(X-Y)^2 + \sigma^2\right]}{2\tau^2} + \frac{1}{2} \ln(2\pi\tau^2) - \frac{1}{2} - \frac{1}{2} \ln(2\pi\sigma^2) \nonumber \\
    & = \frac{\left[(X-Y)^2 + \sigma^2\right]}{2\tau^2} - \frac{1}{2} \left[1 +  \ln\left(\frac{\sigma^2}{\tau^2}\right) \right]
 \end{align}
 
 If both models produce a very similar expected number of  photons, i.e. $X\approx Y$, and differ only in their variances, then this result can be simplified to:
 \begin{align}
    \left \langle \Delta \ln{\mathcal{L}} \right \rangle 
    & = \frac{\sigma^2}{2\tau^2} - \frac{1}{2} \left[1 +  \ln\left(\frac{\sigma^2}{\tau^2}\right) \right] \label{eqn:bayesfactor}
 \end{align}
 
 Note however that if $Y$ and $\tau^2$ are allowed to vary within certain limits or while satisfying certain conditions, then it is not guaranteed that the best-fit point lies at $Y=X$; if the global likelihood maximum (at $Y=X$, $\tau^2=\sigma^2$) cannot be attained, then the best-fit value of $Y$ will depend on the value of $\tau^2$ (and vice versa). Most simply, this can occur when the model is Poissonian, in which case $\tau^2$ is fixed to $Y$, but the data has non-Poissonian components and so $\sigma^2$ differs from $X$ in the true underlying model. A related scenario, studied in Refs.~\cite{Leane:2020nmi, Leane:2020pfc}, occurs when the model requires the same value of $Y$ in multiple pixels but the true underlying model varies across those pixels; this leads to a best-fit model variance $\tau^2$ that differs from the true underlying variance $\sigma^2$ (possibly leading to misattribution of the enhanced variance to a PS population).

\subsection{Variance between realizations (Gaussian approximation)}

In addition to working out the expected log likelihood ratio as a measure of sensitivity to incorrect modeling (such as attempting to describe PSs with a Poissonian template), it is helpful to understand the expected variability in this ratio between different realizations. In the limit where the number of pixels is large, the total $\Delta \ln\mathcal{L}$ for the image is the sum of many independent random variables ($\Delta \ln\mathcal{L}$ for each pixel), and so is expected to follow a Gaussian probability distribution by the Central Limit Theorem (even if the probability distribution for $\Delta \ln\mathcal{L}$ in a single pixel is highly non-Gaussian). Consequently, in this limit, we expect the distribution of the total $\Delta \ln\mathcal{L}$ (summed over pixels) to be well-characterized by its expectation value and variance.

As in the previous subsection, we will work out the result initially for general choices of the PDF parameters for the true and alternative hypotheses, $(X,Y,\sigma^2,\tau^2)$. We will then apply these results to specific scenarios, in particular where the true model (described by $(X,\sigma^2)$) includes a PS component but the alternative model (described by $(Y,\tau^2)$) does not.

We can estimate the variance of $\Delta \ln \mathcal{L}$ by evaluating $\text{Var}(\Delta\ln \mathcal{L}) \equiv \langle (\Delta \ln\mathcal{L})^2 \rangle - \langle \Delta \ln\mathcal{L} \rangle^2$. Let us first focus on the case  where $\sigma \gg \tau$ and $\left \langle \Delta \ln{\mathcal{L}} \right \rangle \gg 1$ and so the first term dominates in Eq.~\ref{eqn:bayesfactor}. This can occur, for example, where there is a bright PS population inducing a large variance $\sigma^2 \gg X$, which cannot be replicated by an alternative model based solely on diffuse emission with Poissonian statistics; in that sense this is a high-detectability limit.

Then using the estimates above and again taking $X\approx Y$, we find that:
\begin{align} \langle (\Delta \ln\mathcal{L})^2 \rangle &\approx \int dN \left[-\frac{(N-X)^2}{2\tau^2} + \frac{(N-X)^2}{2\sigma^2}\right]^2 \nonumber \\
& \times P(N|\{X,\sigma^2\}) \nonumber \\
& \approx \frac{3}{4} \left(\frac{\sigma}{\tau}\right)^4,\end{align}
and thus:
\begin{align} \text{Var}(\Delta\ln \mathcal{L}) & \approx \frac{3}{4} \left(\frac{\sigma}{\tau}\right)^4 - \frac{1}{4} \left(\frac{\sigma}{\tau}\right)^2 \nonumber \\
& = \frac{1}{2} \left(\frac{\sigma}{\tau}\right)^4.\end{align}
Thus we expect the standard deviation in this regime to be:
\begin{align}\text{std}(\Delta\ln \mathcal{L}) & \approx \frac{1}{\sqrt{2}}\frac{\sigma^2}{\tau^2} \nonumber \\
&\approx \sqrt{2} \langle \Delta \ln \mathcal{L}\rangle.\end{align}
We see that we generically expect the scatter in $ \Delta \ln\mathcal{L}$ (from a single pixel) to be of the same order as its expected value. When combining $n_\text{pix}$ pixels, the expectation value and variance are both enhanced by a factor of $n_\text{pix}$, so the standard deviation should be suppressed relative to the expectation value by a factor of $1/\sqrt{n_\text{pix}}$.

In this high-detectability, purely Gaussian case, there is actually a simple analytic expression for the full PDF of $\Delta \ln \mathcal{L}$, which we derive in detail in Appendix~\ref{app:PDFderivation}:
\begin{equation} P(\Delta \ln \mathcal{L} = x) = \frac{1}{\sqrt{\pi x \delta}} e^{-x/\delta}, \, x \ge 0, \label{eq:analyticPDF} \end{equation}
where $\delta \equiv (\sigma^2/\tau^2) - 1$. It can be readily checked that this distribution reproduces the expectation value and variance given above for $\delta \gg 1$. Note that this distribution is not at all Gaussian; however, as discussed above, combining a large number of pixels and summing their $\Delta \ln \mathcal{L}$ contributions is expected to give an approximately Gaussian PDF by the Central Limit Theorem.

If we instead consider the low-detectability case where $\tau^2 \approx \sigma^2$, i.e. $\delta = (\sigma^2/\tau^2) -1 \ll 1$, then we instead obtain:
\begin{align} \langle (\Delta \ln\mathcal{L})^2 \rangle & \approx \int dN \left[-\frac{(N-X)^2}{2\tau^2} + \frac{(N-X)^2}{2\sigma^2}\right.\nonumber \\
& \left. - \frac{1}{2}\ln \frac{\tau^2}{\sigma^2} \right]^2 P(N|\{X,\sigma^2\}) \nonumber \\
& \approx \delta^2 \int dN  \left[- \frac{(N-X)^2}{2\sigma^2} + \frac{1}{2} \right]^2 P(N|\{X,\sigma^2\}) \nonumber \\
& \approx \delta^2/2\end{align}
where we have used the approximation $\ln{(1+\delta)}\approx \delta$, and taken the limits of integration to $\pm \infty$.
In the same limit, 
\begin{align} \langle \Delta \ln\mathcal{L} \rangle &\approx \delta^2/4.\end{align}
Thus for $\delta \ll 1$, the first term dominates the variance and we have:
\begin{align} \text{Var}(\Delta\ln \mathcal{L}) & \approx \delta^2/2 \approx 2 \langle \Delta \ln\mathcal{L} \rangle.\end{align}
Thus in this case the square root of the variance is parametrically enhanced (by a factor of $1/\delta$) relative to the expectation value. The variance and expectation value are parametrically similar and will both be enhanced by a factor of $n_\text{pix}$ when multiple pixels are combined, and so in this regime the standard deviation (square root of the variance) should be of the same order as the square root of the expectation value.

Now we will apply these results to estimate the expected log likelihood ratio between a model containing PSs and one that omits them, when a real population of PSs is present in the data. This $\Delta \ln \mathcal{L}$ will tell us the confidence level with which we expect to be able to exclude the model with no PSs, and hence the confidence level for PS detection. It is similar to the metric we will use for sensitivity to a PS population in our numerical studies.

\subsection{Single component (100 \% PS emission)} \label{subsubsec:case1analytic}

Let us begin by assuming that the data is completely described by a PS population (of identical sources, as described above) without any contribution from a smooth background source. The PS emission has a mean and variance approximated by $(X,\sigma^{2}) = (N, N (1+s))$, where $s$ is the number of photons per source and $N$ the total number of photons. 

Let us consider the expected $\Delta \ln \mathcal{L}$ between the correct PS-based model, and a model that includes only smooth emission, but which correctly predicts the expected number of photons $N$. Such a smooth model must have equal mean and variance, so we must have $(Y,\tau^{2}) = (N,N)$.

Using Eq.~\ref{eqn:bayesfactor}, we plug in these parameters and obtain:

\begin{equation}
   \langle \Delta \ln{\mathcal{L}}\rangle \approx \frac{1}{2} \left[s - \ln\left(1+ s\right) \right]
    \label{eqn:case1naturallogbf}
\end{equation}
Note that {\it all} dependence on the total number of photons $N$ has canceled out; only the number of photons per source is relevant. In particular, this property ensures the likelihood ratio will go to 1 when $s \ll 1$ as required (since this corresponds to the limit of many very faint sources, at which point the smooth model is perfectly adequate), even if the number of sources is very large. Specifically, at small $s$ we have $\langle \Delta \ln{\mathcal{L}}\rangle \approx (s/2)^2$.

However, this behavior also has the perhaps-surprising implication that having {\it more} sources (and hence more photons) of fixed brightness in a single pixel neither increases nor decreases the PS sensitivity based on the pixel likelihood, at least once the numbers are large enough that the relevant likelihoods can be approximated as Gaussian. 

The leading order behavior of this function at large $s$ is $\langle \Delta \ln{\mathcal{L}}\rangle \approx s/2$, i.e. the log likelihood in favor of PSs grows linearly with the brightness of the sources. Since the number of photons seen from a given source is directly proportional to the exposure (i.e. time viewing the source multiplied by the effective area of the instrument), we expect that (at least in this background-free case) $\langle \Delta \ln{\mathcal{L}}\rangle$ will also grow linearly with exposure. The normalization factor here is also familiar. $2\Delta \ln\mathcal{L} \approx s$ is often used as a test statistic, whose square root translates to the significance in sigma; thus roughly speaking, we expect the detection significance of the PS population (measured in sigma) from a given pixel to approach $\sqrt{s}$ for large $s$.

If we do not impose the condition that the expected number of photons is $N$, we can maximize the likelihood for this model under the condition $Y=\tau^2$, obtaining:
\begin{equation}Y_\text{optimal} = \frac{1}{2} \left(\sqrt{4 X^2 + 4\sigma^2 + 1} -1\right).\end{equation}
For the case at hand, this yields $Y_\text{optimal} = \frac{1}{2} \left(\sqrt{4 N^2 + 4 N(1+s) + 1} -1\right)$. If $N \gg s$, $N \gg 1$ (i.e. the number of both sources and photons is large, consistent with our Gaussian approximations), then to a good approximation $Y_\text{optimal} \approx N$ and the estimates above should be reasonable. It is also true in practice, in NPTF analyses of the GCE, that the total photon flux associated with the best-fit GCE model is typically very similar when comparing the fits with and without a model for GCE PSs (e.g. \cite{Lee:2015fea}).

\subsection{Generalization to arbitrary ratio of PS and smooth emission}

Now let us consider the scenario in which a fraction $k$ of the emission is associated with PSs and the remainder with smooth emission. We seek to evaluate $\langle \Delta \ln{\mathcal{L}}\rangle$ between the best-fit model (corresponding to the truth) and the model with only smooth emission.

The total number of predicted photons is the sum of the predicted photons associated with each component. The sum of two Gaussian-distributed random variables is also Gaussian-distributed, with mean (variance) given by the sum of the means (variances) for the individual distributions. Thus within our approximations, the best-fit model (matching the truth) has a Gaussian probability distribution for the number of photons $N$ with parameters $(X,\sigma^{2})=(kN+(1-k)N,kN(1+ s)+(1-k)N)$. 

The model with only smooth components that matches the total number of photons has (as in our previous example) $(Y,\tau^{2}) = (N,N)$. 

Using Eq.~\ref{eqn:bayesfactor}, we obtain:

\begin{equation}
   \langle \Delta\ln \mathcal{L}\rangle \approx 
   \frac{1}{2}[ks-\ln{(1+ks)}]
    \label{eqn:lnBFcase2}
\end{equation}

Thus the effect of a non-zero background fraction on the sensitivity is equivalent to rescaling the photon flux of individual sources. In this case, the change in scaling behavior from $\langle \Delta\ln \mathcal{L}\rangle \propto s^2$ to $\langle \Delta\ln \mathcal{L}\rangle \propto s$ will occur parametrically around $k s \sim 1$. It is worth noting that if there are a large number of pixels, a significant detection may be consistent with $k s \ll 1$ from every individual pixel, and in this case we should expect a faster-than-linear scaling of the log likelihood ratio with increasing $s$ (or $k$).

\subsection{A more accurate probability distribution: accounting for rare sources}
\label{sec:accuratetrueP}

We can also compute the expected value of the likelihood ratio and its variance, between two Gaussian models, if the underlying ``true'' probability distribution is given by Eq.~\ref{eqn:doublepoissondistribution}, for the case $n_0 \lesssim 1$ where the Gaussian approximations break down. For a Gaussian distribution that describes the expected counts with mean $X$ and variance $\sigma^2$, we find:

\begin{align} \langle \ln \mathcal{L}\rangle & = \int dN \left[-\frac{(N-X)^2}{2\sigma^2} - \frac{1}{2} \ln(2\pi\sigma^2)\right] P(N|n_0, s) \nonumber \\
& = p_0(n_0) \left[-\frac{X^2}{2\sigma^2} - \frac{1}{2} \ln(2\pi\sigma^2)\right] \nonumber \\
& + \sum_{i=1}^\infty p_i(n_0) \int dN \left[-\frac{(N-X)^2}{2\sigma^2} - \frac{1}{2} \ln(2\pi\sigma^2)\right] \nonumber \\
&\times \frac{1}{\sqrt{2\pi i s}} e^{-(N - i s)^2/2 i s} \nonumber \\
& = -\frac{1}{2} \sum_{i=0}^\infty p_i(n_0) \left[\frac{(i s -X)^2+ i s}{\sigma^2} + \ln(2\pi\sigma^2)\right], \end{align}
where again we have made the approximation of taking the limits of integration to $\pm \infty$, relying on $i s \gg 1$ for $i \ge 1$ (so that the Gaussians are centered well away from the limits of integration). Now the infinite sums over $i$ can be computed by using the fact that the Poisson probabilities $p_i(n_0) = n_0^i e^{-n_0}/i!$ satisfy $\sum_{i=0}^\infty p_i(n_0) = 1$. In particular, by relabeling dummy indices the following identities can easily be proved:
\begin{align} \sum_{j=0}^\infty j p_j(n_0) & = n_0,\nonumber \\
\sum_{j=0}^\infty j^2 p_j(n_0) & = n_0 (n_0 + 1). \nonumber \\
\sum_{j=0}^\infty j^3 p_j(n_0) & = n_0^3 +3 n_0^2 + n_0, \nonumber \\
 \sum_{j=0}^\infty j^4 p_j(n_0) & = n_0^4 + 6 n_0^3 + 7 n_0^2 + n_0. \label{eqn:poissonidentities}
\end{align}

Applying these results we find:
\begin{align} \langle \ln \mathcal{L}\rangle & = - \frac{1}{2} \left[\ln(2\pi \sigma^2) + \frac{X^2 + s^2 n_0 (n_0 + 1) + s n_0 (1 - 2 X)}{\sigma^2}\right] \end{align}

In particular, if we hold the variance constant then the likelihood is maximized for $X=n_0 s$, and if we set $X=n_0 s$ (i.e. the model matches the expected total number of photons), then we obtain:
\begin{align} \langle \ln \mathcal{L}\rangle & = - \frac{1}{2} \left[\ln(2\pi \sigma^2) + \frac{n_0 s(1+s)}{\sigma^2}\right] \end{align}
The likelihood is then maximized for $\sigma^2 = n_0 s (1+s)$, which is the same variance we found when we directly approximated the probability distribution for the point-source population as Gaussian.

If we examine the expected $\Delta \ln \mathcal{L}$ between this best-fit Gaussian model and a Gaussian model with $\sigma^2 = X = n_0 s$ (representing a purely diffuse signal), we find:
\begin{align} \langle \Delta \ln \mathcal{L}\rangle & = - \frac{1}{2} \left[\ln(1+s) + \frac{n_0 s(1+s)}{n_0 s(1+s)} - \frac{n_0 s(1+s)}{n_0 s}\right] \nonumber \\
& = \frac{1}{2} \left[s - \ln(1+s)\right]  \end{align}
Remarkably, this is exactly the same result we found under the Gaussian approximation for the underlying probability distribution (Eq.~\ref{eqn:case1naturallogbf}), suggesting that this result is quite robust even when the assumptions needed to justify the Gaussian approximation break down. We will see in future sections that this result works fairly well to explain scaling relationships for the (numerically computed) sensitivity as we vary the properties of the sources and diffuse background.

As previously, we can generalize to the case where PSs constitute a fraction $k$ of the total emission, so the total expected photon count is $n_0 s/k$ with an expected number of $(1-k) n_0 s/k$ photons originating from diffuse emission. In this case the probability distribution for $N$ given in Eq.~\ref{eqn:doublepoissondistribution} must be updated accordingly. As previously, we approximate the probability distribution for the number of photons from diffuse emission as a Gaussian with mean and variance $(1-k) n_0 s/k$; for each choice $i$ for the number of sources drawn, the emission from sources (mean and variance $s i$) can be added to that from diffuse emission by the usual prescription for the sum of normally-distributed random variables (i.e. the means and variances add). Thus the overall distribution becomes:
\begin{align} & P(N|n_0, s, k) \approx \nonumber \\
& \sum_{i=0}^\infty p_i(n_0) \frac{e^{-\{N- s [i + n_0(1/k- 1)]\}^2/2 s [i + n_0(1/k- 1)]}}{\sqrt{2 \pi s [i + n_0(1/k- 1)]}}, \label{eqn:doublepoissonmixed}\end{align}
where as previously $p_i(n_0)$ is the Poisson probability of drawing $i$ sources when $n_0$ are expected.

Under this distribution, if we compute the expected likelihood of a Gaussian model with mean $X$ and variance $\sigma^2$, we find (by the same methods as previously):
\begin{align} \langle \ln \mathcal{L}\rangle  
& =  - \frac{1}{2} \left[ \frac{1}{\sigma^2} \left(X^2 - \frac{2 X n_0 s}{k} \right. \right. \nonumber \\
& \left. \left. + \frac{n_0 s(k + s (k^2 + n_0))}{k^2}  \right) + \ln(2\pi \sigma^2) \right]  \end{align}
For fixed $\sigma^2$, this is maximized for $X=n_0 s/k$ (as expected, when the model matches the total number of counts); if we fix $X=n_0 s/k$, then we obtain:
\begin{align} \langle \ln \mathcal{L}\rangle = - \frac{1}{2} \left[ \frac{1}{\sigma^2} \left(\frac{n_0 s (ks + 1)}{k}  \right) + \ln(2\pi \sigma^2 ) \right].  \end{align}
The expected log likelihood difference between the purely diffuse Gaussian model with $\sigma^2=X= n_0 s$ and the Gaussian model with $\sigma^2=(n_0 s / k) (k s + 1)$ (matching our previous prescription in the case of mixed PS and smooth emission) is then given by,
\begin{align} \langle \Delta \ln \mathcal{L} \rangle = \frac{1}{2} \left[k s + \ln \frac{1}{1+k s} \right], \end{align}
exactly as previously.

However, while the expected log likelihood is unchanged by shifting to this modified probability distribution, the variance differs. Working in the limit where the log terms in the delta log likelihood can be ignored, let us examine the variance of the delta log likelihood between the Gaussian models with  $\sigma^2= n_0 s$ and $\sigma^2=(n_0 s / k) (k s + 1)$. In both cases we take $X=n_0 s/k$. Then using the identities in Eq.~\ref{eqn:poissonidentities}, we obtain:
\begin{align}
\text{Var}(\Delta \ln \mathcal{L}) & =  \langle (\Delta \ln \mathcal{L})^2 \rangle - \langle \Delta \ln \mathcal{L} \rangle^2 \nonumber \\
 & \approx k^2 \frac{s(s+6) + 3}{4 n_0} \left( \frac{ks}{k s + 1}\right)^2 + \frac{1}{2} (k s)^2 \nonumber \\
 & \rightarrow \langle \Delta \ln \mathcal{L}\rangle^2 \left(\frac{1}{n_0} + 2 \right), \quad k s \gg 1 .  \end{align}

In particular, we observe that there is now an additional term in the variance which scales as $1/n_0$. Consistently with our previous calculation, this term will be negligible when $n_0 \gg 1$ and our original Gaussian approximation holds, but it can lead to a significant enhancement to the variance when $n_0\ll 1$. In particular, for $k s \gg 1$ and $n_0 \ll 1$, we expect that the variance over the full dataset can be approximated as:
\begin{align}\text{Var}(\Delta \ln \mathcal{L})_{\text{overall}} & \approx n_\text{pix} \langle \Delta \ln \mathcal{L}\rangle^2_\text{per pixel} /n_0 \nonumber \\
& \approx \langle \Delta \ln \mathcal{L}\rangle^2_\text{overall} /n_0 n_\text{pix}, \nonumber \\
\Rightarrow \text{std}(\Delta \ln \mathcal{L})_{\text{overall}} & \approx \frac{\langle \Delta \ln \mathcal{L}\rangle_\text{overall}}{\sqrt{n_0 n_\text{pix}}}.\end{align}
Thus we see that in this case, rather than the suppression of $1/\sqrt{n_\text{pix}}$ that we found earlier (for the high-$k s$ case), instead the suppression is only $1/\sqrt{n_\text{tot}}$, where $n_\text{tot}=n_0 n_\text{pix}$ is the total number of PSs in the image.

Broadly speaking, the standard deviation in $\Delta \ln\mathcal{L}$ is always related to $\langle \Delta \ln \mathcal{L}\rangle$ by a factor of $1/\sqrt{A}$, but $A$ can be either the number of pixels, the number of PSs (when the number of sources per pixel is small), or the test statistic $\langle \Delta \ln \mathcal{L}\rangle$ itself (when the contribution to the test statistic per pixel is small). In the examples we have checked, it is always the smallest of these three parameters that dominates the variance, which is intuitively sensible. 

Note that in particular this means the variance can be much larger than one might naively estimate from the square root of the test statistic; if the number of sources is only $\mathcal{O}(100)$, then the variance in the test statistic will be consistently at the $\mathcal{O}(10\%)$ level even if the sources are bright and the significance of detection is very high. Furthermore, we have so far neglected contributions to the variance from the width of the source count function (SCF) (which will modify the effective $s$ entering these calculations from realization to realization, and hence increase the variance), the presence of a non-zero point spread function (likewise), and cross-talk and degeneracies with other background components.  

\begin{table*}
    \begin{tabular}{lcc}
    \hline
        & $\braket{\Delta \ln{\mathcal{L}}}$ & Var$(\Delta \ln{\mathcal{L}})$ \\
    \hline
    \textit{General Gaussian Model Comparison}\\
    \hline
    \vspace{2mm}
    High Detectability $\delta \gg 1 $ & $\displaystyle \frac{1}{2}[\delta-\ln{(1+\delta)}]$   & $2 \braket{\Delta \ln{\mathcal{L}}}^{2}$ \\
    \vspace{2mm}
    Low Detectability $\delta \ll 1 $ & $\displaystyle \frac{\delta^2}{4}$  &   $ 2 \braket{\Delta \ln{\mathcal{L}}}$  \\
    \hline
    \textit{PS Signal + Diffuse Background, $ks \gg 1$}\\
    \hline
    \vspace{2mm}
    Gaussian probability distribution & $\displaystyle \frac{1}{2} [ks-\ln{(1+ks)}]$  & $2 \braket{\Delta \ln{\mathcal{L}}}^{2}$  \\
    \vspace{2mm}
    More accurate probability distribution: accounting for rare sources& $\displaystyle \frac{1}{2} [ks-\ln{(1+ks)}]$  & $\displaystyle \braket{\Delta \ln{\mathcal{L}}}^{2} \left (2+\frac{1}{n_{0}} \right) $  \\
    \botrule

    \end{tabular}
    \caption{Summary of $\braket{\Delta \ln{\mathcal{L}}}$ and Var$(\Delta \ln{\mathcal{L}})$ for a single pixel (as calculated in Sec.~\ref{sec:analyticforms}), where the log likelihood difference is evaluated between two Gaussian models for $P(N)$ with parameters $(X,\sigma^2)$ and $(Y,\tau^{2})$. In the upper part of the table, the true model is assumed to be the Gaussian with parameters $(X,\sigma^2)$, and we consider two regimes characterized by the detectability parameter $\delta \equiv \sigma^2/\tau^2 - 1$. In the lower part of the table, we rewrite $\sigma,\tau,X$ and $Y$ in terms of parameters describing the PS population (see text), where $k$ denotes the proportion of the total emission that is attributed to the PS population, $s$ denotes the expected number of photons per source, and $n_{0}$ is the number of sources per pixel. In this part of the table we assume $\delta = ks \gg 1$. In the final line, we furthermore employ a more accurate approximation for the true model for $P(N)$ rather than assuming it to be Gaussian (see Sec.~\ref{sec:accuratetrueP}).
    \label{tab:likelihoodratio}}
\end{table*}

\subsection{Implications for analysis choices}

If the overall number of photon counts increases, due to increased exposure (i.e. increased observation time or effective area), the signal fraction $k$ remains constant while $s$ varies linearly. Consequently, we expect $\langle \Delta \ln \mathcal{L}\rangle$ to depend linearly on exposure for sufficiently large $s$, with the transition from quadratic to linear scaling beginning around $s \sim 1/k$.

Suppose a non-zero angular resolution for the instrument causes the expected number of photons from a single source in a pixel to be reduced, due to leakage into neighboring pixels. Then $\langle \Delta \ln \mathcal{L}\rangle$ will be reduced by the same factor, in the regime where the delta log likelihood scales linearly with $s$. The signal fraction $k$ should not be affected by this leakage unless the overall distribution of either the signal or background varies rapidly relative to the angular resolution scale; if there is such a rapid variation, there may also be a correction corresponding to the change in $k$. 

The fact that the sensitivity depends only on $ks$ suggests that it is generally more important to have a low background fraction (high $k$) than a high density of sources (since the latter has no effect on the expected sensitivity in the regimes of validity of our analytic approximations). This suggests that the sensitivity is likely to be dominated by pixels where the expected PS signal is brightest as a fraction of all diffuse backgrounds (which may not be the pixels with the largest number of sources).

We have derived these results for the contribution to $\langle \Delta \ln \mathcal{L}\rangle$ from a single pixel, but the overall log likelihood is simply the sum of the results for the individual pixels. We can thus apply these results even to the realistic case where the background and signal models can have quite different spatial distributions: we simply calculate the appropriate $k$-value in each pixel and estimate the contribution to $\langle \Delta \ln \mathcal{L}\rangle$ accordingly. Also note that the smooth model can be arbitrarily complicated; the only information we have used is that it has Poissonian statistics.

\section{Inputs and methodology for numerical calculations} \label{sec:Methodology}
\subsection{Data selection} \label{subsec:DataSelection}
To calibrate our simulations to the real gamma-ray sky, we employ eleven years of the Pass 8 public \textit{Fermi} data. The data were collected over 573 weeks from August 4, 2008 to June 19, 2019. To employ the most stringent cosmic-ray rejection criteria, we restrict our selection to the ULTRACLEANVETO event class. For most tests, except those that varied energy range, we limited the energy range to $2 - 20$ GeV, following the default in \texttt{NPTFit} and previous NPTF analyses. We restrict ourselves to an analysis of the top three quartiles of the data graded by angular resolution, as this provides enough range in angular resolution to explore the tradeoff with exposure, and the angular resolution degrades significantly in the bottom quartile.

\subsection{\texttt{NPTFit} scan setup} \label{subsec:NPTFit}

We employ \texttt{v.0.2} of \texttt{NPTFit}, together with \texttt{MultiNest}, a Bayesian inference tool that implements a nested sampling algorithm \cite{Mishra-Sharma:2016gis,Feroz:2008xx}. For fits of the simulated data, the number of live points is described within the individual procedure sections; when not otherwise specified we used nlive=100. At each experiment, we checked the recovered evidences at different nlive values to determine if the scans were reasonably converged. We found that changes in $\langle \ln \text{BF}\rangle$ at nlive values beyond 100 were consistently very small, and thus the scans are well-converged.

When fitting to simulated data, our region of interest (ROI) is centered on the GC and has a radius of $15^{\circ}$. We exclude the band with galactic latitude $|b| < 2^{\circ}$. This ROI is chosen for computational efficiency and to minimize contamination from background emissions, while preserving sensitivity to the GCE, motivated by a recent study finding that sensitivity to GCE PSs plateaus for ROIs with radii between $15^{\circ}$ and $20^{\circ}$ \cite{Buschmann:2020adf}. Our expectation is that shifting to a modestly different ROI would not significantly affect the scaling with exposure, angular resolution, etc, that we study in this work, although the overall sensitivity would change and so should not be compared directly between analyses with different ROIs.

During \texttt{NPTFit} scans, the non-Poissonian components of the sky maps must be exposure corrected at each pixel (a computationally-costly process) since the exposure map (a map that reflects the duration of observation and the stringency of data selection) is non-uniform \cite{Mishra-Sharma:2016gis,Feroz:2008xx}. In order to optimize computational efficiency (and consistent with the recommendations in \texttt{NPTFit}), we set $\text{nexp} = 5$ to divide the ROI into $5$ distinct regions within which the exposure is treated as uniform. 

\subsection{Modeling the gamma-ray sky}

We conducted a Bayesian \texttt{NPTFit} analysis of the (real data) \textit{Fermi} sky map at nlive = 500, modeling the sky as a linear combination of spatial templates characterized by parameters and prior distributions that are described in Appendix~\ref{app:detailedmethodology}. For this analysis, and subsequent fits to real data (which were used only to choose parameters for the subsequent simulations), we extended the radius of the ROI from $15^\circ$ to $30^\circ$, but retained the mask of the Galactic plane (consistent with defaults in \texttt{NPTFit}). Smooth/diffuse templates were included for the \textit{Fermi} Bubbles (``Bub"), smooth isotropic emission (``Iso"), smooth GCE (``GCE"), and Galactic diffuse emission (``Dif"). Templates were included for PS populations associated with the GCE (``GCE PS"), isotropic / extragalactic sources (``Iso PS"), and the Galactic disk (``Disk PS"). Smooth/diffuse templates each have one associated parameter, $A_{\text{smooth}}$, controlling their overall normalization in the model; PS population templates (hereafter ``PS templates'') are associated with an overall normalization parameter $A_{\text{PS}}$ which controls the number of sources, and with a SCF which describes the number of sources as a function of their flux. We use a singly-broken power law model for the SCF, as is the default in \texttt{NPTFit}:
\begin{equation}
    \frac{dN}{dS} = A_{\text{PS}} T_{\text{PS}}
    \begin{cases}
    \left ( \frac{S}{S_{b}} \right )^{-n_{1}}  & S \geq S_{b} \\
    \left ( \frac{S}{S_{b}} \right )^{-n_{2}} & S \le S_{b}
    \end{cases},
    \label{eqn:scf}
\end{equation}
where $A_{\text{PS}}$ is an overall normalization factor, and $T_{\text{PS}}$ is the position-dependent template with the fixed normalization given in the \texttt{NPTFit} code (see Appendix \ref{app:detailedmethodology} for details). Note the parameter $S_b$ controls the expected number of photon counts per source at the position of the break in the power law. The expected number of sources in a given pixel is then set by:
\begin{equation} N_\text{tot} = A_{\text{PS}} T_{\text{PS}} S_b \left(\frac{1}{n_1 - 1} + \frac{1}{1 - n_2}\right)\label{eq:nsource} \end{equation}
whereas the expected number of photons is set by:
\begin{equation} S_\text{tot} = A_{\text{PS}} T_{\text{PS}} S_b^2 \left(\frac{1}{n_1 - 2} + \frac{1}{2 - n_2}\right). \label{eq:nphotons} \end{equation}

Note that in the main text of the paper we employ the default Galactic diffuse emission model from \texttt{NPTFit}, constructed from the {\it Fermi} Collaboration's \texttt{p6v11} diffuse model. This model is known to have features that can bias the results \cite{Buschmann:2020adf} when it is used directly to reconstruct PS populations from the real data; however, it should provide a reasonable description of the data when we are only interested in constructing and analyzing simulations (where the model is correct by construction). To check this assertion, in Appendix~\ref{app:modela} we recalculate our results with a different Galactic diffuse emission model and comment on the differences. Either of these Galactic diffuse emission models reconstruct the GCE as being $100\%$ PSs, with the smooth GCE component being negligible: consequently, our simulations will generally explore the sensitivity of NPTF methods to a PS population bright enough to explain the full GCE. (However, note that there are other models of the Galactic foregrounds where the flux attributed to the smooth GCE component is not negligible \cite{Buschmann:2020adf,Leane:2020nmi}.)

After performing this fit, we extracted the posterior median parameters associated with each template, which were then used as the baseline inputs to  simulations for the rest of the paper. These simulation parameters are displayed in Table \ref{tab:fitparam} in Appendix~\ref{app:detailedmethodology}. Note in particular that (consistent with previous NPTF studies) the inferred shape of the SCF for the GCE PS is quite sharply peaked around $S_b$, so we will generally be simulating GCE PS populations where the PSs all have roughly the same brightness as observed at Earth (fixed by $S_b$). This is likely {\it not} a realistic luminosity function, but serves as a convenient basis for understanding the sensitivity of NPTF methods. We discuss the sharpness of the SCF peak further in Appendix~\ref{app:sharpSCF}.

\subsection{Producing simulated sky maps}
\label{subsec:skymapsimulations}

For each template, we generated realizations based on the posterior median parameters from the real data. For the smooth/diffuse emission components, we performed a Poisson draw from the associated template (with normalization given by the simulation parameters taken from the fit to real data).  To obtain realizations of PS populations, we employed \texttt{NPTFit-Sim}.

To obtain the full skymaps, the individual components were summed. All skymaps were binned using \texttt{HEALPix}, a package designed to allow equal-area pixelization of the sky \cite{Gorski:2004by}. The nside value controls the pixel size, with the sky having a total of $12\times \text{nside}^2$ equal-area pixels. By default we set nside to 128, which is also the \texttt{NPTFit} default, and corresponds to roughly a $~0.5^{\circ}$ mean spacing between individual pixel centers in the region toward the GC. For nside=128, there are 2808 pixels within our ROI.

\subsection{Sensitivity figure of merit} \label{subsec:sensitivity}

A \texttt{NPTFit} analysis returns posterior probability distributions for each of the parameters, and an estimate of the overall Bayesian evidence for the model. Comparing two \texttt{NPTFit} analyses, with different template choices, allows us to evaluate the Bayes factor (BF) between the two scenarios, as the ratio of their evidences. In particular, we can define the sensitivity to a GCE PS population in terms of the BF in favor of a model that contains the complete set of templates (Dif, Bub, Iso, GCE, GCE PS, Disk PS, Iso PS) compared with a model that excludes the GCE PS template. A high value of this BF corresponds to a high-significance detection of the GCE PS template, over and above the smooth GCE template. For convenience, we will generally work with $\ln{\text{BF}}$ rather than the BF itself. Where $\text{BF} \lesssim 1$ and so $\ln\text{BF}$ is negative, there is no detection of GCE PSs. 

The BF directly gives the ratio of Bayesian probabilities that the model with the GCE PS template is correct, compared to the model without that contribution. For those more accustomed to frequentist statistics, it may be helpful to think of the BF as comparable to a likelihood ratio $\mathcal{L}_1/\mathcal{L}_2$, with additional terms that penalize models with more degrees of freedom. In this sense $2 \ln{\text{BF}}$ is broadly analogous to the commonly-used test statistic $2 \Delta \ln{\mathcal{L}}$, which for a likelihood that is Gaussian near its maximum ($\mathcal{L}(x) \propto e^{-x^2/(2\sigma^2)}$) can be written as $2 \Delta \ln{\mathcal{L}} \approx (x/\sigma)^2$, and thus can be thought of as the ``number of sigma'' squared associated with the deviation from the best-fit point.

The $\ln \text{BF}$ in favor of a GCE PS population can vary widely between realizations. For our main figure of merit for sensitivity, we will use the expected value of $\ln \text{BF}$ obtained by taking the average across realizations, $\langle \ln \text{BF} \rangle$, although we will also show the scatter between realizations. 

\section{Procedures for parameter variation}
\label{sec:varyingdiffparams}

Within each subsection below, we describe the general procedure for varying different inputs: exposure, angular resolution, source brightness, and pixel size. We describe the method for adjusting parameters in the simulation of skymaps as well as how to account for these variations through the priors when analyzing the skymaps using \texttt{NPTFit}. If the test involves combinations of these variations, then the priors must be modified by simultaneously implementing the adjustment factors to the priors for each alteration performed. 

\subsection{Exposure}
\label{subsec:varyingexposue}

Although \textit{Fermi} is a space-based telescope, it does not observe every part of the sky simultaneously. As a result, an exposure map is needed to keep track of how long \textit{Fermi} observed a particular region of the sky and with what effective area. The exposure map provided by the \textit{Fermi}-LAT Collaboration and implemented in \texttt{NPTFit} has units of $\text{cm}^{2}\text{s}$ \cite{Mishra-Sharma:2016gis}.

Increasing the amplitude of the exposure map could describe longer observations with \textit{Fermi} or less stringent cuts on photons as part of the data selection. We define an exposure rescaling factor $\chi$, which allows us to vary the intensity of the exposure map through a scalar multiplicative factor, hence rescaling the expected number of photons present in simulated data. In our baseline case, $\chi=1$. In general, the exposure rescaling could be position-dependent (e.g. corresponding to longer observations of only part of the ROI). However, we expect such position-dependent variations to be modest for the inner Galaxy region, as the size of our ROI is smaller than the field of view of {\it Fermi}.

We implemented the variation of exposure in the simulated data by modifying the template parameters as follows. For smooth/diffuse templates, the template normalization $A_{\text{smooth}}$ is multiplied by the rescaling factor $\chi$, since $A_{\text{smooth}}$ determines the mean photon counts within each pixel. For non-Poissonian templates, when the other parameters are held fixed, $A_{\text{PS}}$ determines how {\it many} sources are present, which is not a function of exposure. Therefore, we instead multiply the counts break $S_b$ by $\chi$, as $S_{b}$ controls the expected number of photons emitted by a source lying at the break in the SCF; as discussed previously, for the fits we perform, $S_b$ corresponds to the typical number of photons per source. To ensure that the total number of sources does not change, we divide $A_{\text{PS}}$ by $\chi$ following Eq.~\ref{eq:nphotons}. 

When performing \texttt{NPTFit} analyses on these modified skymaps, the input exposure map must be multiplied by $\chi$. Furthermore, we adjust the range of priors governing $A_{\text{smooth}}$ for Poissonian sources and $A_{\text{PS}}$ and $S_{b}$ for non-Poissonian sources, so that they correspond to the same underlying physical emission parameters as in the original $\chi=1$ analysis. For example, the boundaries of a uniform prior on $\log A_{\text{smooth}}$ are shifted by $+\log\chi$; the boundaries of a log prior on $\log A_{\text{PS}}$ are shifted by $-\log\chi$; and the boundaries of the linear prior on $S_b$ are multiplied by $\chi$. (The original values of all priors are displayed in Table~\ref{tab:priors} in Appendix~\ref{app:detailedmethodology}.) For simulated data, the number of live points we utilized in the scans is nlive=300. We checked that the relative changes in the recovered evidences (under variations to nlive) are negligible for all individual realizations.

\subsection{Angular resolution}
\label{subsec:methodangularresolution}

Angular resolution, characterized by the Point Spread Function (PSF), represents how well a telescope such as \textit{Fermi} is able to reconstruct the original direction of a detected photon.  A non-delta-function PSF represents an uncertainty in the direction of a photon's origin. As a result, the image produced of a photon source is ``smeared" across one or more pixels. Since the \texttt{NPTFit} implementation does not account for correlations between neighboring pixels (see e.g. \cite{Collin:2021ufc} for a discussion), this smearing has the potential to bias the recovered SCF. 

Modifications to the photon direction reconstruction, or construction of future gamma-ray telescopes, may allow for better angular resolution (equivalently, a narrower PSF) than {\it Fermi} can currently achieve. However, even within the {\it Fermi} dataset photons can be separated by the quality of their directional reconstruction, allowing us to improve angular resolution at the cost of exposure. Specifically, {\it Fermi} photons are divided into four quartiles ranked by angular resolution, and separate PSF estimates are provided for each of the quartiles. Furthermore, lower-energy photons have intrinsically worse angular resolution, so a cut on photon energy has the effect (among others) of modifying the effective PSF. 

\textit{Fermi}'s PSF is modeled by a pair of King functions (defined in Eq.~\ref{eqn:kingfunction}) and is characterized by a set of several parameters. The PSF is approximately Gaussian near the core, with larger non-Gaussian tails. Eq.~\ref{eqn:fermipsf} displays the full functional form of  \textit{Fermi}'s PSF. \footnote{\url{https://fermi.gsfc.nasa.gov/ssc/data/analysis/documentation/Cicerone/Cicerone_LAT_IRFs/IRF_PSF.html}} 

\begin{equation}
    P(x,\vec{\alpha}_{P})=f_{\text{core}}K(x,\sigma_{\text{core}},\gamma_{\text{core}})+(1-f_{\text{core}})K(x,\sigma_{\text{tail}},\gamma_{\text{tail}})
    \label{eqn:fermipsf}
\end{equation}
\begin{equation}
    K(x,\sigma,\gamma) = \frac{1}{2 \pi \sigma ^{2}} \left (1 - \frac{1}{\gamma}\right) \left [1+\frac{x^{2}}{2 \gamma \sigma^{2}} \right ]^{-\gamma}
    \label{eqn:kingfunction}
\end{equation}
Here $x$ is a rescaled distance from the center of the source, with an energy-dependent scale factor $S_p(E)$: 
\begin{align}
    x & = \frac{\delta p}{S_{p}(E)} \nonumber \\
    \delta p & = 2 \sin^{-1} \left(\frac{|\hat{p}^\prime - \hat{p}|}{2}\right),
    \label{eqn:xspepsf}
\end{align}
where $\hat{p}$ and $\hat{p}^\prime$ are the unit vectors corresponding respectively to the true and reconstructed directions of the photon. The parameters that define the PSF ($S_p(E)$, $f_\text{core}$, and $\gamma$ and $\sigma$ for the two King functions in Eq.~\ref{eqn:fermipsf}) are provided with the \textit{Fermi} dataset as functions of energy and event selection. 

Note that because of the rather complex form of the {\it Fermi} PSF, different event selections may have PSFs that are not related simply by an overall shift in scale (e.g. by a modification to $S_p(E)$), but are different in shape. To maximize the practical applicability of our work, rather than simply rescaling the PSF, we test the effect of using the true PSFs for different quartiles of the {\it Fermi} data ranked by PSF, and for different energy ranges. However, we will show that within the range of event selections we study, the effect on sensitivity of changing the PSF can be quite well described by the variation of a summary parameter such as the $68\%$ containment angle, suggesting that the detailed form of the {\it Fermi} PSF is not a crucial ingredient.

We divide the dataset into 40 log-spaced energy bins spanning the range from 0.2 GeV to 2000 GeV (i.e. 10 bins per decade). For each quartile and energy bin, we re-simulate the data with the same underlying model parameters but different PSF parameters. We stack these simulations together where appropriate (e.g. when testing multiple quartiles simultaneously, or when considering a broad energy range). When we analyze the simulated data, we use the worst PSF for any subset of the simulated data, which is consistent with what has been done in previous studies on the real data \cite{Lee:2015fea, Leane:2020pfc}. For example, if the simulated data involved photons from the top three PSF quartiles and a range of energies from $E_\text{min}$ to $E_\text{max}$, the PSF parameters used will correspond to the third-best PSF quartile and the energy $E_\text{min}$ (since the angular resolution of {\it Fermi} improves monotonically with increasing energy). 

\subsection{Energy range}

Varying the energy range of the data selection has multiple effects. Including a wider range of energies effectively increases the exposure; including lower-energy photons worsens the angular resolution. As mentioned above, we use the real PSF of {\it Fermi} for different energy ranges as one way to probe the effects of varying angular resolution. However, changing the energy range has additional effects that are not reducible to changes to angular resolution and exposure: low-energy photons are more abundant than high-energy ones in general, but also the spectra of the various emission components are different. Consequently, changing the energy range will modify the flux fraction associated with the GCE (and all other components). The choice of energy range thus needs to be optimized depending on the signal of interest.

To address the specific question of the optimal energy range for GCE studies, we change our energy cut on the real data and then repeat the analysis described in Sec.~\ref{sec:Methodology}. That is, we re-fit the templates to the real data with the new energy range, simulate the data based on these new template parameters, and determine the sensitivity to PSs as a function of energy range.

\subsection{Pixel size}
\label{subsec:pixelsizevarmethod}

Another somewhat ad-hoc choice in the standard NPTF analysis is the choice of spatial binning for the photons, i.e. the size of the equal-area \texttt{HEALPix} pixels (or equivalently, their number). Following previous studies such as \cite{Lee:2015fea,Mishra-Sharma:2016gis,Leane:2019xiy}, we utilized $\text{nside}=128$ for the majority of our analysis. The reason for this choice is the similarity between the $\text{nside}=128$ pixel radius and \textit{Fermi}'s angular resolution in the energy range of interest. 

If the pixel size chosen is substantially smaller than the angular resolution, PSs will always occupy multiple pixels, and the fact that \texttt{NPTFit} does not model correlations between neighboring pixels could lead to a loss of sensitivity. In the extreme limit of small pixel size, where all pixels contain either 0 or 1 photons, all sensitivity to PS populations would be lost. On the other hand, pixels much larger than the angular resolution increase the background from diffuse signals in any given pixel, and again this might be expected to reduce the sensitivity to PSs.

To examine these effects in detail, we use the same template parameters derived from the data for $\text{nside}=128$ (without any adjustment of parameters or priors) but simulate skymaps at $\text{nside} = 512$ using the procedure described in Sec.~\ref{sec:Methodology}.
After generating simulated skymaps at this higher resolution, we can increase the pixel size for each realization as desired, using a HEALPix package that combines ``children" pixels to create a superpixel, while accounting for proper normalization of photon counts \cite{Gorski:2004by}. This preserves the distribution of sources in the individual realizations as we explore different pixel sizes.

For variations in pixel size, priors do not need to be adjusted. Instead, the templates must be properly normalized within the ROI to ensure accurate scaling during parameter retrieval (see Appendix \ref{app:detailedmethodology}). 

\subsection{Source brightness}

The previously-discussed parameters describe instrumental properties and analysis choices. In this section, we discuss how the template model parameters are adjusted to describe a genuinely different source population. In analyses of sensitivity as a function of source brightness, our goal is to understand the potential of the \texttt{NPTFit} algorithm to detect faint sub-threshold sources.

The brightness of PSs can be varied by adjusting the $S_{b}$ and $A_{\text{PS}}$ parameters of the SCF (Eq.~\ref{eqn:scf}). We tested the effect of varying the brightness of individual PSs while keeping the total flux in the PS population fixed (as appropriate for a PS population making up most or all of the GCE). Using Eq.~\ref{eq:nphotons}, this requirement can be satisfied by simultaneously varying $S_b$ and $A_{\text{PS}}$ as follows:
\begin{equation}
    S_{b} \to n S_{b}, A_{\text{PS}} \to \frac{1}{n^{2}} A_{\text{PS}}
    \label{eqn:countsbrightness}
\end{equation}
For example, if $n = 1/2$, the number of photons each source emits is reduced by half, however, the number of sources increases by a factor of 2 to compensate. This equates to a factor of 4 increase in the template normalization factor $A_{\text{PS}}$, since the number of sources scales as $A_{\text{PS}} S_b$.

As for the exposure tests, when we rescale the simulated parameters we also rescale the priors in the fit to simulated data. For example, if $S_b \to n S_b$ and there was initially a linear prior on $S_b$ in the range $[0.05,80]$, the new prior is linear with range $[0.05n, 80n]$. If the prior on $\log_{10} A_{\text{PS}}$ is initially $[-6, 1]$, it is adjusted to $[-6- 2\log_{10}{n},1 - 2 \log_{10}{n}]$. We also checked the effect of keeping the priors fixed; except in situations where the true parameters approached the edge of the prior (or fell outside it), the effect was minimal.

For all simulations involved in the variation of source brightness, the number of live points we used to scan the data was set to nlive=100 for computational efficiency.

\section{A simplified isotropic scenario}
\label{sec:isotropic}

Before analyzing the results with all templates, we perform a simplified analysis including only isotropic components in our simulations, and approximating the exposure map as uniform. This analysis serves as a test of our analytic predictions. We describe some additional studies under this simplified scenario in Appendix \ref{app:isotropic}.

In this case, our normalization convention for the emission templates requires that the templates $T$ and $T_\text{PS}$ are both 1 in all pixels within the ROI. The normalization of the simulated signals was determined by matching the parameters for the PS component ($A_\text{PS}$, $n_1$, $n_2$ and $S_b$) to the isotropic PS component extracted from the real {\it Fermi} data.
For our baseline analyses, the smooth component normalization was chosen such that the total flux contributions of the smooth isotropic and PS components are equal. Explicitly, given our normalization convention for the templates $T$ and $T_\text{PS}$, this means that $A(\theta)_\text{smooth}$ is given by:
\begin{equation}
    A(\theta)_\text{smooth}=A(\theta)_{\text{PS}}S_{b}^{2} \left [\frac{(n_{1}-n_{2})}{(n_{1}-2)(2-n_{2})} \right ]
    \label{eqn:normalization}
\end{equation}
where (as previously) $A(\theta)_{\text{PS}}$ is the template normalization for the emission associated with the isotropic PS population, $S_b$ is the break of the source-count function, 
and $n_1, n_2$ are the slope of the source count functions defined by a singly-broken power law.

As previously, we use \texttt{NPTFit-Sim} to simulate PSs and a Poisson draw to simulate the smooth component. The priors on the various parameters are set as discussed in Sec.~\ref{sec:varyingdiffparams} and Appendix \ref{app:detailedmethodology}. (Note that if the simulated value of $A_\text{smooth}$ was outside the prior range on $A_\text{iso}$ in the main analysis, we would need to adopt different priors for this isotropic study, but in fact it lies well within the prior range so this is not a problem.)

\subsection{Variation of exposure (narrow PSF)}
\label{subsec:isotropicpsf}

\begin{figure*}[ht]
    \centering
    \includegraphics[width=0.49\linewidth]{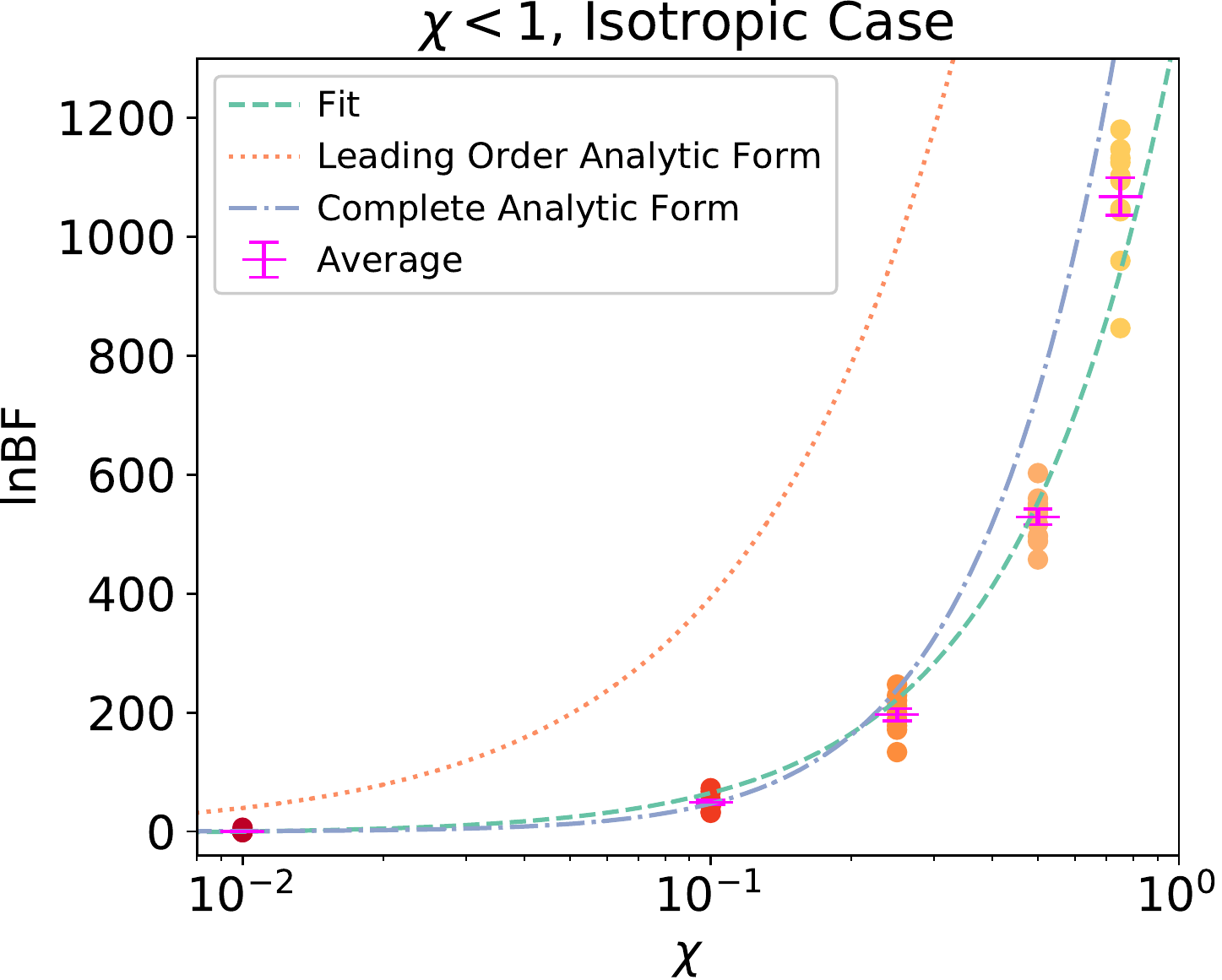}
    \hspace{0.4em}
    \includegraphics[width=0.49\linewidth]{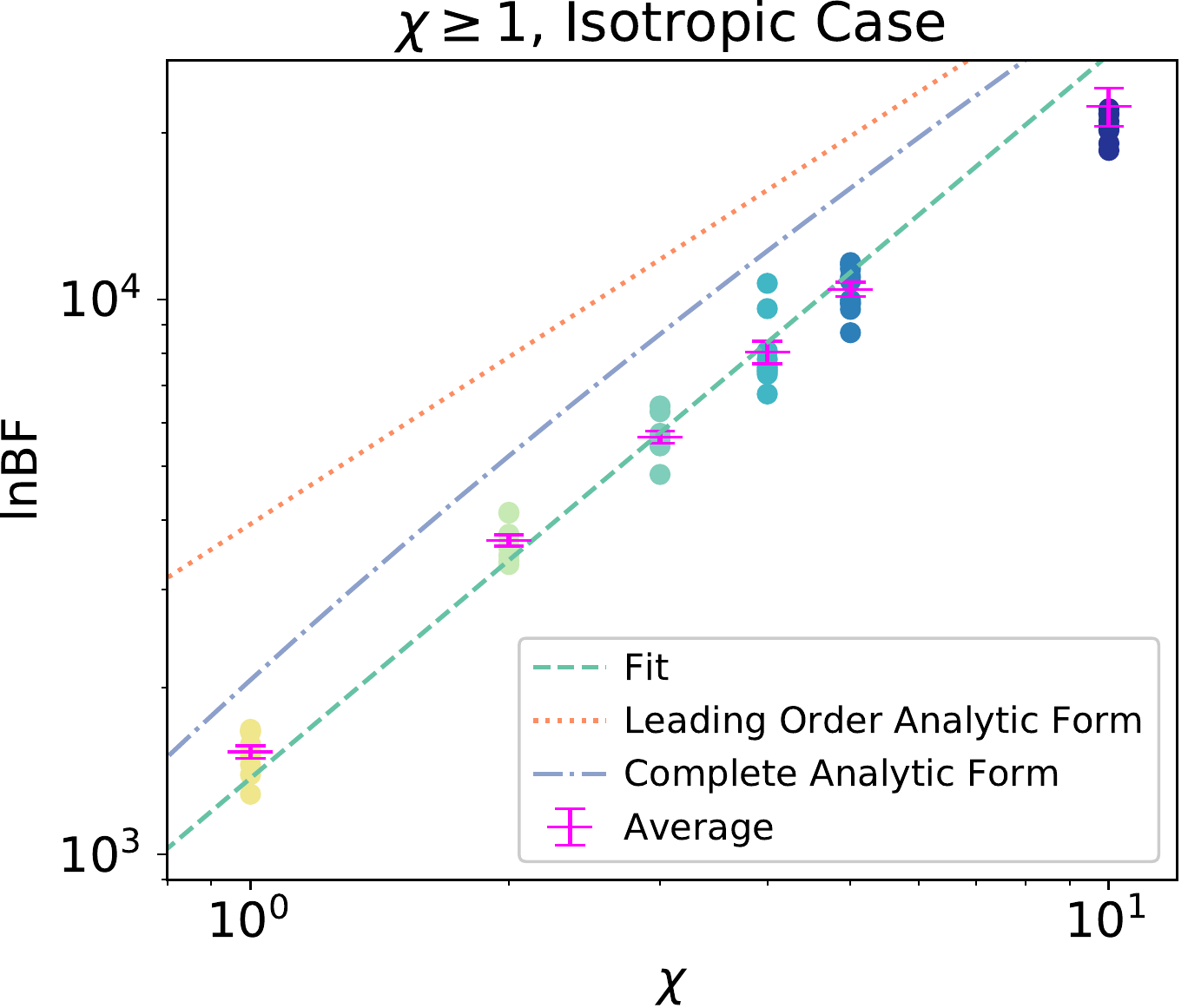}
    \caption{$\ln{\text{BF}}$ across 10 realizations (circle markers), and $\langle \ln{\text{BF}} \rangle $ with error bars obtained from the $\sigma/\sqrt{10}$ standard error of the mean (magenta), for varying values of $\chi$/exposure. Signal and background are isotropic and both simulations and scans employ a narrow Gaussian PSF. The green dashed line denotes the modified power law fit defined in Eq.~\ref{eqn:powerlawshift}. The orange dotted line denotes the first (linear) term of the analytic form Eq.~\ref{eqn:lnBFcase2}, while the blue dash-dot line denotes the full analytic solution described in Eq.~\ref{eqn:lnBFcase2}.  }
    \label{fig:analyticcase2comparisonpsf}
\end{figure*}

\begin{figure*}[ht]
    \centering
    \includegraphics[width=0.49\linewidth]{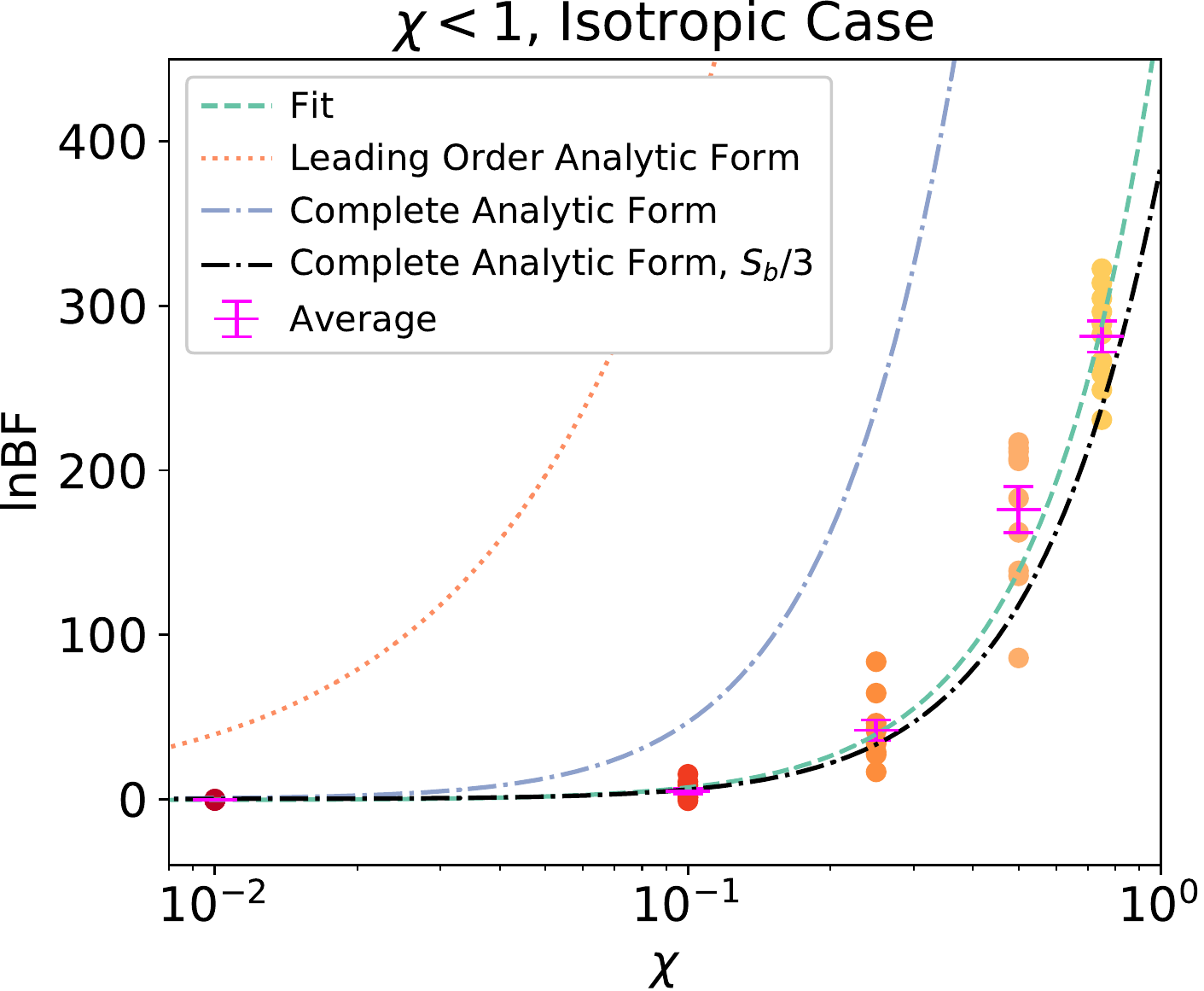}
    \hspace{0.4em}
    \includegraphics[width=0.49\linewidth]{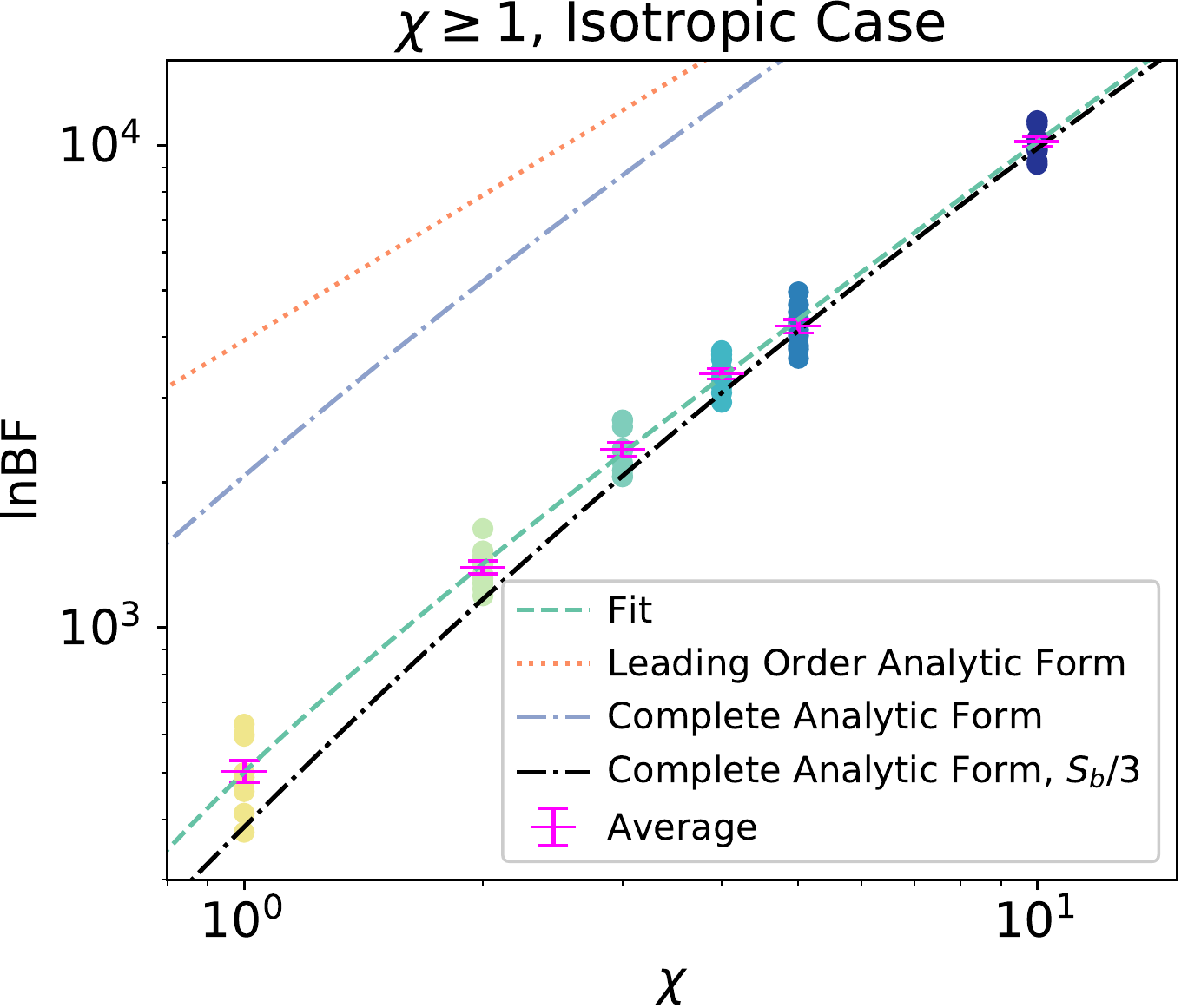}
    \caption{
    $\ln{\text{BF}}$ across 10 realizations (circle markers), and $\langle \ln{\text{BF}} \rangle $ with error bars obtained from the $\sigma/\sqrt{10}$ standard error of the mean (magenta), for varying values of $\chi$/exposure. Signal and background are isotropic and both simulations and scans employ a realistic PSF model. The green dashed line denotes the modified power law fit defined in Eq.~\ref{eqn:powerlawshift}. The orange dotted line denotes the first (linear) term of the analytic form in Eq.~\ref{eqn:lnBFcase2}, while the blue dash-dot line denotes the full analytic approximation from the same equation with $s \rightarrow S_b$. The black dash-dot line shows the analytic approximation with $s \rightarrow S_b/3$.}
    \label{fig:analyticcase2comparison}
\end{figure*}

The analytic approximations we derived in Sec.~\ref{sec:analyticforms} assumed that the PSs were not smeared by the PSF. Thus as a cross-check of the scaling behavior estimated from our analytic results, we performed an initial set of simulations where the PSF was taken to be extremely narrow (i.e. the angular reconstruction was effectively perfect), covering a range of exposure levels $\chi$. Specifically, we sampled exposure rescaling factors $\chi$ between $10^{-2}$ and $10$ (recall $\chi=1$ corresponds to the baseline exposure), and for each case generated and scanned skymaps that employed a Gaussian PSF with a tiny variance $10^{-20}$. For each choice of $\chi$ we ran the analysis for 10 simulated realizations, and for each realization evaluated the $\ln{\text{BF}}$ between models with and without isotropically distributed PSs (both models allow for an isotropically distributed smooth component).

Fig.~\ref{fig:analyticcase2comparisonpsf} plots the Bayes factor preference in favor of PSs as a function of exposure, together with the analytic solution for the likelihood ratio in the case where PSs and background are equally bright (Eq.~\ref{eqn:lnBFcase2} with $k=1/2$ and $s= S_b$). We also show the result of including only the linear $ks/2$ term in the analytic estimate of Eq.~\ref{eqn:lnBFcase2}. 

For each choice of exposure, we evaluate $\langle \ln \text{BF}\rangle(\chi)$ by taking the average of $\ln\text{BF}$ across realizations at each exposure level; we also compute the standard error of the mean across the realizations for this quantity at each $\chi$ value (indicated by magenta vertical bars in Fig.~\ref{fig:analyticcase2comparisonpsf}).

We work by default (here and in the remainder of this work) with the log of the Bayes factor between models, rather than the likelihood ratio; however, in this specific example we also evaluated the log likelihood ratio and found that it was generically quite close to the log Bayes factor (and in particular the difference between the two was not responsible for the difference between the numerical results and the analytic approximation for $\langle \Delta \ln \mathcal{L}\rangle$). Thus we treat our analytic approximation as a rough estimate for $\langle \ln \text{BF}\rangle$.

For these parameter choices, we observe that the analytic form mildly overestimates the sensitivity, by a factor of roughly 20-30\% in $\langle \ln \text{BF}\rangle$ at high exposure, but accurately captures the fall-off of the detection sensitivity at low exposure, and the scaling at high exposure. The remaining discrepancy is likely due to the approximations we have made in deriving our analytic results (e.g. relating to the shape of the probability distribution, and assuming we can treat all integrals as having limits $\pm \infty$, as well as approximating the SCF as a delta-function).

We observe a consistent scatter at the $\mathcal{O}(10\%)$ level in $\langle \ln \text{BF}\rangle$ between different realizations, which does not obviously decrease at large $\chi$. (Note that here we are discussing the standard deviation across realizations, not the standard error of the mean; the latter is smaller by a factor of $1/\sqrt{n_{\text{realizations}}}$.) This can be understood in terms of our variance calculations in Sec.~\ref{sec:analyticforms}. The parameters we have simulated correspond to $5.61\chi$ photons/source, 2808 pixels, and an average of $0.11$ sources/pixel; thus we expect a total number of sources in the ROI around $280$, and a standard deviation in the log likelihood ratio that is of order $\langle \ln \text{BF}\rangle/\sqrt{280} \sim 0.06 \langle \ln \text{BF}\rangle$ or $\sqrt{\langle \ln \text{BF}\rangle}$, whichever is larger. This is consistent with the $\mathcal{O}(10\%)$ scatter we observe at high exposure.

In addition to the comparison to the analytic prediction, we can parameterize the scaling of the sensitivity with $\chi$ as a power law and fit for the parameters (although power-law behavior should be expected to break down at sufficiently small $\chi$, where $\ln \text{BF}$ can attain negative values). The fitting function we use is:
\begin{equation} \langle \ln \text{BF}\rangle(\chi) = \alpha \chi^{\beta} + \gamma,
    \label{eqn:powerlawshift}\end{equation}
where the offset parameter $\gamma$ serves to correct the behavior at small $\chi$ where there is not enough data to detect a significant signal. For each value of $\chi$ we took the central value of $\langle \ln \text{BF}\rangle(\chi)$ to be the average over realizations, with an error bar determined by the standard error of the mean, and performed a least-squares fit. The resulting best-fit model is also plotted in Fig.~\ref{fig:analyticcase2comparisonpsf}.

\subsection{Variation of exposure (realistic PSF)}

In a realistic scenario, we will always have to deal with a PSF that is not arbitrarily narrow. We repeat the simulation and analysis described above using the full PSF appropriate to the real  {\it Fermi} dataset (for the top PSF quartile), and show results in Fig.~\ref{fig:analyticcase2comparison}. We compare these results to the same analytic solution (i.e. with no allowance for the PSF) as in the previous analysis, and again perform a least-squares fit to a simple power law fitting function.

We observe that the analytic solution still describes the shape of $\langle \ln \text{BF}\rangle$ as a function of $\chi$ quite well, but now the discrepancy in our sensitivity metric is more pronounced (a factor of a few at high $\chi$). At least qualitatively, this discrepancy can be largely absorbed by taking $s/S_b$ to be a constant other than unity; Fig.~\ref{fig:analyticcase2comparison} shows the effect of using the analytic approximation with $s$ replaced by $ S_b/3$. The variance remains $\mathcal{O}(10\%)$ at high $\chi$, which can be understood as discussed above.

In this more realistic case, we thus recommend using the analytic estimate only to understand scaling behavior rather than as a quantitative estimate of the expected sensitivity, although a reasonably good description can be obtained by fitting a constant rescaling factor to be applied to $s$.

\subsection{Variation of relative flux contributions}
\label{app:subdominantcomp}

\begin{figure}[ht]
    \centering
    \includegraphics[width=0.98\linewidth]{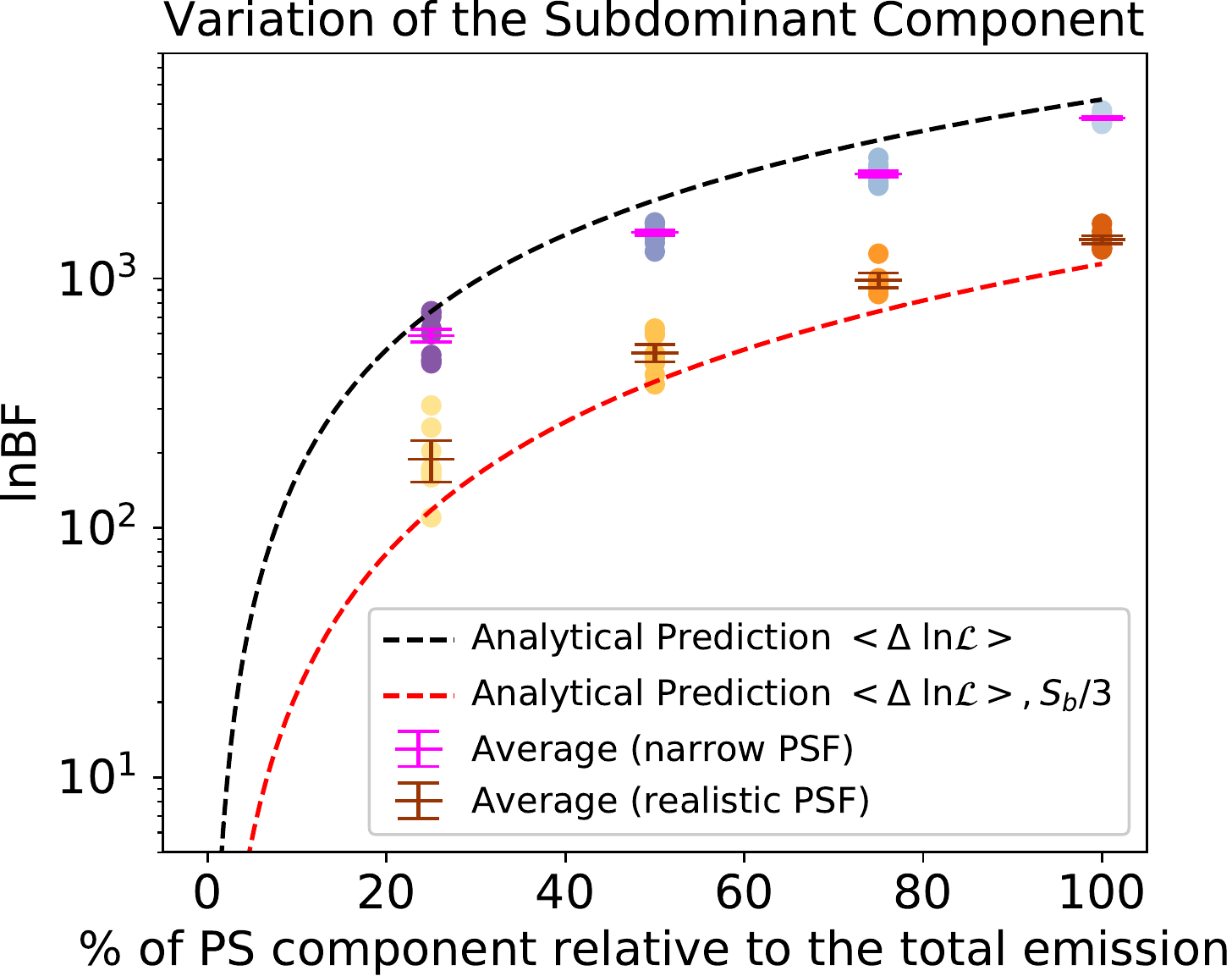}
    \caption{$\ln{\text{BF}}$ across 10 realizations (circle markers), and $\langle \ln{\text{BF}} \rangle $ with error bars obtained from the $\sigma/\sqrt{10}$ standard error of the mean (magenta, brown), as the relative contributions between a smooth and PS component within an isotropic map are varied, for realistic and narrow PSF prescriptions. Dashed lines indicate the analytic prescription of Eq.~\ref{eqn:lnBFcase2} with the replacements $s\rightarrow S_b$ (black) and $s\rightarrow S_b/3$ (red).
  }
    \label{fig:isotropicfluxfrac}
\end{figure}

We can also test the effects of varying the relative flux contributions of the smooth and PS components while allowing the total flux to remain constant. 
Fig.~\ref{fig:isotropicfluxfrac} demonstrates how the sensitivity changes as the PS flux fraction is varied, for both the narrow PSF and realistic PSF (top quartile) cases. For comparison, we also overlay the predictions given by our analytic approximations, Eq.~\ref{eqn:lnBFcase2}, with $s \rightarrow S_b$ and $s\rightarrow S_b/3$. We find that as the relative contribution of the PS component increases, the sensitivity of \texttt{NPTFit} to PSs naturally increases with a shape consistent with the analytic prediction provided by Eq.~\ref{eqn:lnBFcase2}, and in the narrow-PSF case the $s\rightarrow S_b$ substitution provides quantitatively accurate results. Although the case where the map is simply produced with a smooth component is not shown in the figure due to the log-scaling, the result averages to $-0.54 \pm 0.067$ in the realistic-PSF case and $-0.56 \pm 0.14$ in the narrow-PSF case, both of which are small, as expected. In the regime where there is no significant preference for PSs, we expect the Bayes factor in favor of the model without PSs to be highly dependent on the choice of priors (as also discussed in e.g.~\cite{Chang:2019ars, Collin:2021ufc}). We explore this point further in Appendix \ref{sec:priors}.

\section{Results of simulated parameter variations in the full inner Galaxy analysis}
\label{sec:results}

In this section we now proceed to a numerical analysis using simulated {\it Fermi} data for the inner Galaxy, employing the complete set of templates discussed in Sec.~\ref{sec:Methodology}. The results of this section can thus be used directly to optimize \texttt{NPTFit}-based approaches to studies of the inner Galaxy and GCE.

\subsection{Varying the exposure}
\label{sec:exposuretest}

As in the isotropic case, we sampled exposure rescaling factors $\chi$ between $10^{-2}$ and $10$. For each choice of $\chi$ we ran the analysis for 20 simulated realizations, and for each realization evaluated the $\ln{\text{BF}}$ between models with and without the GCE PS template. 

\begin{figure*}
    \centering
    \includegraphics[width=0.49\linewidth]{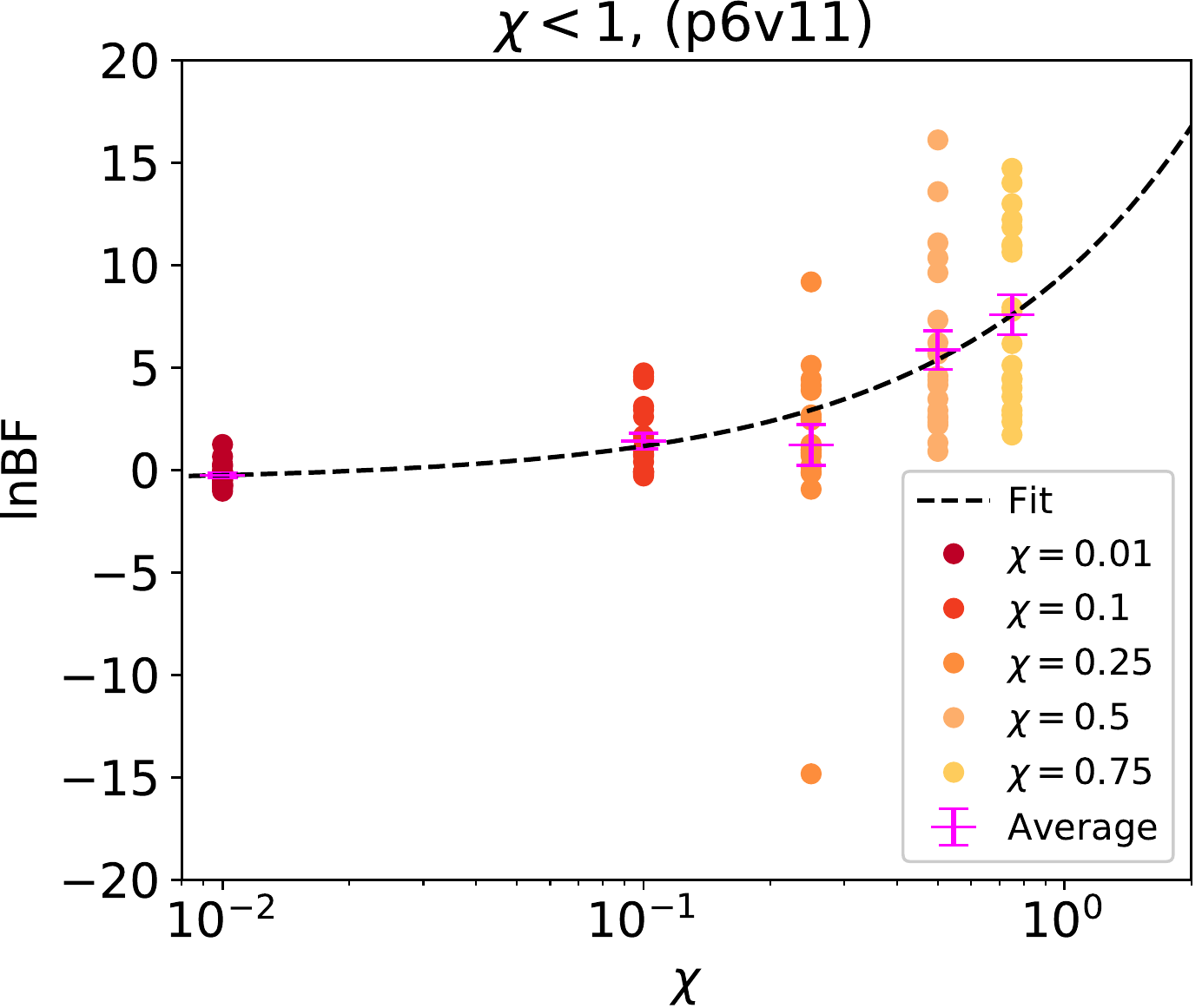}
    \hspace{0.4em}
    \includegraphics[width=0.49\linewidth]{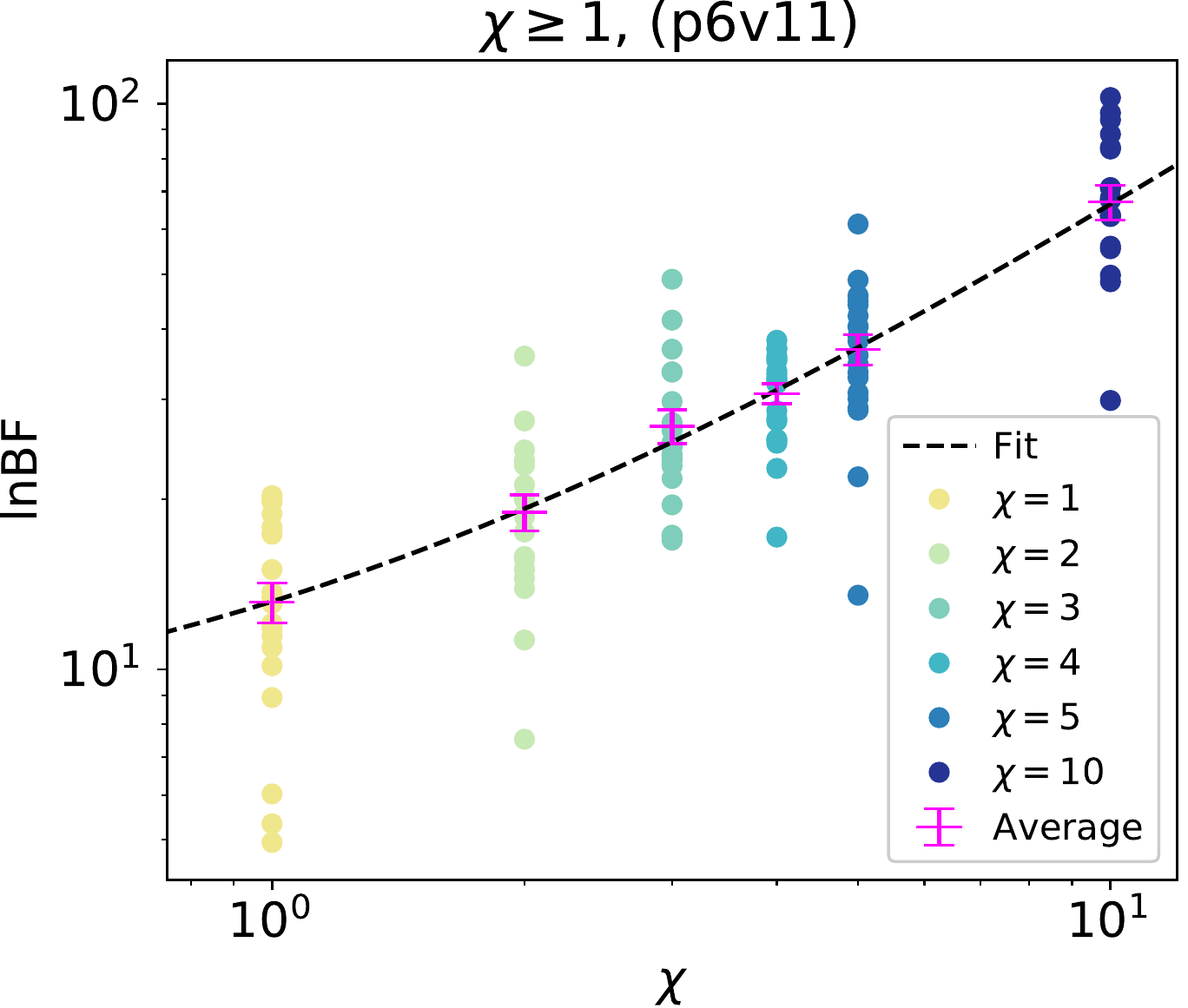}
    \caption{$\ln{\text{BF}}$ across 20 realizations (circle markers), and $\langle \ln{\text{BF}} \rangle $ with error bars obtained from the $\sigma/\sqrt{20}$ standard error of the mean (magenta), for varying values of $\chi$/exposure.
    \textit{Left}: realizations with $\chi<1$. \textit{Right}: realizations with $\chi \geq 1$. The best-fit line is a standard power law with an additive shift defined in Eq.~\ref{eqn:powerlawshift}.}
    \label{fig:exposuretest}
\end{figure*}

Fig.~\ref{fig:exposuretest} shows the resulting values of $\ln\text{BF}$ for each exposure level. We evaluate $\langle \ln \text{BF}\rangle(\chi)$ by taking the average of $\ln\text{BF}$ across realizations at each exposure level (indicated by magenta vertical bars in the figure along with error bars that denote the $\sigma/N_{\text{realizations}}$ (where $N_{\text{realizations}}$ is the number of realizations in a given sample) standard error of the mean across all the realizations within a particular case). 

As discussed in Sec.~\ref{sec:isotropic}, we fit a power-law function (Eq.~\ref{eqn:powerlawshift}) to the data for $\langle \ln \text{BF}\rangle(\chi)$. The resulting best-fit parameters are given in Table \ref{tab:exposurefit}, and the best-fit model is plotted in Fig.~\ref{fig:exposuretest} (solid blue line). We find approximately that $\langle \ln \text{BF} \rangle \propto \chi^{0.8}$ at large BF. This is broadly consistent with our expectation from the analytic estimate in Sec.~\ref{sec:analyticforms} that $\langle \ln \text{BF} \rangle$ should scale $\sim$linearly in $\chi$.

\begin{table}[H]
\centering
\begin{tabular}{r r r }
\hline
Recovered Parameters & \texttt{p6v11}\\
\hline
$\alpha$ (coefficient) & $11.30 \pm 0.69$ \\
$\beta$ (power) & $0.76 \pm 0.04$  & \\
$\gamma$ (shift) & $-0.62 \pm 0.20$  & \\
\botrule
\end{tabular}
\caption{Best-fit parameters obtained using least-squares regression method along with the $1\sigma$ error for the power law fit of $\langle \ln{\text{BF}}\rangle$ to $\chi$, as defined in Eq.~\ref{eqn:powerlawshift}.}
\label{tab:exposurefit}
\end{table}

\subsection{Varying the angular resolution}
\label{sec:varyingangexp}

\subsubsection{PSF models for different quartiles} \label{subsec:varyingpsf}

We begin by examining how the sensitivity of \texttt{NPTFit} varies when the top three quartiles of data by PSF are analyzed separately, keeping the energy range fixed at its default value of $2-20 \text{ GeV}$. Quartiles are labeled in order of decreasing angular resolution (so e.g. ``PSF 01'' represents the best quartile). 

\begin{figure}
    \centering
    \includegraphics[width=0.98\linewidth]{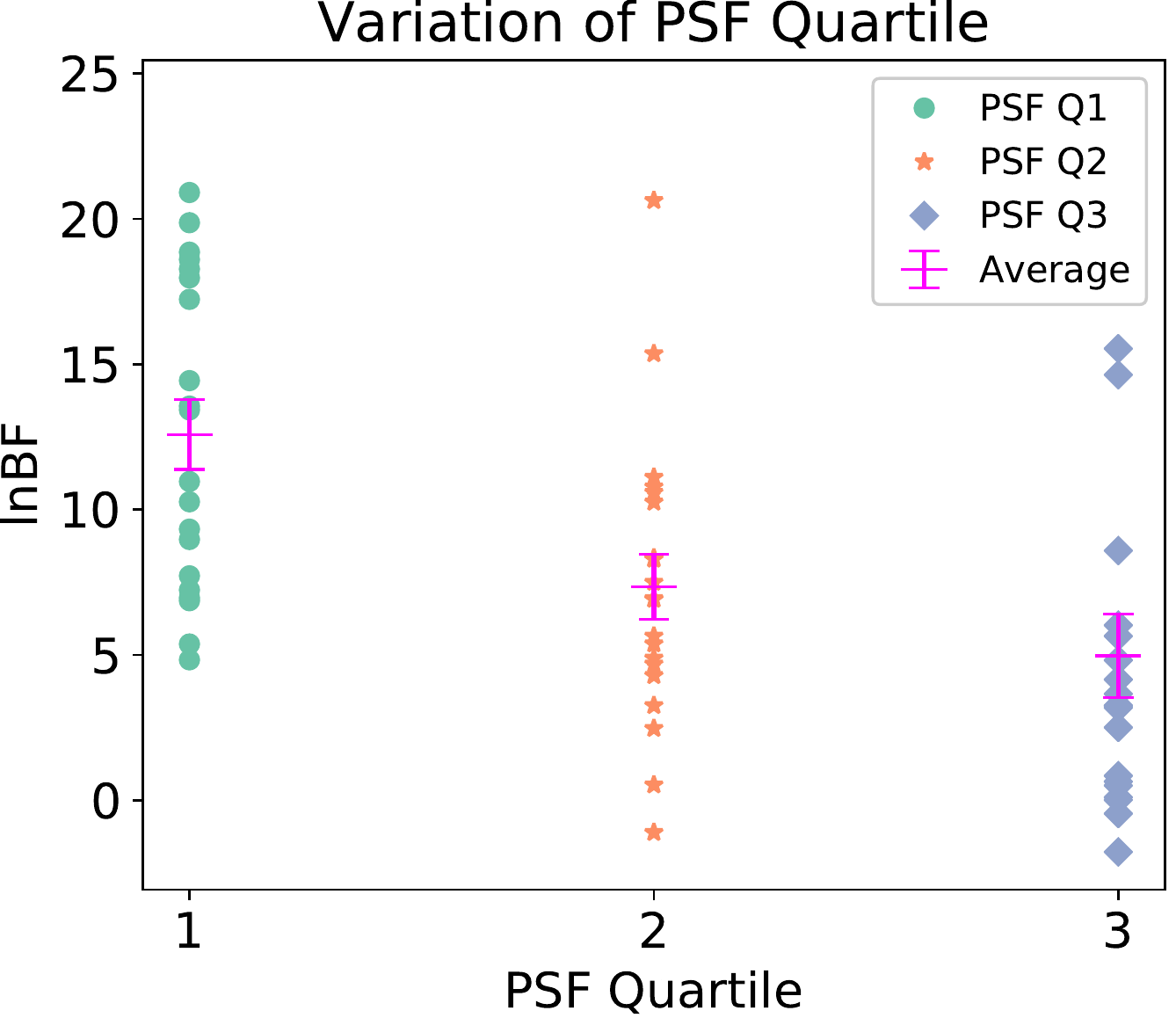}
    \caption{
    $\ln{\text{BF}}$ across 20 realizations (circle/star/diamond markers), and $\langle \ln{\text{BF}} \rangle $ with error bars obtained from the $\sigma/\sqrt{20}$ standard error of the mean (magenta), for the top three PSF quartiles.
   }
    \label{fig:psfquartiletest}
\end{figure}

Within each quartile, we simulated and analyzed 20 realizations. All simulations are generated at a rescaling factor of $\chi = 1$.  Fig.~\ref{fig:psfquartiletest} shows a striking decline in sensitivity in the quartiles with worse angular resolution. As previously, we display both the scatter between realizations and the average $\langle \ln \text{BF}\rangle$ across realizations for a given quartile.

\begin{figure*}
    \centering
    \includegraphics[width=0.98\linewidth]{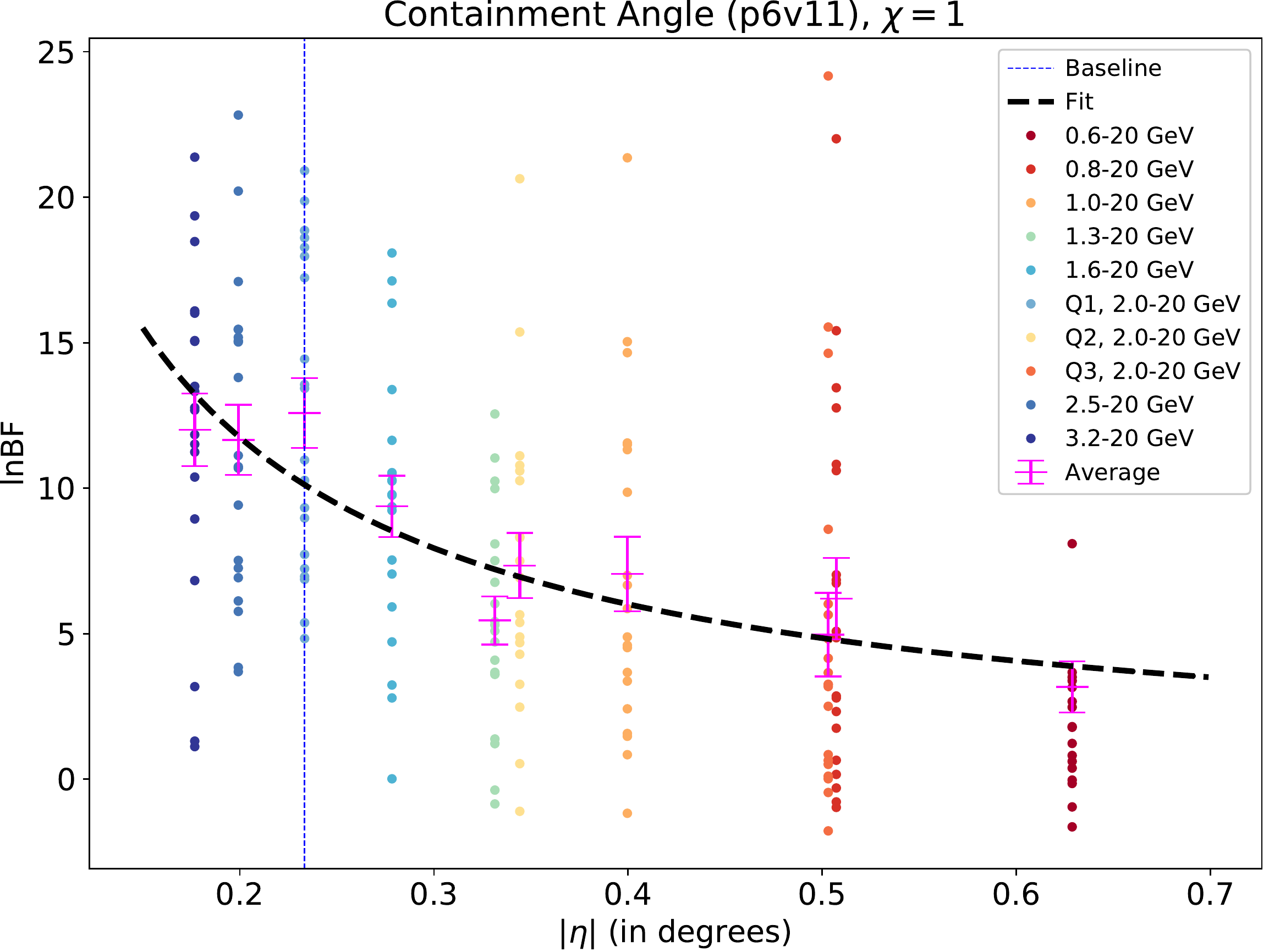}
    \caption{$\ln{\text{BF}}$ across 20 realizations (circle markers), and $\langle \ln{\text{BF}} \rangle $ with error bars obtained from the $\sigma/\sqrt{20}$ standard error of the mean (magenta), as a function of the $68 \%$ containment angle of the PSF. Realizations are obtained by considering the top PSF quartile for different energy ranges; for the baseline $2-20$\,GeV range, we test each of the top three quartiles individually. 
    The best-fit line (black) is the power-law fit to the $\langle \ln{\text{BF}}\rangle$ obtained using least-squares. The baseline case (blue vertical line) denotes the baseline case 2-20\,GeV for quartile 1, typically used in previous \textit{Fermi} analyses. }
    \label{fig:containmentanglep6v11}
\end{figure*}
\subsubsection{PSF models for different energies} \label{subsec:varyingpsfenergy}

Another practical way to vary the angular resolution in {\it Fermi} data is to modify the energy window. We first examined the (theoretical) case where the angular resolution is varied while keeping all other parameters constant. We re-simulated the data with the original exposure map ($\chi=1$) but with PSF corresponding to the appropriate {\it Fermi} PSF for energies between 0.6 GeV and 3.2 GeV (recall that the baseline analysis uses the {\it Fermi} PSF at 2 GeV), in the top PSF quartile. In each case we performed 20 realizations.

In Fig.~\ref{fig:containmentanglep6v11} we plot the resulting values of $\ln\text{BF}$, against the value of the $68 \%$ containment angle associated with each PSF model (in degrees), which we denote $\eta$. We also include the results for the 2 GeV PSF in all three quartiles (discussed above).

To summarize the results, we fit the data with a power law, $\langle \ln \text{BF}\rangle(\eta) = a \eta^{b}$, using the same least-squares analysis as described above for the case of $\langle \ln \text{BF}\rangle(\chi)$.
Tab.~\ref{tab:containmentanglefit} displays the resulting best-fit parameters. We find that $\langle \ln \text{BF}\rangle(\eta) \propto \eta^{-1}$, i.e. the sensitivity appears to scale approximately inversely with the containment radius, at least while holding the pixel size constant at $\text{nside}=128$.

\begin{table}[H]
\centering
\begin{tabular}{r r c}
\hline

Parameter & \texttt{p6v11} \\
\hline
$a$ (coefficient) & $2.5 \pm 0.6$ \\
$b$ (power) & $-1.0 \pm 0.2$  & \\

\botrule
\end{tabular}
\caption{Recovered parameters and the $1\sigma$ error for the power-law fit to $\langle \ln \text{BF}\rangle$ as a function of $68\%$ containment angle.}
\vspace{-10pt}
\label{tab:containmentanglefit}
\end{table}

Thus as a rule of thumb, we expect an increase in the exposure by a factor of $n$ to be approximately compensated by an increase in the containment radius (not the containment area) by a factor of $n$; if the exposure can be more than doubled while worsening the containment angle by less than a factor of two, this will generally be a beneficial tradeoff. 

\begin{figure*}
    \centering
    \includegraphics[width=0.49\linewidth]{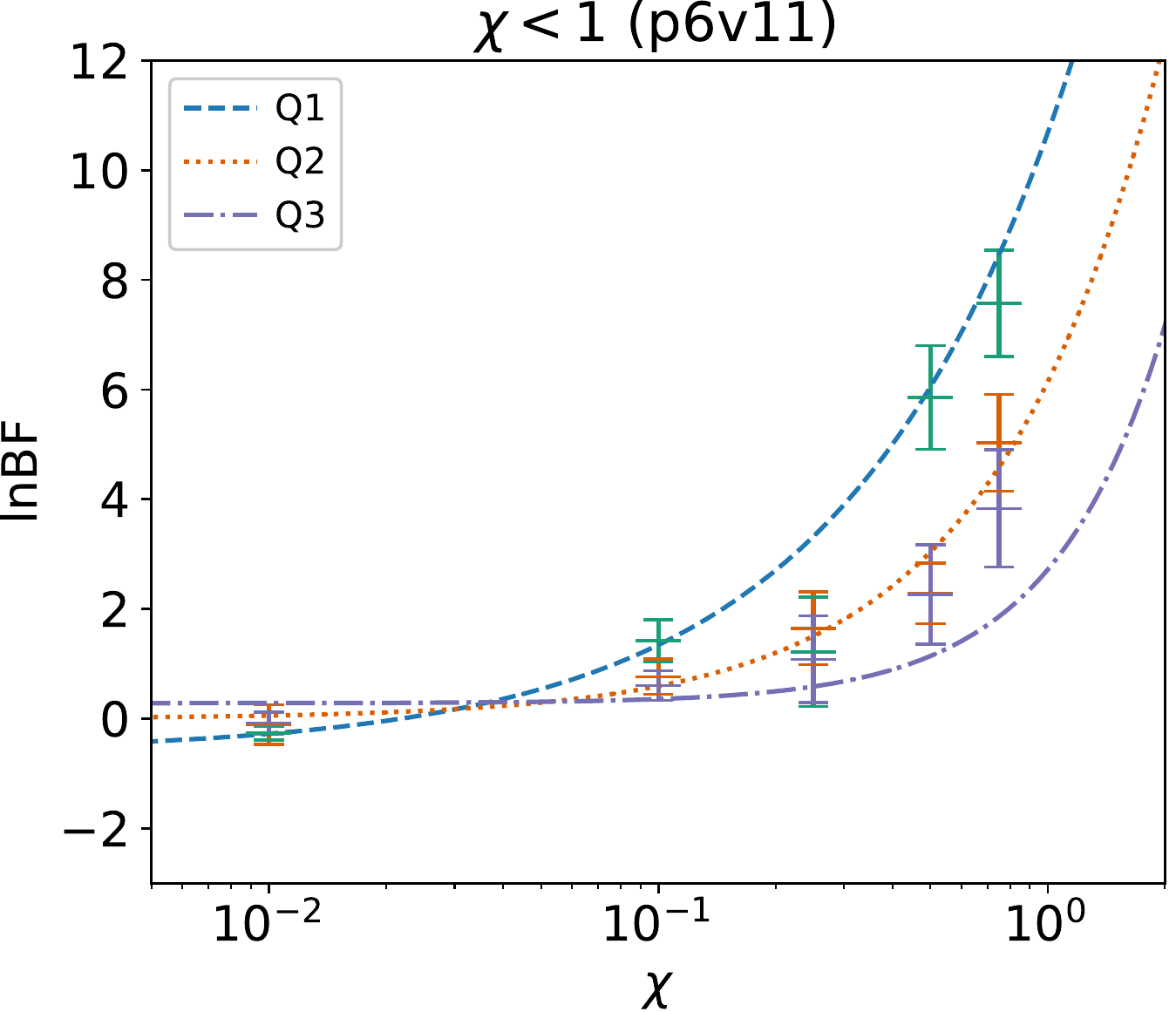}
    \hspace{0.4em}
    \includegraphics[width=0.49\linewidth]{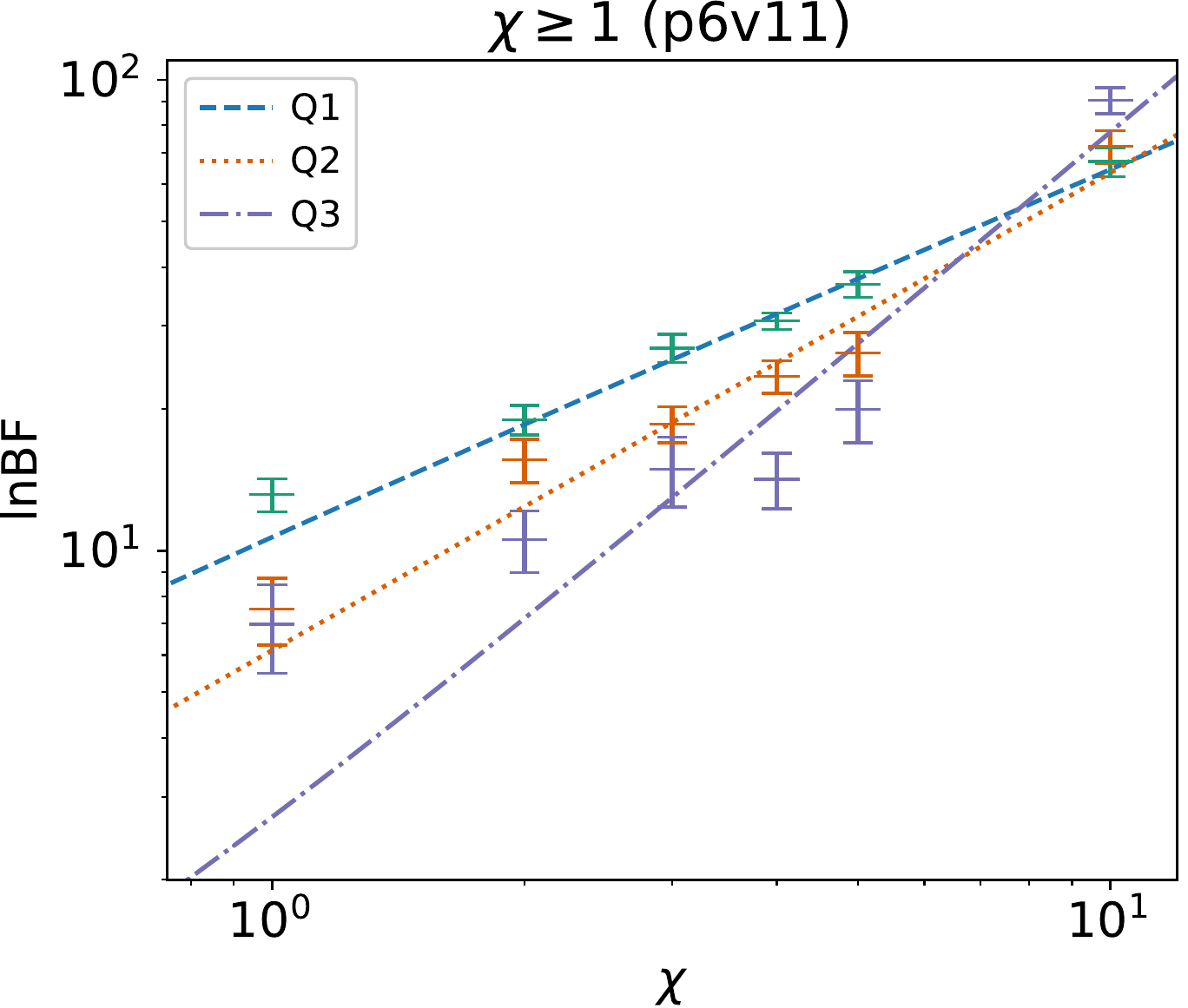}
    \caption{$\langle \ln{\text{BF}}\rangle$ and the $1
    \sigma$ standard error of the mean across 20 realizations for the top three quartiles graded in angular resolution sampling different values of $\chi$. \textit{Left}: realizations with $\chi<1$. \textit{Right}: realizations with $\chi \geq 1$. The best-fit lines show the Eq.~\ref{eqn:powerlawshift} fit to the data.}
    \label{fig:angularexp1sb}
\end{figure*}

Note that our choice of $\text{nside}=128$ corresponds to a mean pixel spacing ($0.46^\circ$) that exceeds or is comparable to the $68\%$ containment angle for all but the widest energy range (0.6-20 GeV) that we consider. We will show in Sec.~\ref{sec:pixelsizevar} that decreasing the pixel size below the PSF does not appear to have large effects on the expected sensitivity (although it can increase the variance), but studies focusing on a broader energy range might still wish to test smaller nside values to reduce leakage of PSs into neighboring pixels.

\subsubsection{Simultaneous variation of exposure and angular resolution} 
\label{subsec:tradeoffs}
To check the stability of the scaling rules we have found so far and the validity of this simple estimate, we now explicitly test the effect of simultaneously varying the angular resolution and the exposure. In many realistic situations, and in particular for \textit{Fermi} data, relaxing cuts on photon quality will simultaneously increase the effective exposure and worsen angular resolution.

We repeat the analysis described in Sec.~\ref{sec:exposuretest} for simulated data using the appropriate PSF model for PSF quartiles 2 and 3, with 20 realizations for each combination of $\chi$ and quartile. We scanned the realizations at nlive = 300. Our results for $\langle \ln \text{BF} \rangle(\chi)$ for each quartile are summarized in Fig.~\ref{fig:angularexp1sb}. As in Eq.~\ref{eqn:powerlawshift}, we fit the data for each quartile with a power law with a constant offset, and provide the best-fit parameters in Tab.~\ref{tab:tradeoffp6v11}.

In general we observe that the slope appears to become steeper (more rapid increase in sensitivity with exposure) in quartiles with worse angular resolution. This reflects that significant detection of PSs requires a higher $\chi$ value when the angular resolution is worse, but for sufficiently large $\chi$ factors, the significance becomes almost independent of angular resolution. This may be related to the pixels surrounding a PS becoming bright enough to be individually detected as significant PSs.

We can also test the effect of stacking together the simulated data corresponding to the different quartiles, which has the effect of increasing the effective exposure $\chi_{\textrm{eff}} >\chi$ relative to the one-quartile case. The sum of the first and second quartile has $\chi_{\text{eff}} = 2$, and the combined top three quartiles have $\chi_{\textrm{eff}} = 3$.

\begin{figure}
    \centering
    \includegraphics[width=0.98\linewidth]{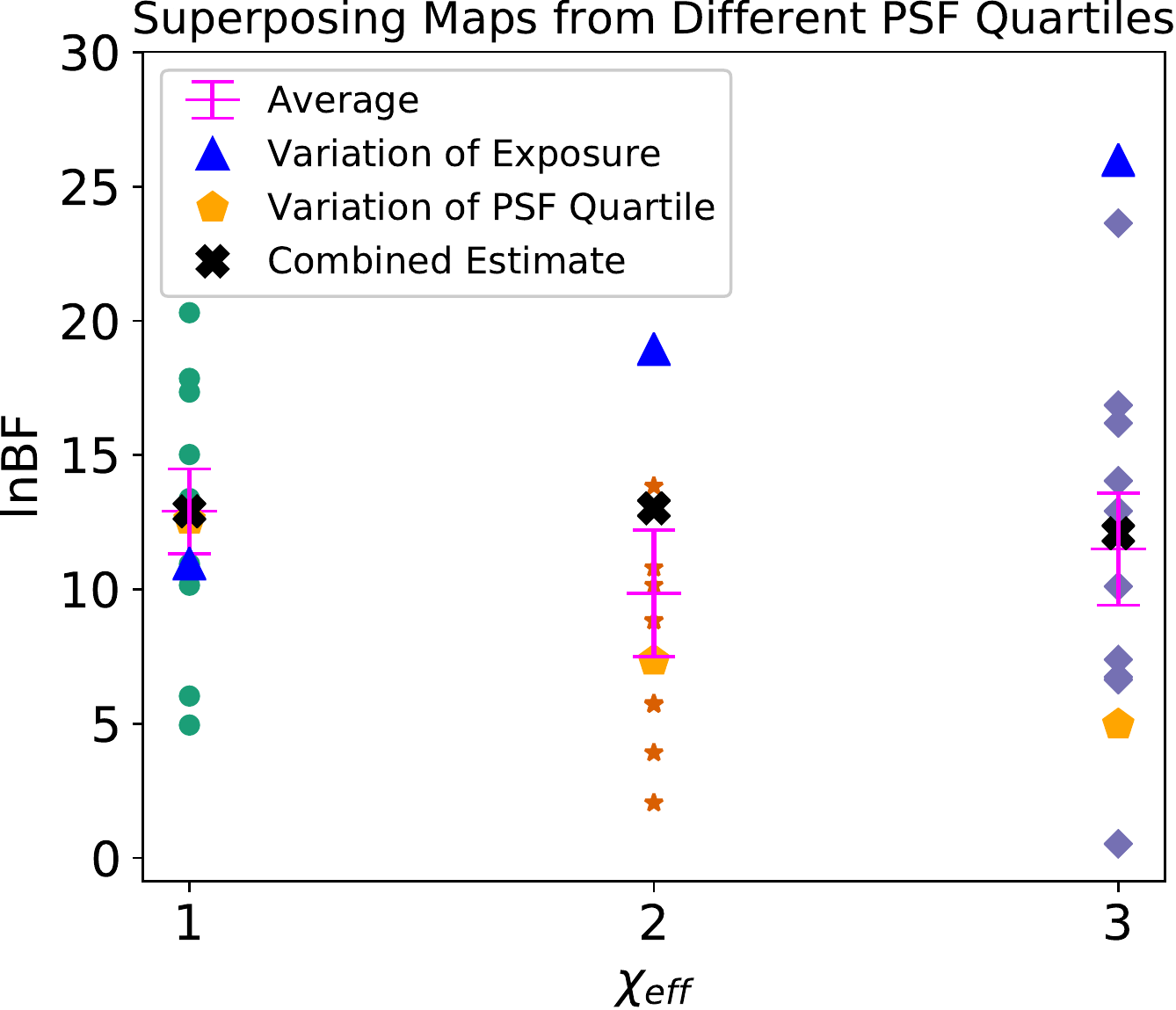}
    \caption{$\ln{\text{BF}}$ across 10 realizations (circle/star/diamond markers), and $\langle \ln{\text{BF}} \rangle $ with error bars obtained from the $\sigma/\sqrt{10}$ standard error of the mean (magenta), stacking increasing numbers of skymaps generated with different angular resolutions (corresponding to the PSF quartiles). All scans assumed the angular resolution of the worst included quartile. The blue triangles indicate the increased sensitivity as predicted by varying the exposure, while the orange pentagons indicate the expected worsening of sensitivity due to angular resolution degradation. The black filled ``X" symbols display an estimate for the overall sensitivity change from combining the two (see text for details).}
    \label{fig:multiquartilesuperposed}
\end{figure}

 Fig.~\ref{fig:multiquartilesuperposed} shows the sensitivity based on 20 realizations for each of these three cases scanned at nlive=300. We find that to quite a good approximation, the increased number of photons simply cancels out the effects of worsening the angular resolution on average. As a simple estimate we calculated the combined effects of the predictions of varying the exposure and angular resolution. To do so, we define a rescaling factor $r=r_{\text{exposure}}(\chi)r_{\text{PSF}}(\text{Quartile})$, where $r_{\text{exposure}}(\chi)$ is the ratio of the expected log BF at exposure $\chi$, denoted $\langle\ln{\text{BF}}\rangle(\chi)$, to the baseline expected log BF $\langle\ln{\text{BF}}\rangle(\chi=1)$, as obtained from Eq.~\ref{eqn:powerlawshift} and Table~\ref{tab:exposurefit}. Thus $r_\text{exposure}(\chi)$ characterizes the increase in sensitivity with enhanced exposure. $r_{\text{PSF}}(\text{Quartile})$ is the ratio of the expected log BF for a specified single quartile, denoted $\langle\ln{\text{BF}}\rangle(\text{Quartile})$, to the baseline expected log BF $\langle\ln{\text{BF}}\rangle(\text{Quartile}=1)$, obtained from Fig.~\ref{fig:psfquartiletest}. Thus $r_\text{PSF}(\text{Quartile})$ characterizes the decline of sensitivity with worsening angular resolution. $\langle\ln{\text{BF}}\rangle(\chi)$ and $\langle\ln{\text{BF}}\rangle(\text{Quartile})$ are denoted as blue stars and orange pentagons on Fig.~\ref{fig:multiquartilesuperposed}, respectively. To obtain the combined estimate denoted by the black filled ``X", we multiplied the calculated $r$ factor with the baseline value $\langle \ln{\text{BF}}\rangle(\text{Quartile 1})$ obtained from the realizations for the baseline case in Fig.~\ref{fig:multiquartilesuperposed}. We find that this estimate agrees with the simulation results that on average adding quartiles with worse angular resolutions to gain exposure does not yield large increases (or decreases) in the average sensitivity to PSs.
 
 Quantitatively, Q3 has a containment angle slightly more than twice that of Q1 (see Fig.~\ref{fig:containmentanglep6v11}), while including Q2 and Q3 triples the exposure. The scaling of the sensitivity with exposure is slightly sublinear whereas for containment angle it is linear to a good approximation, and in practice we find that these two effects almost completely cancel out. Thus the overall sensitivity is (perhaps surprisingly, and somewhat coincidentally) insensitive to the inclusion of additional quartiles.
 
 We might wonder if by approximating the PSF in the stacked dataset as the PSF of the worst quartile, we introduce biases in the recovered parameters. We checked this explicitly over our sample of 20 realizations. On average, we find that the median parameter deviations were rather small, mostly at $<1\sigma$ level, with some exceptions for components such as the isotropic emission. However, we also checked cases where we stacked different realizations of the same quartile, so that the PSF was the same between different subpopulations and was thus modeled in the same way for the simulated data and the fit. We found, on average, that the biases were similar in these cases; they were not obviously worsened by stacking maps with different PSFs, while modeling with the worst PSF. Thus, the mis-reconstruction cannot be attributed to mismodeling of the PSF in a subset of the data. 
 
\subsubsection{Varying the energy range} \label{subsec:energyrangeevenquartile}

While we have previously explored the effect of changing the PSF to one appropriate for other energy ranges, we now  explore the effect of changing the energy range itself. We kept the upper limit of the energy range fixed at 20 GeV, since high-energy photons are rare and their inclusion/exclusion is unlikely to qualitatively change the results. We varied the low-energy limit of the energy range between 0.6-3.2 GeV, spanning the peak of the GCE, by including or excluding low-energy bins. As discussed previously, the bin boundaries are log-spaced in energy, with 10 bins per decade, starting at 0.2 GeV.

As a first test, we sought to understand how the sensitivity could be expected to vary just as a result of the modified angular resolution combined with the larger number of photons in low-energy bins. To explore this question, we held the underlying physical model fixed, and treated the enhanced number of photons as an effective exposure factor $\chi_{\text{eff}}$, while using the appropriate PSF for the lowest-energy photons in the analysis. Specifically, we took $\chi_\text{eff}$ to be the ratio of the total number of photons in the real data (over the whole sky) in the modified energy range, to the total number of photons in the original energy range.

\begin{figure}
    \centering
    \includegraphics[width=0.98\linewidth]{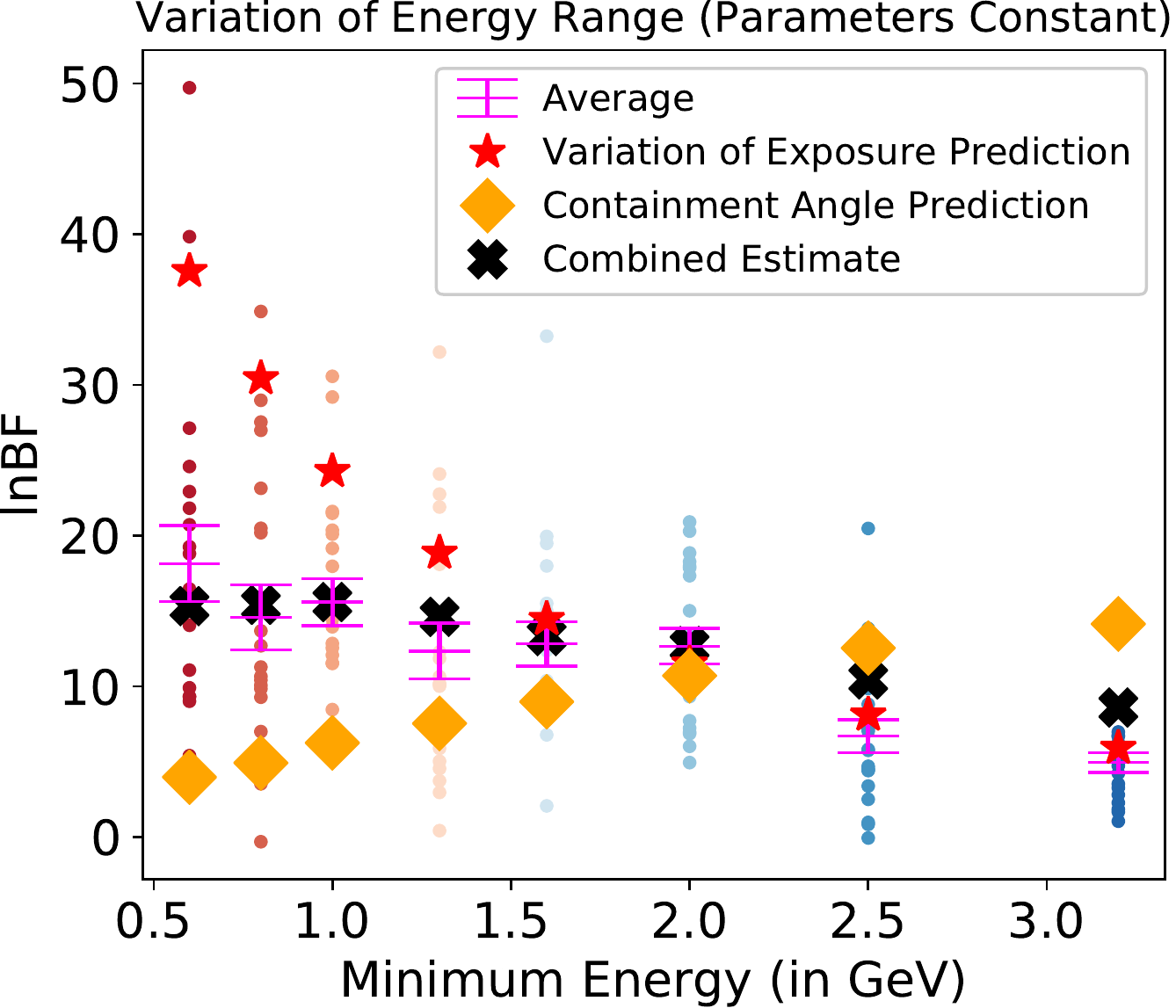}
    \caption{$\ln{\text{BF}}$ across 20 realizations (circle markers), and $\langle \ln{\text{BF}} \rangle $ with error bars obtained from the $\sigma/\sqrt{20}$ standard error of the mean (magenta), varying minimum energy while holding the parameters constant. All scans employ the angular resolution of the lowest-energy photons in the range.
    The increase in photon counts is captured through an effective exposure factor $\chi_{\text{eff}}$. The red stars display the predicted effect of this exposure variation, while the orange diamonds predict the effect of changing the containment angle. The black filled ``X'' symbols display an estimate for the overall sensitivity change from combining the two (see text for details). }
    \label{fig:energyrangevaryparamconst}
\end{figure}

Fig.~\ref{fig:energyrangevaryparamconst} shows the results of this test. The results indicate that due to the worse angular resolution obtained by including data from lower energy ranges, we expect at best a mild increase in the (expected) sensitivity, compared with a substantial increase in the case where only the exposure is varied. As a first-order comparison to our simulated results, we analyzed the combination of the effects of varying the exposure and PSF as discussed in Sec.~\ref{subsec:varyingexposue}, \ref{subsec:varyingpsf}. Similar to Sec.~\ref{subsec:energyrangeevenquartile}, we define a rescaling factor $r=r_{\text{exposure}}(\chi)r_{\text{PSF}}(\eta)$, where $r_{\text{exposure}}(\chi)$ is the ratio of $\langle\ln{\text{BF}}\rangle(\chi)$ to $\langle\ln{\text{BF}}\rangle(\chi=1)$, obtained from Eq.~\ref{eqn:powerlawshift} and Table~\ref{tab:exposurefit}. $r_{\text{PSF}}(\eta)$ is the ratio of $\langle\ln{\text{BF}}\rangle(\eta)$ to $\langle\ln{\text{BF}}\rangle(\eta[2-20\,\text{GeV}])$ (baseline case), obtained from Fig.~\ref{fig:containmentanglep6v11}. $\langle\ln{\text{BF}}\rangle(\chi)$ and $\langle\ln{\text{BF}}\rangle(\eta)$ are denoted as red stars and orange diamonds on Fig.~\ref{fig:energyrangevaryparamconst}, respectively. To obtain the combined estimate denoted by the black filled ``X", we multiplied the calculated $r$ to the $\langle \ln{\text{BF}}\rangle [2-20\,\text{GeV}]$ obtained from the simulations in Fig.~\ref{fig:energyrangevaryparamconst}. The estimates indicate a fairly flat scaling behavior of sensitivity across different energy ranges. This suggests that the beneficial effects of increasing sensitivity are canceled out by the worsening of angular resolution at lower energy ranges. 

As an example, consider varying the minimum energy of the event selection from 2.0 to 1.0 GeV. The $68 \%$ containment angle of the PSF increases from the baseline $~0.23^{\circ}$ to $~0.40^{\circ}$. As shown in Fig.~ \ref{fig:containmentanglep6v11}, this change in PSF induces a decrease in $\braket{\ln{\text{BF}}}$ by a factor of $0.58$. However, the larger number of photon counts with a minimum energy of 1.0 GeV corresponds to an effective rescaling factor of $\chi_{\text{eff}} =2.75$ relative to the case with minimum energy 2.0 GeV (ignoring differences in the spectrum between the different components). Therefore, based on Tab. \ref{tab:exposurefit}, the value of $\ln{\text{BF}}$ should increase by a factor of $\sim 2.10$, if this exposure change were the only factor. The combined effect would correspond to only a $\sim 22\%$ increase in $\ln{\text{BF}}$. Thus in this case, we would expect the increase in sensitivity from additional photons to come close to offsetting the loss of angular resolution, leading to very little net change in sensitivity (with perhaps a slight advantage for a 1.0 GeV minimum energy). This resembles the roughly flat behavior with energy we actually observe in Fig.~\ref{fig:energyrangevaryparamconst}.

This is how the sensitivity would behave if the signal and backgrounds had identical spectra, but of course this is not the case for the GCE. For a specific signal, such as the GCE in this case, we need to either input a theoretical spectrum for each component, or re-fit the model parameters from the real data in each energy band. We take the latter approach here, and then repeat the sensitivity analysis on data simulated using these updated, energy-dependent parameters.
\begin{figure}
    \centering
    \includegraphics[width=0.98\linewidth]{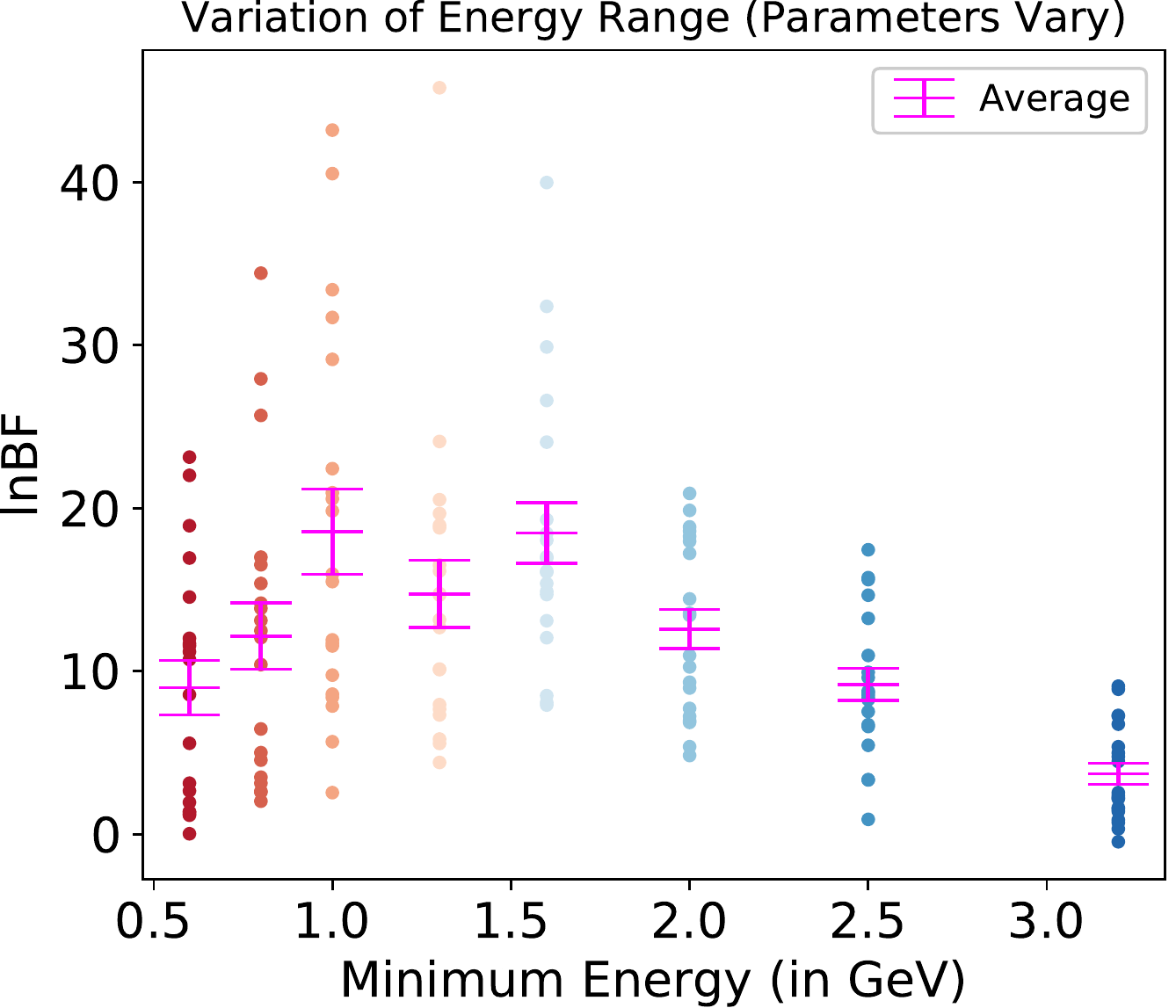}
    \caption{$\ln{\text{BF}}$ across 20 realizations (circle markers), and $\langle \ln{\text{BF}} \rangle $ with error bars obtained from the $\sigma/\sqrt{20}$ standard error of the mean (magenta), varying the minimum energy and updating the template parameters to match the posterior median parameters from fits to the real data in the same energy range. All scans employ the angular resolution of the lowest-energy photons in the range.} 
    \label{fig:energyvariationrefit}
\end{figure}

Fig.~\ref{fig:energyvariationrefit} shows the result of these simulations and analyses. If only photon number and angular resolution were relevant, there would be a strong argument for extending the energy range for the analysis all the way down to 0.6 MeV (or lower), but for the actual GCE spectrum we observe that the highest expected sensitivity is obtained for a minimum energy of 1.0 or 1.6 GeV. This energy scale roughly coincides with the peak of the GCE distribution. 

One might ask if features in Fig.~\ref{fig:energyvariationrefit} simply reflect fluctuations in the total GCE flux inferred from the real data in different energy ranges (used to fix the simulation parameters). We checked this explicitly and found no evidence of such an association; the parameters controlling the simulated GCE PS flux vary smoothly over the relevant range of threshold energies, and the fluctuations in Fig.~\ref{fig:energyvariationrefit} are thus likely to be statistical.

\subsection{Pixel size variation} \label{sec:pixelsizevar}
We examined a wide range of pixel size to determine an optimal value for analysis. We started at an nside value of 512 and downgraded to nside values of 256, 128, 64, 32. At each pixel level, we computed $\ln{\text{BF}}$ across 20 realizations.

\begin{figure}
    \centering
    \includegraphics[width=0.98\linewidth]{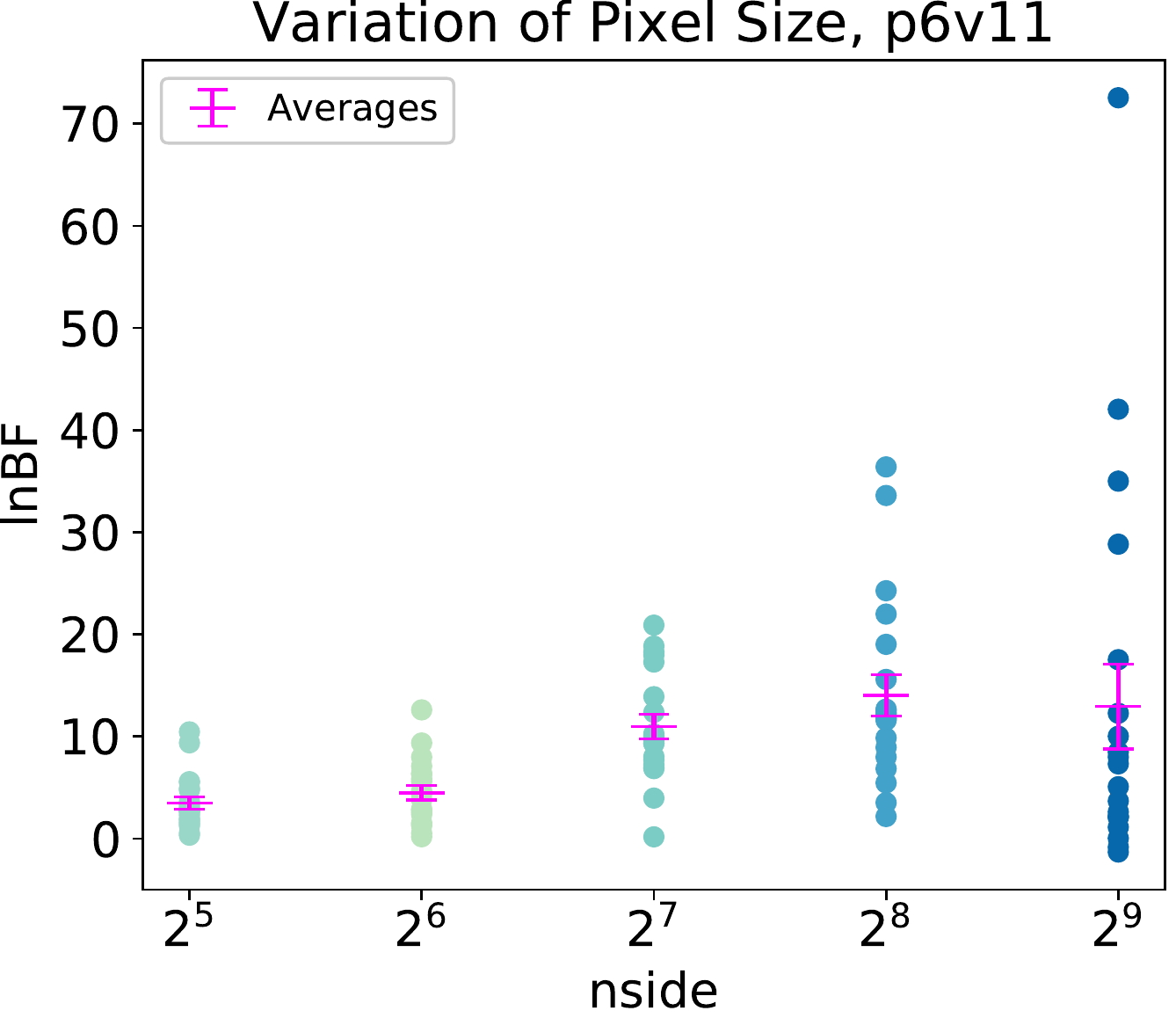}
    \caption{$\ln{\text{BF}}$ across 20 realizations (circle markers), and $\langle \ln{\text{BF}} \rangle $ with error bars obtained from the $\sigma/\sqrt{20}$ standard error of the mean (magenta), varying the nside parameter. The pixel area varies inversely with nside.
    }
    \label{fig:pixelvar}
\end{figure}

Fig. \ref{fig:pixelvar} shows the recovered sensitivity as a function of nside. 
We find that there is a slight increase in the sensitivity to a population of PSs as resolution is improved. However, the scatter between individual realization is also increased. It is plausible that this occurs because with small pixel sizes there is a greater risk that a relatively-bright source happens to land near a pixel boundary and consequently loses significance. In realizations where this behavior happens to be rare, the significance is naturally higher than for smaller nside (as the likelihood contributions from a larger number of pixels are summed), but in other realizations this effective dimming of the sources will markedly decrease the inferred significance of the population. (An alternative way to think about this is that pixels much smaller than the PSF are not independent data points, and so by treating them as independent we may artificially enhance the apparent significance of the result \cite{Collin:2021ufc},\footnote{We thank Nicholas Rodd for pointing out this effect.} but may also miss correlations that could reveal a signal.) In the regime where the pixel size is significantly larger than the PSF, we expect the significance to be reduced because we have reduced the number of independent data points (pixels), discarding information in the process.

It appears that nside 128 and 256 are likely the optimal values: nside 256 has a slightly higher average sensitivity but with considerably more scatter between realizations. The relative insensitivity of NPTF methods to pixel size, in a simpler context, was previously studied in Ref.~\cite{2011ApJ...738..181M}.

\begin{figure*}
    \centering
    \includegraphics[width=0.49\linewidth]{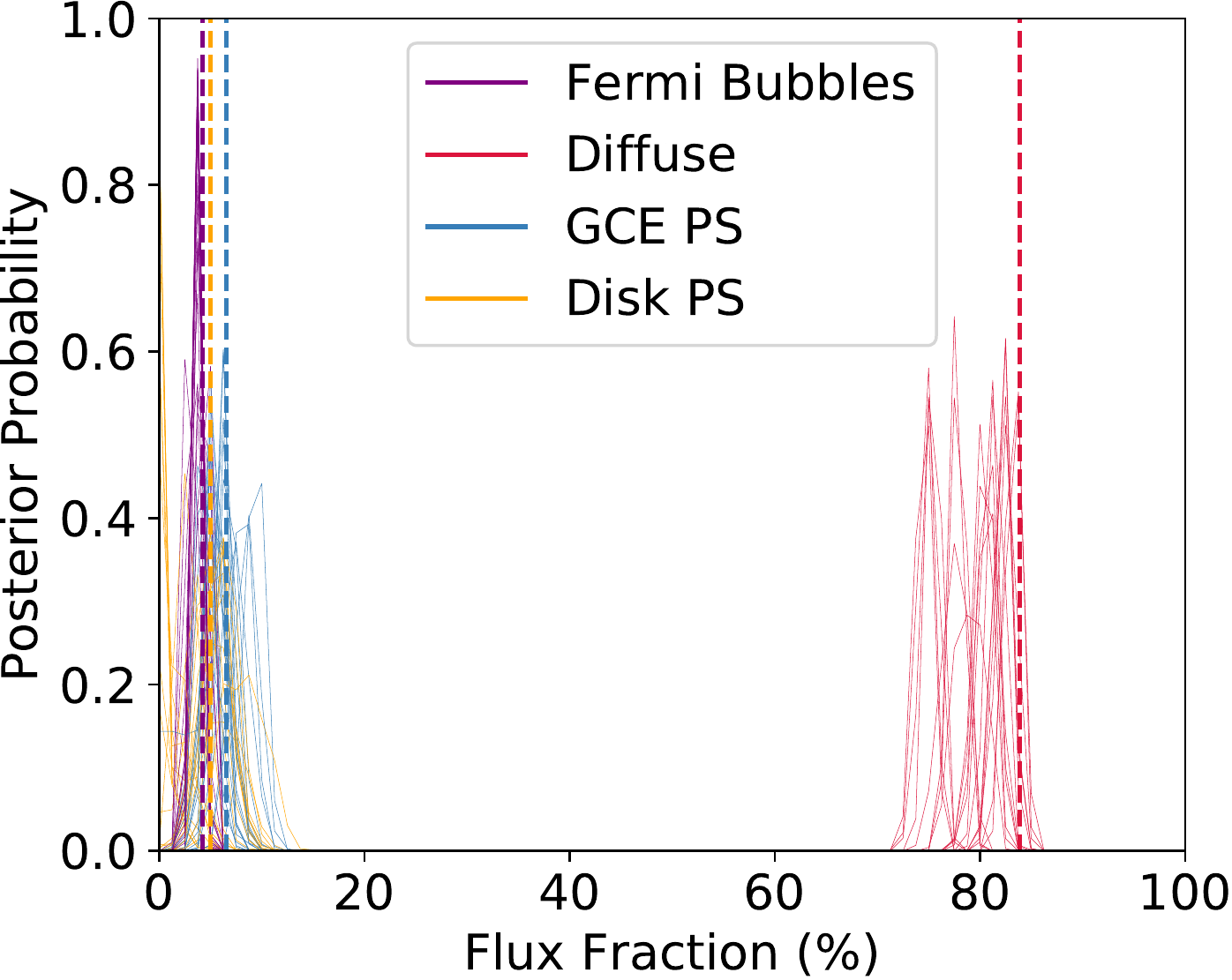}
    \includegraphics[width=0.49\linewidth]{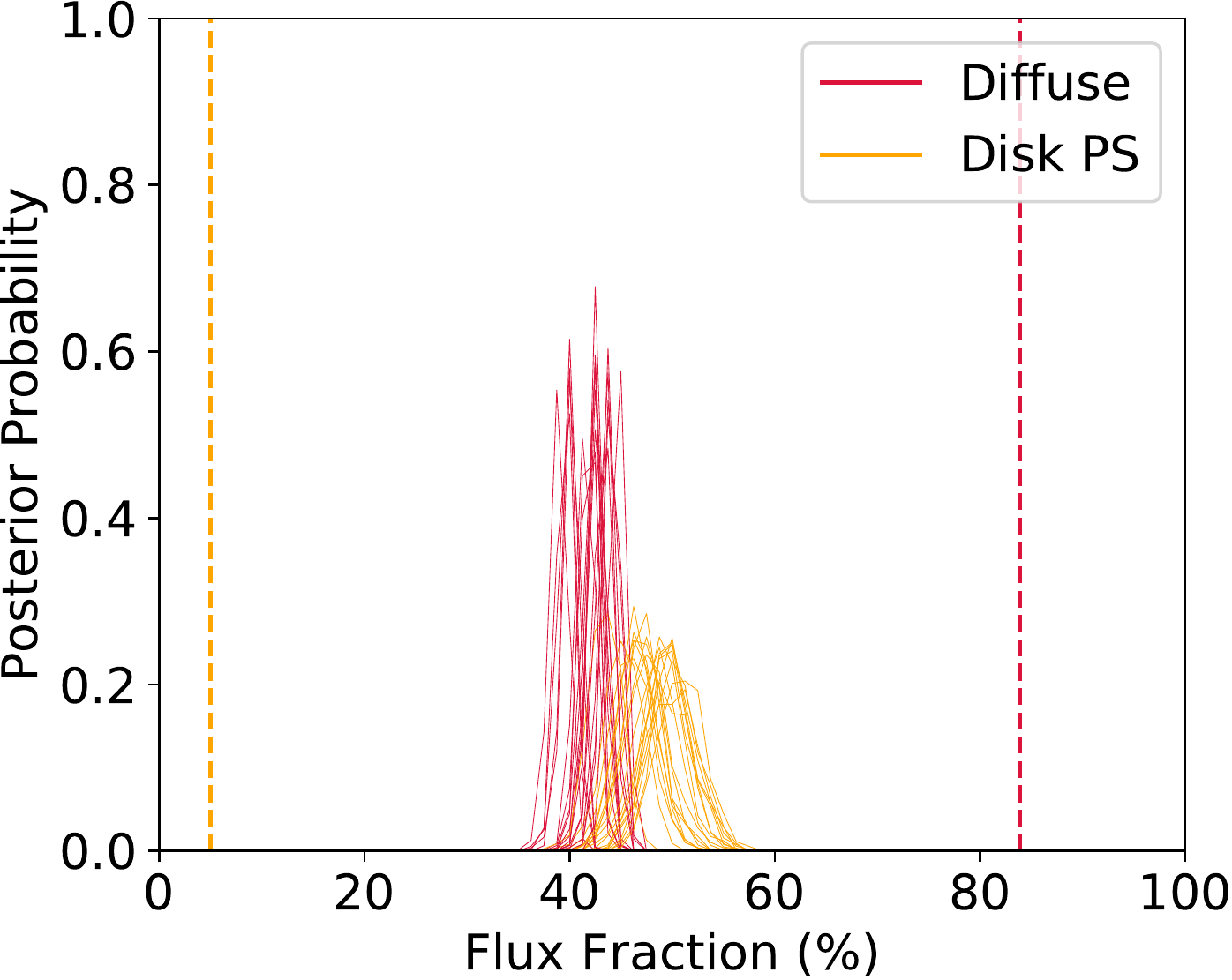}
    \caption{\textit{Left}: Flux fraction plot at the common pixel size employed in previous analyses (nside = 128) of certain emissions: \textit{Fermi} bubbles (purple), diffuse emission (red), GCE PS (blue), disk PS (orange). \textit{Right}: Flux fraction plot demonstrating mis-reconstruction of the diffuse (red) and disk PS (orange) components at a low nside value (nside = 32). The recovered flux fraction are obtained from 20 simulations.  The vertical dashed lines denote the flux fraction injected into the simulations.}
    \label{fig:pixelvarintensityp6v11}
\end{figure*}

On the other end of the spectrum, low nside values cause severe discrepancies in the recovered flux fraction for Galactic diffuse emission and PS populations associated with the Galactic disk (Disk PS). At an nside value of 32, for example, a significant portion of the injected Galactic diffuse emission is absorbed into the Disk PS template. In contrast, at the standard choices of pixel size (such as nside = 128), the recovered flux fractions are fairly consistent with the injection values. The left panel of Fig.~\ref{fig:pixelvarintensityp6v11} shows the recovered flux fractions and the injected values (dashed lines) for nside = 128 for the \textit{Fermi} bubbles, diffuse, GCE PS, and disk PS contributions. The right panel of Fig.~\ref{fig:pixelvarintensityp6v11} displays the mis-reconstruction of the diffuse and disk PS emissions due to a large pixel size (nside = 32). The recovered flux fractions are systematically biased across all realizations, even in this case where the model is correct by construction. Thus low-nside analyses should be treated with considerable caution, in addition to their lack of sensitivity.

\subsection{Source brightness} \label{subsec:nptfitdetectionlimit}

\begin{figure*}
    \centering
    \includegraphics[width=0.49\linewidth]{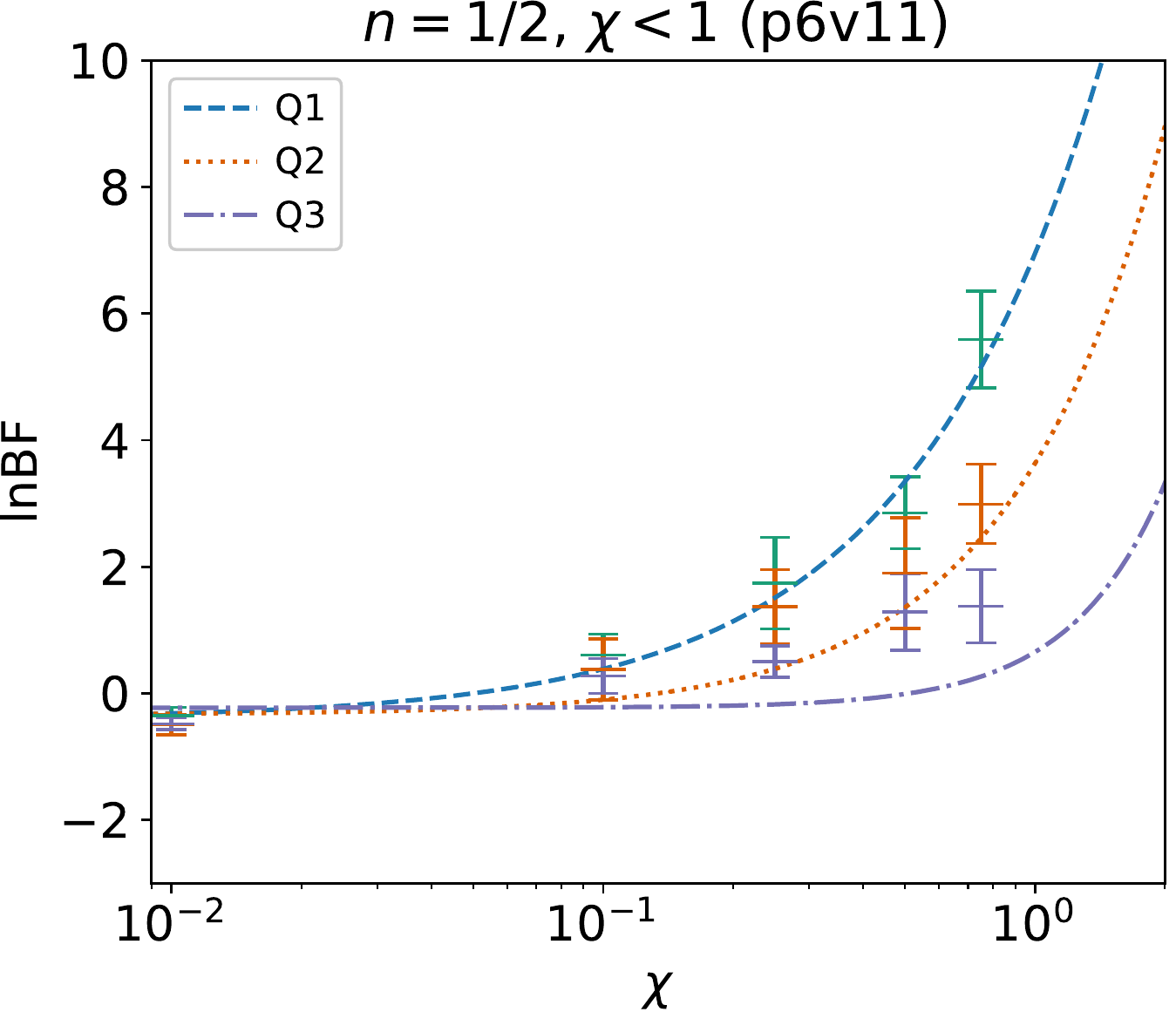}
    \hspace{0.4em}
    \includegraphics[width=0.49\linewidth]{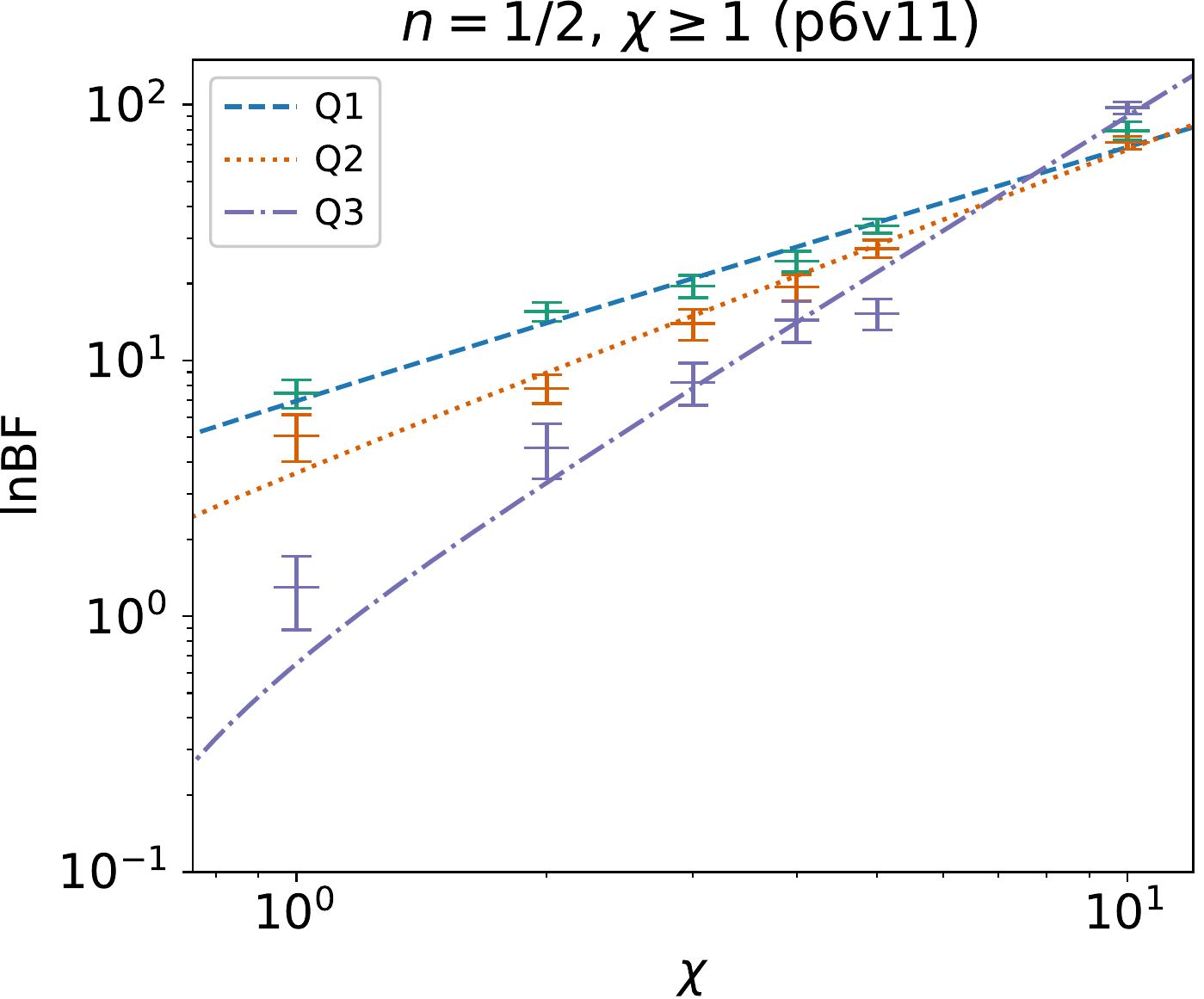} \\
    \includegraphics[width=0.49\linewidth]{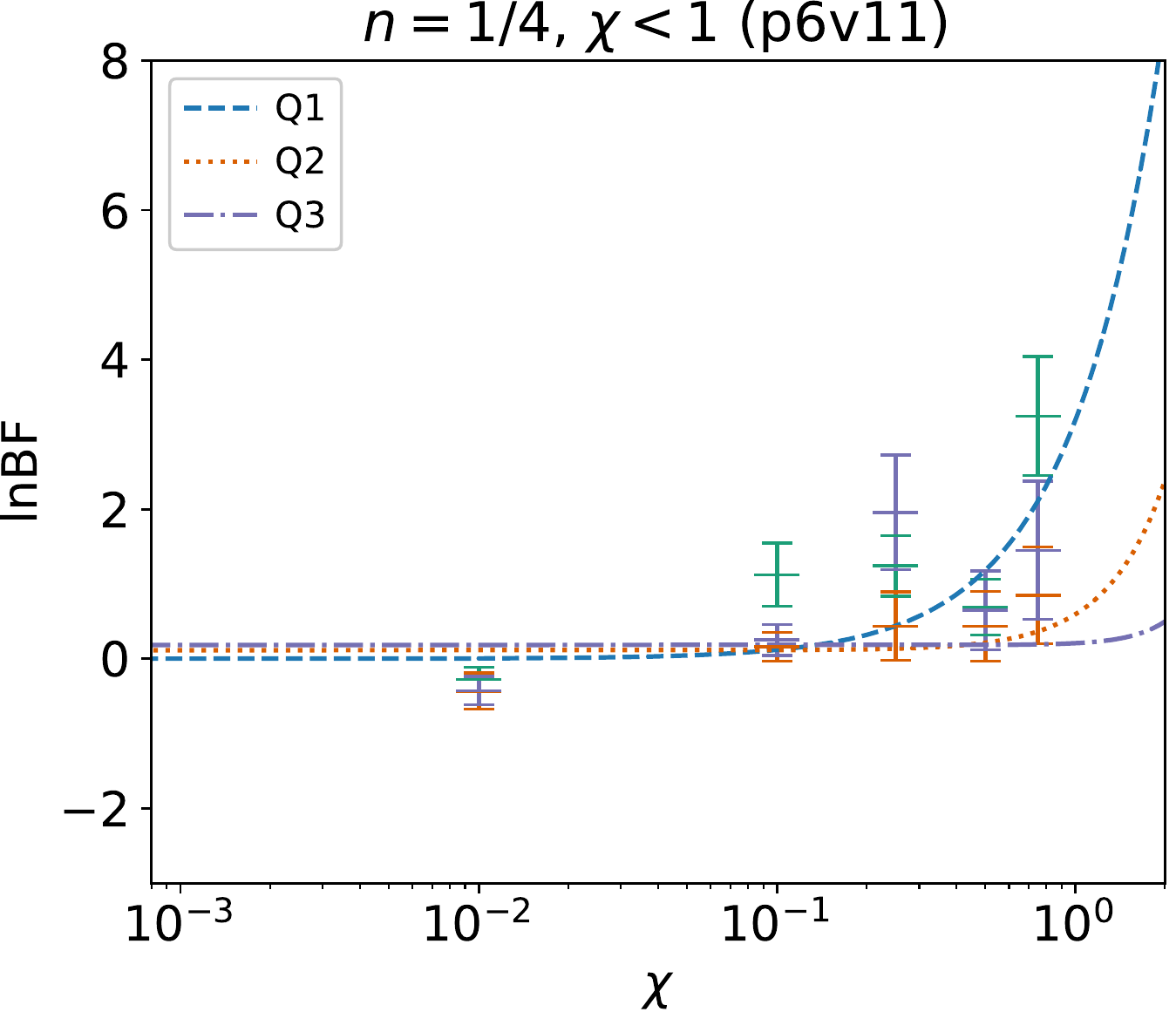}
     \hspace{0.4em}
     \includegraphics[width=0.49\linewidth]{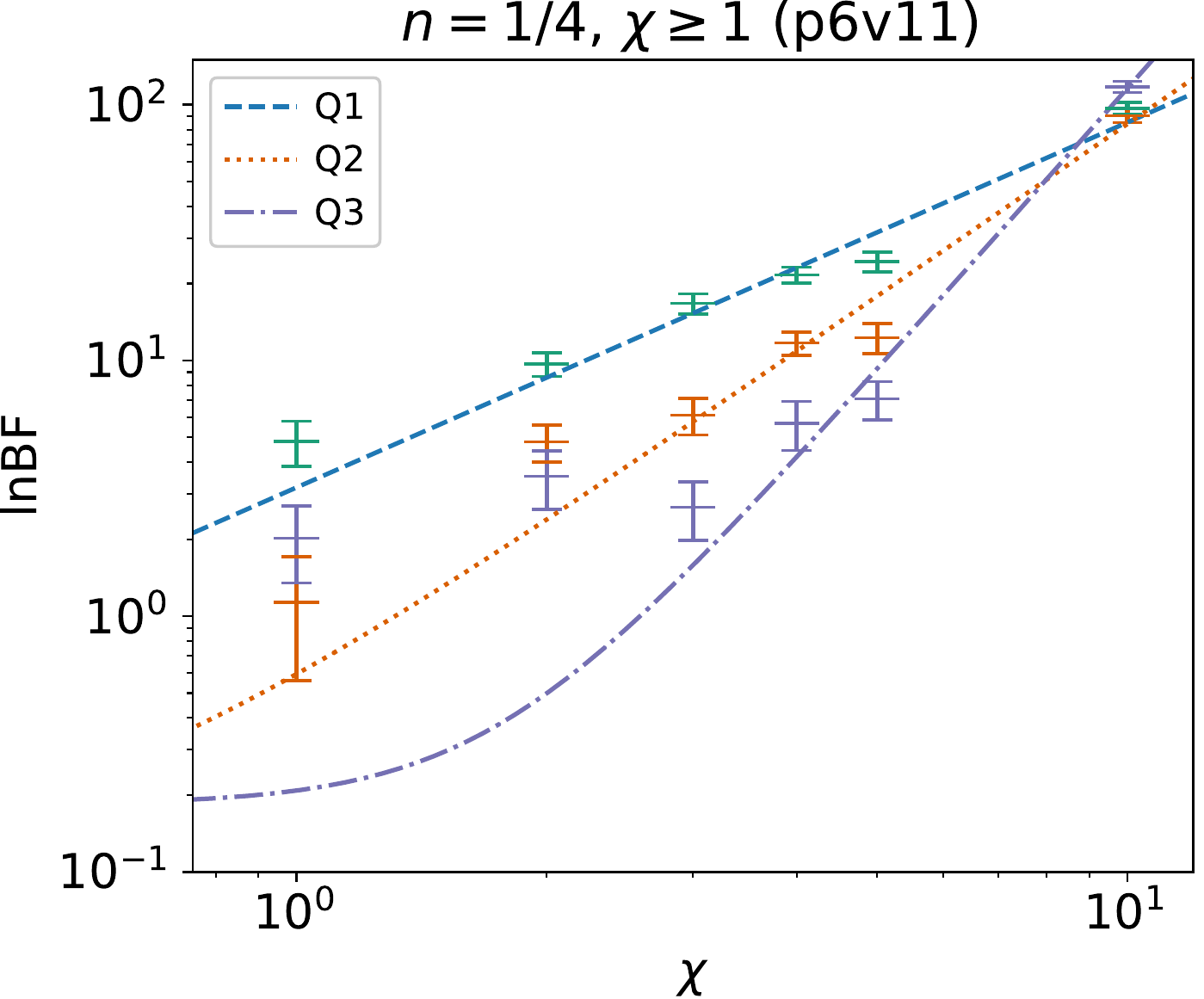}
    \caption{$\langle \ln{\text{BF}}\rangle$ and the $\sigma/\sqrt{20}$ standard error of the mean across 20 realizations at different $\chi$ values for the top three quartiles graded by angular resolution. \textit{Left}: realizations with $\chi<1$. \textit{Right}: realizations with $\chi \geq 1$. In the \textit{upper panels} we simulated the sources at half of the baseline brightness ($S_{b}=17.11$ from Table~\ref{tab:fitparam}), but doubled the number of sources, in accordance with Eq.~\ref{eq:nphotons}; in the \textit{lower panels} we simulated the sources at $1/4$ of the baseline brightness but quadrupled the number of sources. The dashed, dotted, and dash-dotted lines correspond to the shifted power law (Eq.~\ref{eqn:powerlawshift} for quartiles 1, 2, and 3, respectively).}
    \label{fig:angexp0.5sb}
\end{figure*}

\begin{table*}
\centering
\begin{tabular}{l r r r r  }
\hline
 &  $n=1/4$ & $n = 1/2$ & $n=1$ & \\
\hline
\textit{Quartile} 1\\
\hline
$\alpha$ (coefficient) & $3.18 \pm 0.70$ & $7.35 \pm 0.49$ & $11.30 \pm 0.69$ \\
$\beta$ (power) & $1.43 \pm 0.13$  & $0.97 \pm 0.05$ & $0.76 \pm 0.04$ \\
$\gamma$ (shift) & $0.0003 \pm 0.31$ & $-0.40 \pm 0.14$& $-0.62 \pm 0.20$\\
\hline
\textit{Quartile} 2\\
\hline
$\alpha$ (coefficient) & $0.48 \pm 0.18$ & $3.97 \pm 0.48$& $6.14 \pm 0.78$  \\
$\beta$ (power) & $2.24 \pm 0.20$  & $1.22 \pm 0.07$& $1.01 \pm 0.08$ \\
$\gamma$ (shift) & $0.11 \pm 0.26$  & $-0.34\pm 0.18$& $-0.002 \pm 0.37$ \\
\hline
\textit{Quartile} 3\\
\hline
$\alpha$ (coefficient) & $0.02 \pm 0.02$ & $0.88 \pm 0.39$ & $2.44 \pm 0.96$ \\
$\beta$ (power) & $3.67 \pm 0.40$  & $ 2.01 \pm 0.22$& $1.50 \pm 0.21$ \\
$\gamma$ (shift) & $0.18 \pm 0.30$  & $-0.23 \pm 0.19 $& $0.30 \pm 0.37$ \\
\hline
\botrule
\end{tabular}
\caption{Best-fit parameters obtained using least-squares regression method for the power law fit to the $\ln{\text{BF}}$ values (based on $20$ realizations for each point) as we varied the source brightness of the GCE PS component in \texttt{p6v11}. Parameter uncertainties obtained from the standard error of the mean across 20 realizations are also displayed.}
\label{tab:tradeoffp6v11}
\end{table*}

Finally, we would like to understand how sensitivity scales with the brightness of the sources. This is important both for understanding the prospects for future detection of a bulge PS population with {\it Fermi}, and in understanding the likely sensitivity of future telescopes with different exposure and angular resolution.

\begin{figure}
    \centering
    \includegraphics[width=0.98\linewidth]{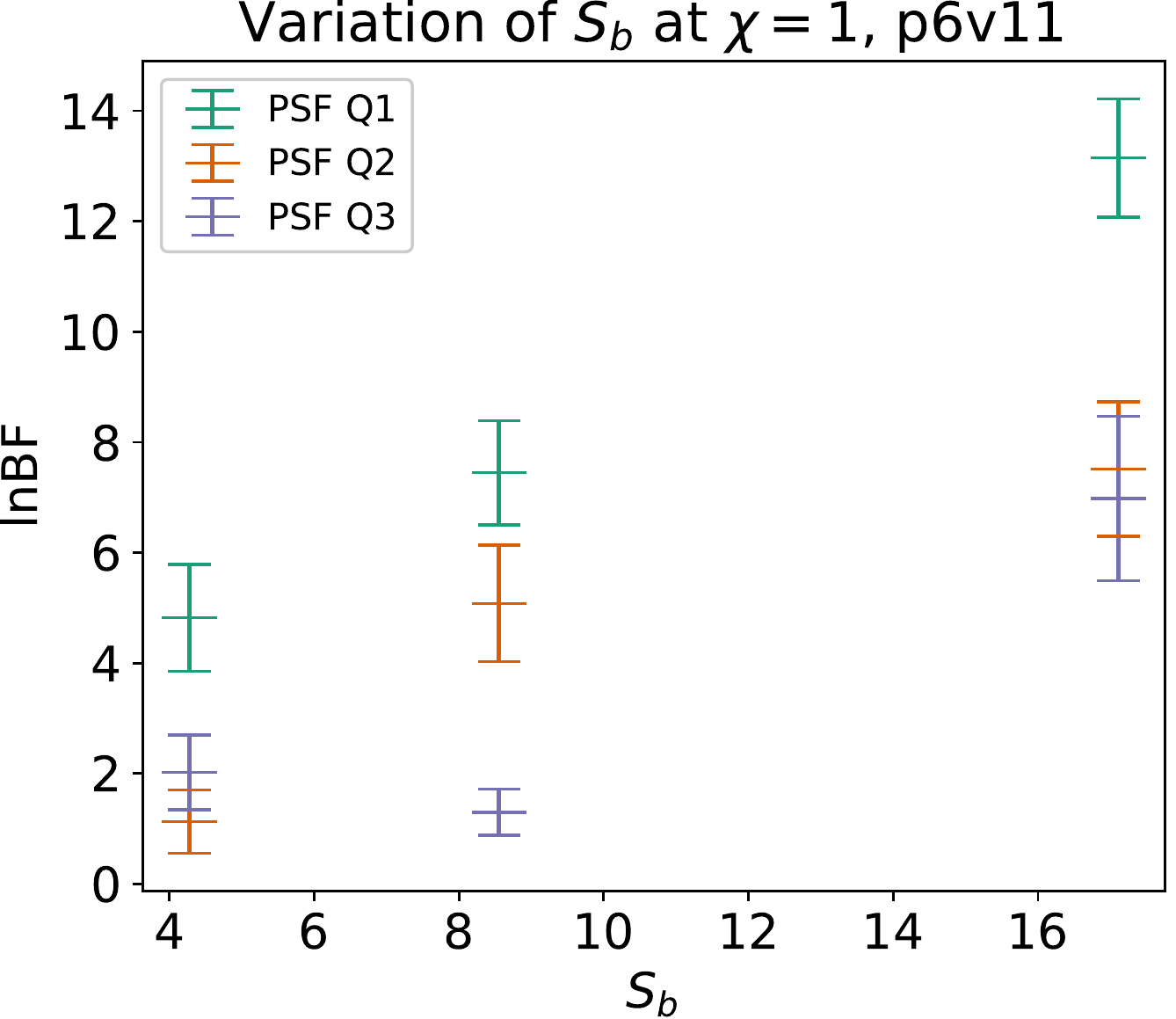}
    \caption{$\langle \ln{\text{BF}}\rangle$ and the $\sigma/\sqrt{20}$ standard error of the mean (obtained from 20 realizations at each point) across the top three quartiles at varying brightness level for the full \textit{Fermi} case, with the baseline exposure. The best-fit $S_b$ in the real data is 17.11; we test the effect of reducing $S_b$ by a factor of 2 or 4. The baseline case was scanned using nlive=300, while the other cases were performed using nlive=100 for computational efficiency.
    }
    \label{fig:sensitivitychi1}
\end{figure}

We perform the same test as described in Sec.~\ref{subsec:tradeoffs}, where the default source brightness parameter for the GCE PS component $S_b = 17.11$ is multiplied by a factor of $n=1/2$ or $1/4$ (and the number of sources is modified to keep the total flux constant). Fig.~\ref{fig:angexp0.5sb} shows the results, while Table~\ref{tab:tradeoffp6v11} shows the corresponding best-fit parameters and uncertainties obtained from the power-law fit in Eq.~\ref{eqn:powerlawshift}, using the least-squares method.

Our results indicate that the parameter (of the power-law fit) $\alpha$ increases at higher values of $n$. Since $\alpha$ corresponds to the sensitivity at $\chi=1$, it is expected to observe an increase as the brightness of the sources increase. On the other hand, $\beta$ decreases modestly as $n$ increases. Assuming that $\braket{\ln{\text{BF}}}$ has the same functional behavior as $\braket{\Delta\ln{\mathcal{L}}}$, our expectation from the analytic results is $\beta\approx 1$, at least if the fit is dominated by the region where the expected number of counts/source and hence $\braket{\ln{\text{BF}}}$ is large. However, at low brightness levels, as predicted by the analytic equations, we expect a quadratic scaling, $\beta\approx 2$. Thus it is reasonable to see a stronger scaling for smaller values of $n$, where the quadratic behavior is relevant for a larger range of $\chi$. In other words, increased exposure is more important for fainter sources.

To clearly demonstrate the effect of varying the source brightness on the overall sensitivity, we plotted  $\langle \ln{\text{BF}}\rangle(\chi=1)$ across 20 realizations as a function of $S_{b}$, for the top three PSF quartiles, in Fig.~\ref{fig:sensitivitychi1}. Within the $S_{b}$ range we tested, we find that, on average, sensitivity increases as the PS population brightens. The decrease in sensitivity qualitatively matches expectations based on the theoretical results presented in Sec.~\ref{sec:analyticforms}, where the expected $\Delta\ln{\mathcal{L}}$ monotonically decreases as $s$ is lowered, although we observe significant scatter across the different quartiles.

We eventually expect to lose all sensitivity to point sources as their brightness becomes sufficiently small; as $S_{b}\to 0$, it becomes impossible to distinguish the PS flux and the smooth emission, and we expect the effects of priors to dominate the results (see Appendix \ref{sec:priors} for a study of the prior dependence). Roughly speaking, we expect the onset of this regime to occur around $S_b \sim 1$ (as for sources with counts $s \ll 1$ it will be rare to observe multiple photons from a single source); more quantitatively, we observe from the analytic approximations (Eq.~\ref{eqn:lnBFcase2}) that the log likelihood difference per pixel becomes rapidly smaller once $ks$ drops below 1. (Nonetheless, given a sufficiently large number of pixels, it may still be possible to probe the properties of sources with $s \lesssim 1$, as discussed in e.g.~\cite{List:2021aer}.)

Approximating the sensitivity-exposure relation by the power law with an additive shift as in Eq.~\ref{eqn:powerlawshift}, we can estimate the $\chi$ value that corresponds to  $\ln{\text{BF}} \geq 1$ for Quartile 1, as a function of the brightness of the source population (indicating some sensitivity to the sources; we could increase this threshold to require a more significant detection). Using the best-fit parameters in Table~\ref{tab:tradeoffp6v11}, we find that for $n=1$, this threshold corresponds to $\chi \geq 0.08$; for $n=1/2$, to $\chi \geq 0.18$; and for $n=1/4$, to $\chi \geq 0.45$. As expected, a higher level of exposure is required to detect fainter populations of PSs. 

\section{Conclusions} \label{sec:conclusion}

We have investigated the statistical behavior of non-Poissonian template fitting, as implemented in the \texttt{NPTFit} public code, when characterizing unresolved point sources. In particular we have explored the sensitivity to point sources both analytically and numerically, in a simplified isotropic-emission scenario and a realistic scenario relevant to gamma rays from the inner Galaxy. We define the sensitivity to point sources (or detectability of point sources) as the ratio of the maximum likelihood (analytic case) or Bayesian evidence (numerical case) between the true underlying model versus a model that excludes point sources associated with the signal component. 
We first derived analytic estimates of the sensitivity for point sources with a delta-function SCF, where all sources have the same expected number of photons per source $s$. We found that the expected contribution to the log likelihood ratio from a given pixel is a function only of $ks$, where $k$ is the fraction of the emission that is attributed to point sources; the scaling of the log likelihood with $k s$ is linear for $k s \gg 1$ and quadratic for $k s \ll 1$. We also examined the variance in this sensitivity, reflecting the expected scatter between realizations. By exploring a range of scenarios, we found that the standard deviation of our sensitivity metric was generically smaller than the expectation value by a factor of $\sqrt{A}$, where $A$ can be the total number of sources in the ROI, the total number of pixels in the ROI, or the log likelihood ratio itself; in general, whichever parameter is smallest dominates the variance. This behavior can lead to a relatively large scatter between the sensitivity inferred from different simulations, which we indeed observe in the numerical data. We tested our analytic predictions using numerical simulations in a simplified case where both point sources and smooth emission are isotropic (detailed in Sec.~\ref{sec:isotropic} and Appendix \ref{app:isotropic}), and found that the analytic results were quantitatively quite accurate in the case where the PSF is very narrow, and provide a good description of various scaling relations even with a more realistic PSF. The analytic and isotropic results may be relevant to other analyses employing non-Poissonian template fitting, e.g. the analysis of the neutrino background presented in \cite{Aartsen:2019mbc}.  

We then numerically investigated the role of several key parameters in the \texttt{NPTFit} analysis of a population of point sources associated with the \textit{Fermi} gamma-ray skymap. The parameters we tested included exposure, angular resolution, source brightness and pixel size. This analysis was performed within the full \textit{Fermi} scenario using both \texttt{p6v11} templates and \texttt{Model A} templates for the Galactic diffuse emission (detailed in Appendix \ref{app:modela}). The results we quote below are based on simulations with the default \texttt{p6v11} template from the public \texttt{NPTFit} code, but we found consistent results using an alternative background model denoted \texttt{Model A}. 

For the cases we tested, we found the following general relationships between exposure, angular resolution, and sensitivity: 
\begin{itemize}
\item Gaining exposure alone induces an increase in sensitivity that is roughly linear for the exposure range and analysis choices we focused on. The analytic approximation predicts a scaling between linear and quadratic; our results are broadly consistent with this expectation, but in complex/realistic sky models involving multiple templates with different morphologies, we have observed both slightly sub-linear scaling and stronger-than-quadratic scaling, the latter in quartiles with poor angular resolution.  
\item Worse angular resolution results in lower sensitivity (as expected) such that the third-highest-quality angular resolution quartile has average sensitivity approximately $60\%$ lower than of the highest quality quartile (although the degree of degradation can vary depending on the exposure level and the brightness of the sources, with the loss of sensitivity being more pronounced for fainter sources and lower exposure).
\item As a simplified parametrization of the angular resolution (which varies according to quartile selection and included energy range), we examined how the sensitivity varies according to the $68\%$ containment angle radius $\eta$, and found that sensitivity has an approximate inverse proportionality relation to $\eta$. 
\item With these results, we tested the effects of increasing exposure at the expense of angular resolution in practical situations relevant to {\it Fermi} data, by varying the quartile selection and energy range. We find, on average, that the increase in sensitivity from a larger exposure is offset by the degraded angular resolution to a good approximation. 

\item Hence, this tradeoff only produces a small net change in sensitivity; there is a broad range of possible analysis choices that yield similar expected sensitivity. The large scatter in sensitivity between realizations means that some caution is needed when interpreting changes in the Bayes factor in real data upon addition of extra quartiles, energy bins, ROI, etc; it appears possible for changes in the analysis choice that modify the dataset to have a large apparent effect on the Bayes factor purely due to this scatter.
\end{itemize}

Note that the properties (for the exposure vs angular resolution tradeoff) identified above are specific to the current implementation of \texttt{NPTFit}, which assumes that all photons share the same angular resolution (fixed by the worst angular resolution in the dataset). It is possible that a more sophisticated approach that tracks angular resolution separately for different categories of photons could lead to a greater benefit from including lower-angular-resolution quartiles or lower-energy photons.

We also examined the role of source brightness to the sensitivity of \texttt{NPTFit} to point sources to understand how the sensitivity falls off as the source brightness declines. For sufficiently faint point sources, as expected, \texttt{NPTFit} is unable to distinguish point sources from background smooth emission. More specifically, in the top graded quartile for angular resolution, the minimum exposure (in relation to the baseline exposure) required to achieve any hint of detection (defined somewhat arbitrarily as an average Bayes factor $\ln{\text{BF}}\geq 1$) scales as $O(S_{b}^{-1.28})$, where $S_{b}$ is the peak number of photons per Galactic Center Excess point source.

We explored simulations with different pixel sizes, and found that pixel size was not a crucial factor in determining the sensitivity of \texttt{NPTFit}, except in the case of extremely large or small pixel sizes. Very large pixel sizes induced inaccurate recovery of the various physical emission components; small pixel sizes led to higher variance across realizations.

These results serve as a systematic demonstration of the behavior of \texttt{NPTFit} under a range of conditions relevant to analyses of {\it Fermi} data from the inner Galaxy (and potentially more broadly).

\section*{Acknowledgements}

The authors thank Siddharth Mishra-Sharma, Nicholas Rodd, and the anonymous referee for helpful feedback and suggestions. The work of T.R.S. was supported by the US Department of Energy, Office of Science, Office of High Energy Physics of the U.S. Department of Energy under grant Contract No. DE-SC0012567 through the Center for Theoretical Physics at MIT; the National Science Foundation under Cooperative Agreement No. PHY-2019786 (The NSF AI Institute for Artificial Intelligence and Fundamental Interactions, \url{http://iaifi.org/}); and a grant from the Simons Foundation (Grant Number 929255, T.R.S). The work of L.G.C.B. was funded by the Massachusetts Institute of Technology Undergraduate Research Opportunities Program (UROP) office.

\appendix

\section{Detailed Methodology} \label{app:detailedmethodology}

\subsection{Analysis templates} \label{subsec:templateandsmoothing}
In this appendix we detail the spatial templates we used in our analyses, and the physical processes they attempt to capture. For our primary (\texttt{p6v11}) analysis we largely used the templates that were publicly released with the \texttt{NPTFit} code package, as we summarize below. In Appendix~\ref{app:modela} we test the effects of using an alternative pair of templates to describe the Galactic diffuse emission (vs a single template in the \texttt{p6v11} analysis), collectively denoted \texttt{Model A}.

\begin{itemize}
    \item \textbf{\textit{Fermi} Bubbles} \\
    The \textit{Fermi} Bubbles are large lobes of extended hard gamma-ray emission that appear to be emanating from the vicinity of the GC \cite{2010ApJ...724.1044S}. The exact origin of this large structure is currently unknown. The default template in \texttt{NPTFit} is based on Ref.~\cite{2010ApJ...724.1044S} and has a uniform intensity within the Bubbles region (prior to exposure effects). This template has a single associated degree of freedom, its  normalization parameter $A_\text{bub}$.
    \item \textbf{Isotropic Background} \\
    An appreciable fraction of the overall gamma-ray emission is expected to originate from extragalactic sources and possess an approximately isotropic distribution \cite{1972ApJ...177..341K,1975ApJ...198..163F,Sreekumar:1997un,DiMauro:2015tfa}. We thus include a template for emission that is isotropic across the full sky (prior to applying exposure effects). This emission includes photons from both relatively bright sources and much fainter sources, so we include two isotropic components, one representing smooth emission and one representing PSs. 
    
    The smooth isotropic template has a single degree of freedom: its template normalization parameter $A_\text{iso}$. The non-Poissonian template for isotropic PSs has one degree of freedom for its normalization ($A_\text{ISO}^\text{PS}$) and three degrees of freedom associated with its SCF, as described by Eq.~\ref{eqn:scf},  $\{n^\text{ISO-PS}_{1},n^\text{ISO-PS}_{2},S^\text{ISO-PS}_{b}\}$. Note that we label PS templates with capital letters and Poissonian templates with lower-case letters. 
    \item \textbf{Galactic Center Excess (NFW Profile)} \\
    The template for the GCE is constructed from the line-of-sight projection of the square of a generalized Navarro-Frenk-White (NFW) profile \cite{Navarro:1996gj, Calore:2015bsx, Daylan:2016tia}. The explicit form of the density profile of the generalized NFW for the Milky Way is:
    \begin{equation}
        \rho(r) \propto \frac{(r/r_{s})^{-\gamma}}{(1+r/r_{s})^{3-\gamma}}
        \label{eqn:densityprofnfw}
    \end{equation}
    where $r_{s} = 20 \text{ kpc} $, $\gamma = 1.25$. The line of sight projection to determine the 2D spatial flux distribution is:
    \begin{equation}
        \label{eqn:nfwlos}
        J(\psi) = \int_0^\infty \rho^{2}(r) ds
    \end{equation}
    where $\psi$ is the angle away from the GC, $s$ is the line-of-sight distance, and $r^2 = R^2 + s^2 - 2 R s \cos\phi$ where $R$ is the distance between the Earth and the GC.
     
     As for the isotropic emission, we use this spatial morphology for two different emission components, one smooth and one representing a PS population. The smooth isotropic template has a single degree of freedom: its template normalization parameter $A_\text{gce}$. The non-Poissonian template for GCE-distributed PSs has one degree of freedom for its normalization ($A_\text{GCE}^\text{PS}$) and three degrees of freedom associated with its SCF, as described by Eq.~\ref{eqn:scf},  $\{n^\text{GCE-PS}_{1},n^\text{GCE-PS}_{2},S^\text{GCE-PS}_{b}\}$.

    \item \textbf{Galactic Diffuse Emission}\\
    The Galactic diffuse emission is the dominant contributor to the total gamma-flux in the energy range relevant to {\it Fermi}. There are three main contributors to this emission: (1) Proton collisions with the gas, which produce pions that decay to photons, $pp \to X + \pi^{0} \to X + \gamma \gamma$, (2) Inverse Compton Scattering (ICS) driven by cosmic ray electrons upscattering abundant low-energy photons in the interstellar radiation field, and (3) cosmic ray electrons scattering on the ambient gas, which produces photons via bremsstrahlung. The first two processes generally dominate the total emission.
    
    In our default analysis, the diffuse emission is modeled as a single smooth template similar to the \texttt{p6v11} diffuse model packaged with the public version of \texttt{NPTFit}, but accounting for 573 weeks of data. This template has a single degree of freedom: its template normalization parameter $A_\text{dif-p6v11}$. This template is quite old and is known to have significant deficiencies for modeling the real Galactic sky (e.g. \cite{Buschmann:2020adf}), so in Appendix~\ref{app:modela} we tested the effect of replacing this template with two separate templates based on the \texttt{Model A} of Ref.~\cite{Calore:2014xka}. One template corresponds to the $\pi^{0}$ and bremsstrahlung emission (which traces the gas), the other to the ICS component. Each template has a single degree of freedom corresponding to its normalization parameter; these parameters are denoted $A_\text{pibrem}$ and $A_\text{ics}$ respectively.
    
    \item \textbf{Disk-Correlated Sources} \\
    The galactic disk of the Milky Way contains many gamma-ray PSs, which we model as having a doubly exponential thick-disk source distribution \cite{Lee:2015fea}. The 3D number density of sources is approximated as:
    \begin{equation}
        n(z,R) \propto \exp{\left[\frac{-R}{5 \text{ kpc}} \right]}\exp{\left[\frac{-|z|}{1 \text{ kpc}} \right]}
    \end{equation}
    where $R$ and $z$ are respectively the radial distance from the GC and the height above the Galactic disk. This number density function is then integrated along the line of sight as described for the GCE template.

\item \textbf{Template Normalization}\\
We normalize each of these templates according to the standard \texttt{NPTFit} convention. For smooth/Poissonian components, we store counts templates denoted by $T$, which include a factor of the exposure map (i.e. they are in counts not flux), are smoothed by the PSF, and are normalized to have an average of one count/pixel in the region within $30^\circ$ of the GC where $|b| > 2^\circ$. The maps of expected counts per pixel for a smooth/Poissonian component are thus given by $A_\text{smooth} T$ for each component, and $A_\text{smooth}$ describes the average number of counts/pixel in the normalization region. For PS components, the templates $T_{\text{PS}}$ denote the number of sources per pixel rather than the number of photon counts per pixel, and so do not include the instrument response functions (i.e. PSF smoothing and multiplication by the exposure map). The $T_{\text{PS}}$ templates are normalized so that when $T_{\text{PS}}$ is multiplied by the exposure map, the mean of the resulting map is equal to the mean of the exposure map (in the same normalization region as discussed above). The map of expected sources per pixel is proportional to this normalized $T_{\text{PS}}$ map, with the normalization controlled by $A_{\text{PS}}$ and the properties of the SCF, as in Eq.~\ref{eqn:scf}.

\end{itemize}
\vspace{22mm}
\subsection{Analysis parameters} \label{subsec:analysisparam}

\begin{figure}
    \centering
    \includegraphics[width=0.98\linewidth]{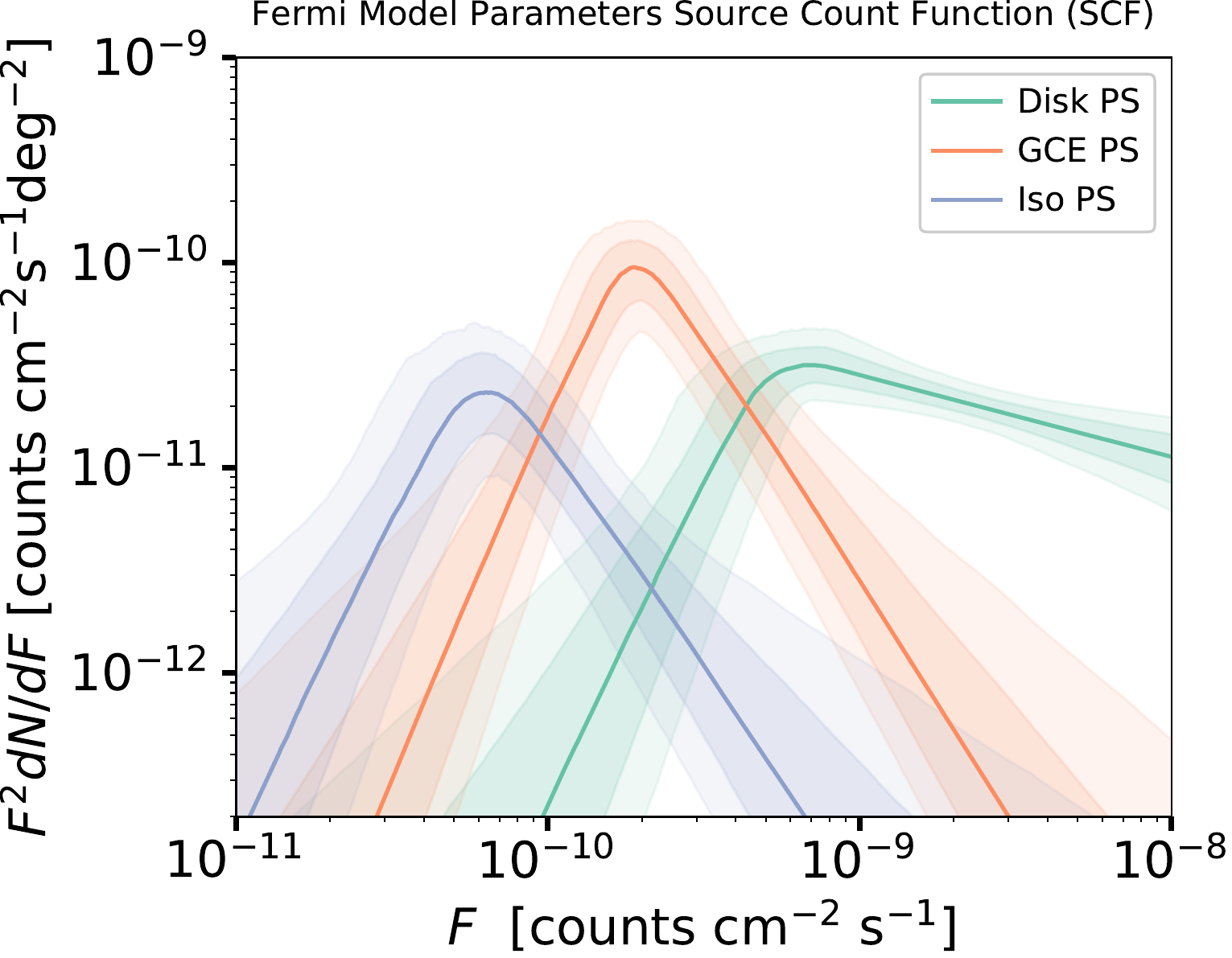}
    \caption{The singly-broken power law form (median and the $68\%$ and $95\%$ containment bands) that characterizes the source count functions for the Disk PS, GCE PS, and Isotropic PSs components obtained from \textit{Fermi} data collected until 2019. These functions were also used to describe and generate simulated data.}
    \label{fig:bestfitscf}
\end{figure}

The baseline analysis parameters we used to create the sky map simulations were obtained by fitting the real \textit{Fermi} data (with the data selection described in Sec.~\ref{subsec:DataSelection}) using \texttt{NPTFit}. We perform the fit at nlive = 500 for greater accuracy, and we choose the ROI as described in Sec.~\ref{subsec:NPTFit}. The priors used in this fit, which are also used to analyze simulation data, are given in Table~\ref{tab:priors}. In Fig.~\ref{fig:bestfitscf}, we plot the best fit source count functions (SCF) we obtained, which describe PSs we injected into our skymap simulations. Note, however, that these source count functions are quite peaked in flux, and the GCE SCF in particular can be treated approximately as a delta function (as we discuss in Appendix~\ref{app:sharpSCF}). 

The resulting posterior median parameters, which are used to generate simulations, are given in Table~\ref{tab:fitparam}. As we will discuss shortly, we repeated the analysis for two different choices of Galactic diffuse emission model, labeled \texttt{p6v11} and \texttt{Model A}, which include different numbers of templates; consequently, some templates are only relevant for one of the two diffuse models.

\begin{table}[H]
\centering
\begin{tabular}{r c c c c c}

\toprule
\hline
Parameters & \texttt{p6v11} & \texttt{Model A} \\

\hline
\hline

$\log_{10} A_\text{bub}$ & $0.03$ & $-0.03$ \\ [2pt]
$\log_{10} A_\text{iso}$ & $-2.16$ & $-1.17$  \\ [2pt]
$\log_{10} A_\text{gce}$& $-2.31$ & $-2.09$ \\ [2pt]
$\log_{10} A_\text{dif-p6v11}$& $1.22$ & - \\ [2pt]
$\log_{10} A_\text{pibrem}$& - & $0.97$ \\ [2pt]
$\log_{10} A_\text{ics}$& - &$0.84$  \\ [2pt]
$\log_{10} A_\text{GCE}^{\text{PS}}$& $-2.47$ & $-1.32$ \\ [2pt]
$n_{1}^\text{GCE-PS}$& $4.39$ & $4.48$ \\ [2pt]
$n_{2}^\text{GCE-PS}$& $-1.57$ & $-1.37$ \\[2pt]
$S_{b}^\text{GCE-PS}$& $17.11$ & $5.56$  \\ [2pt]
$\log_{10} A_\text{DSK}^\text{PS}$ & $-3.73$ & $-2.53$ \\ [2pt]
$n_{1}^\text{DSK-PS}$& $2.40$ & $2.28$  \\ [2pt]
$n_{2}^\text{DSK-PS}$& $-1.23$ & $0.08$  \\ [2pt]
$S_{b}^\text{DSK-PS}$& $48.64$ & $12.86$ \\ [2pt]
$\log_{10} A_\text{ISO}^\text{PS}$& $-1.60$ & $-4.92$ \\ [2pt]
$n_{1}^\text{ISO-PS}$& $4.25$ & $3.69$ \\ [2pt]
$n_{2}^\text{ISO-PS}$& $-1.24$ & $-0.57$ \\ [2pt]
$S_{b}^\text{ISO-PS}$& $5.61$ & $27.41$ \\ [2pt]

\botrule
\end{tabular}
\caption{Posterior median parameters extracted from data collected until 2019 by \textit{Fermi}, and used to generate simulated data. The two columns correspond to (left) use of a single diffuse Galactic emission template, labeled ``dif-p6v11'', and (right) use of two diffuse Galactic emission templates, labeled ``pibrem'' and ``ics''.}
\label{tab:fitparam}
\end{table}

\begin{table}[H]
\centering

\begin{tabular}{r c c c c c}

\toprule

\hline

Parameter  & \texttt{p6v11} & \texttt{Model A} \\

\hline
\hline

$\log_{10} A_{\text{bub}}$ & {[$-3$,$ 1$]} & {[$-3$, $1$]} \\ [2pt] 
$\log_{10} A_\text{iso}$ & $[-3,1]$ &$[-3,1]$  \\ [2pt] 
$\log_{10} A_\text{gce}$& $[-3,1]$ & $[-3,1]$ \\ [2pt] 
$\log_{10} A_\text{dif-p6v11}$& $[-3,1]$ & - \\ [2pt] 
$\log_{10} A_\text{pibrem}$& - & $[-2,2]$ \\ [2pt] 
$\log_{10} A_\text{ics}$& - &$[-2,2]$  \\ [2pt] 
$\log_{10} A_\text{GCE}^\text{PS}$& $[-6,1]$ & $[-6,1]$ \\ [2pt]  
$n_{1}^\text{GCE-PS}$& $[2.05,5]$ & $[2.05,5]$ \\ [2pt] 
$n_{2}^\text{GCE-PS}$& $[-3,1.95]$ & $[-3,1.95]$ \\ [2pt] 
$S_{b}^\text{GCE-PS}$& $[0.05,80]$ & $[0.05,80]$  \\ [2pt] 
$\log_{10} A_\text{DSK}^\text{PS}$& $[-6,1]$ & $[-6,1]$ \\ [2pt] 
$n_{1}^\text{DSK-PS}$& $[2.05,5]$ & $[2.05,5]$  \\ [2pt] 
$n_{2}^\text{DSK-PS}$& $[-3,1.95]$ & $[-3,1.95]$  \\ [2pt] 
$S_{b}^\text{DSK-PS}$& $[0.05,80]$ & $[0.05,80]$ \\ [2pt] 
$\log_{10} A_\text{ISO}^\text{PS}$& $[-6,1]$ & $[-6,1]$ \\ [2pt] 
$n_{1}^\text{ISO-PS}$& $[2.05,5]$ & $[2.05,5]$ \\ [2pt] 
$n_{2}^\text{ISO-PS}$& $[-3,1.95]$ & $[-3,1.95]$ \\ [2pt] 
$S_{b}^\text{ISO-PS}$& $[0.05,80]$ & $[0.05,80]$ \\ [2pt] 

\botrule
\end{tabular}
\caption{Prior ranges employed for analyses of both real and simulated data, unless specified otherwise in the text.}
\label{tab:priors}
\end{table}

\subsection{Simulation procedure} \label{subsec:appendixsimprocedure}
We generated exposure maps for our event selection using the {\it Fermi} Science Tools, and then simulated realizations of the gamma-ray sky using the parameters of Table~\ref{tab:fitparam}, modified by any variations that we wanted to test. When we performed variations that shifted the simulated values of individual parameters, we shifted the relevant priors to ensure they were still appropriate; see the sections relevant to individual analyses for details.

We worked in \texttt{Python 3.6.4}. For templates representing smooth emission, we performed a standard random Poisson draw to obtain the counts per pixel; for non-Poissonian templates, we simulated sky realizations using the public code package \texttt{NPTFit-Sim}.

All simulations were done using \texttt{HEALPix} with nside = 128, with the exception of tests described in Sec. \ref{sec:pixelsizevar} and analogous tests in the appendices). We simulated and saved realizations of each individual template, and then summed them to obtain mock realizations of the \emph{Fermi} dataset.

\section{Degree to which the SCF approximates a delta-function}
\label{app:sharpSCF}

\begin{figure*}
    \centering
    \includegraphics[width=0.49\linewidth]{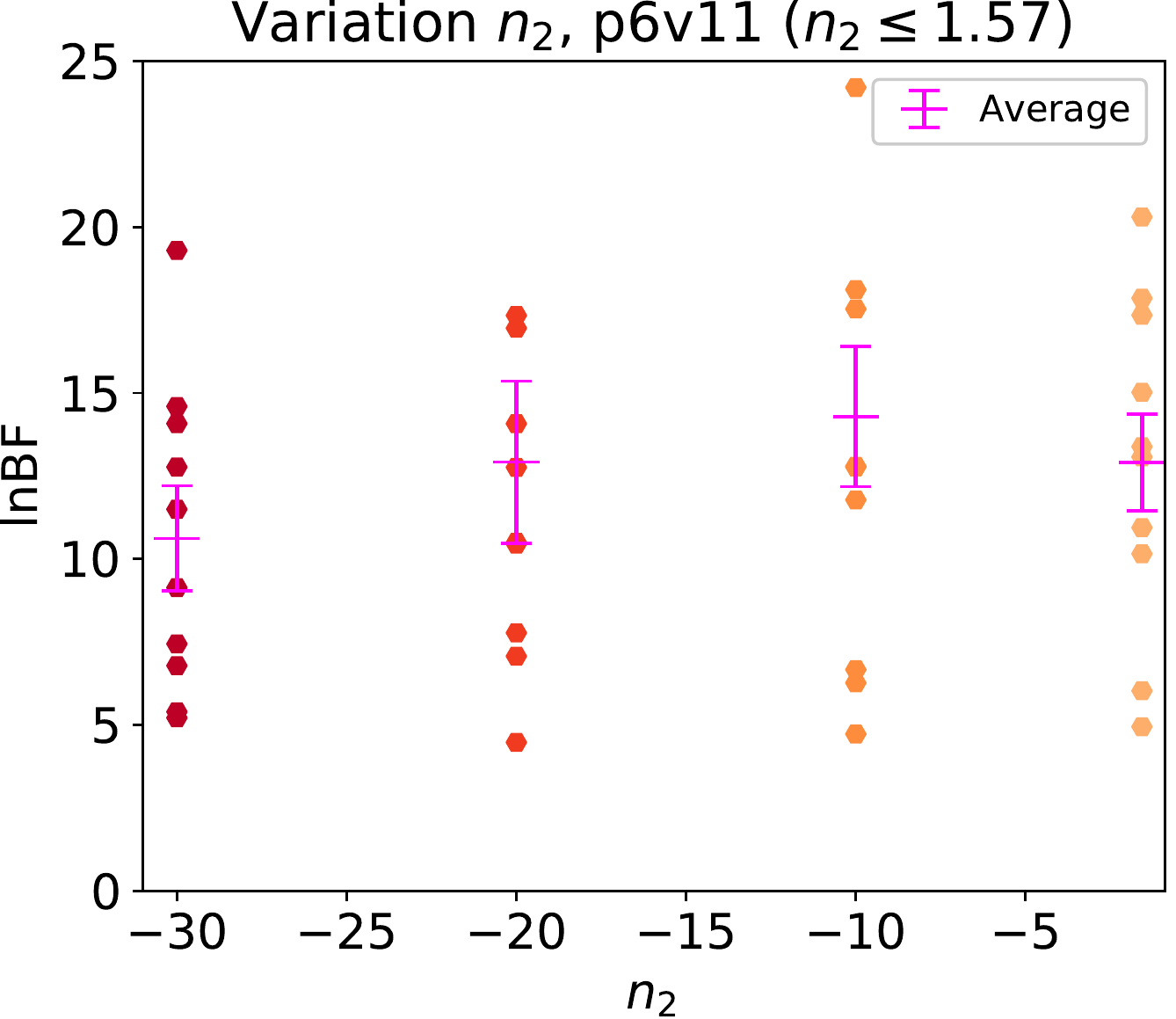}
    \includegraphics[width=0.49\linewidth]{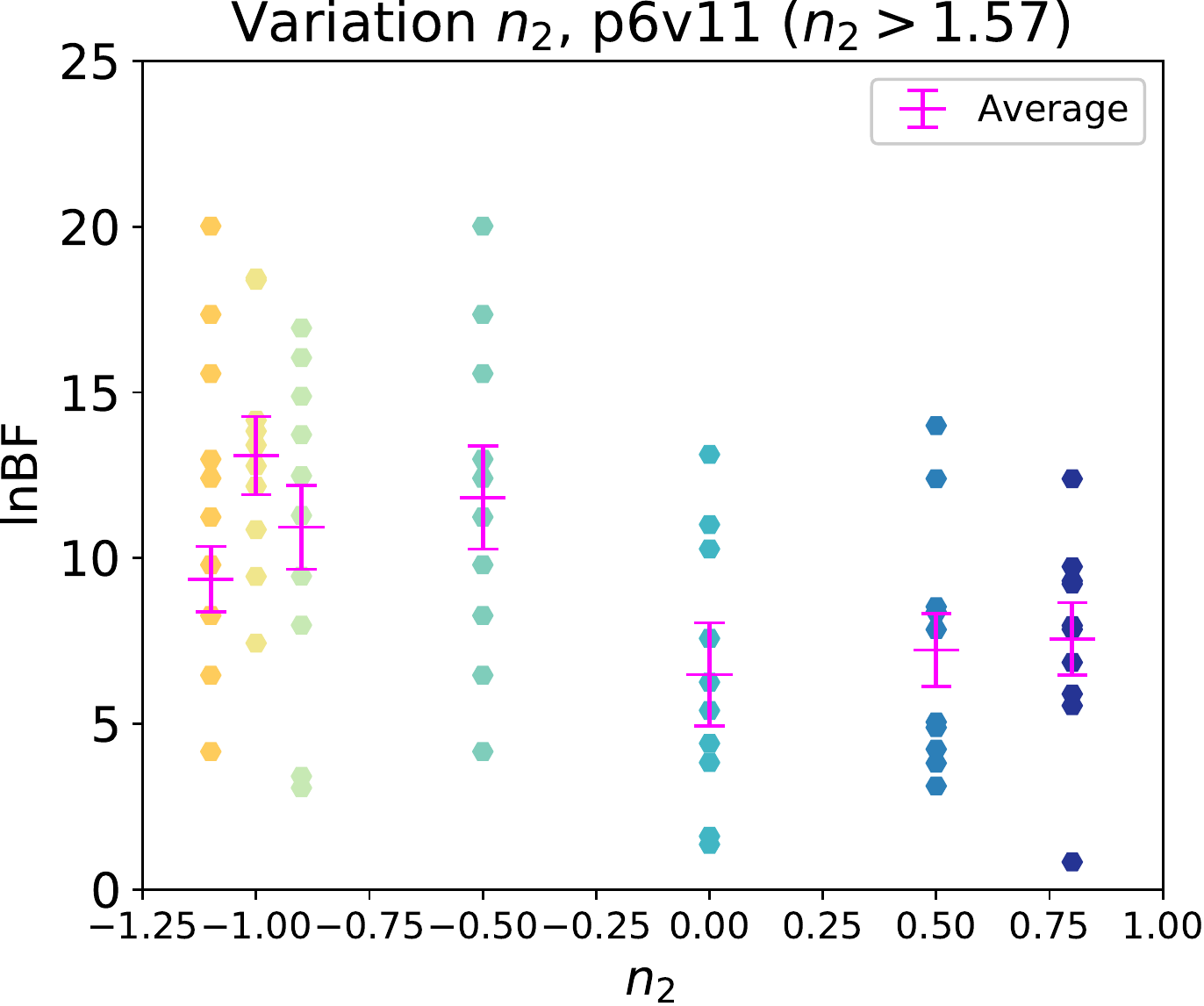}
    \caption{$\ln{\text{BF}}$ across 10 realizations (circle markers), and $\langle \ln{\text{BF}} \rangle $ with error bars obtained from the $\sigma/\sqrt{10}$ standard error of the mean (magenta), varying $n_{2}$ of the SCF for the GCE PS component in the case with all \textit{Fermi} templates. 
    }
    \label{fig:scftest}
\end{figure*}

To demonstrate the sharpness of the peak of the SCF, we studied how the sensitivity of our \texttt{NPTFit}-based pipeline to GCE PSs varies as the SCF width of the GCE PS component narrows. In the SCF described by Eq.\,\ref{eqn:scf}, we varied the value of $n_{2}$ while constraining $n_{1}$ to its posterior median value (4.39), which corresponds to a steep drop at high flux. We adjusted $A_{\text{PS}}$ to preserve the total number of photons across each scenario.

We began by testing very negative values of $n_2=-30, -20, -10$, corresponding to near-delta-function forms for the SCF. The left panel of Fig.\,\ref{fig:scftest} displays the resulting $\ln{\text{BF}}$ values (with 10 realizations at each $n_2$ value and nlive=300). The sensitivity results were very stable as we varied $n_2$ in this range, indicating that our baseline choice of SCF is not meaningfully different from a delta function for the purpose of sensitivity calculations.  We then continued increasing $n_2$ to larger values (up to 0.8), as shown in the right panel of Fig.\,\ref{fig:scftest}; the results remained relatively stable for $n_2 < 0$, but the sensitivity dropped off after that point, by approximately a factor of 2. 

Our ability to test even broader SCFs is limited by our functional form for the SCF, which only has one break in the power law; increasing $n_2$ above 1 would mean the sources are predominantly low-luminosity, but in the absence of a low-end break in the SCF, this would yield an infinite number of sources. We attribute the lack of a precipitous sensitivity decline to the fact that we are restricted to $n_2 < 1$, where consequently both the number of sources and the overall photon flux are dominated by relatively bright sources, with fluxes in the neighborhood of $S_b$. For example, at $n_{2}=0$, $~45\%$ of flux comes from sources above $S_{b}$, and even at the highest $n_{2}$ value we tested $n_{2}=0.8$, $~33\%$ of flux is drawn from sources above $S_{b}$. To see a very large decrease in sensitivity from increasing $n_2$ (rather than lowering $S_b$), we would likely need to allow for more flexible SCF prescriptions with a large $n_2$ but a low-luminosity cutoff.

\section{Effects of Priors on $\Delta \ln{\mathcal{L}}$ and $\ln{\text{BF}}$ } \label{sec:priors}

Since we work in a Bayesian framework, we expect our results for the sensitivity to depend on the priors to some degree. Especially where the PSs are faint enough that the data cannot effectively discriminate between PSs and diffuse emission, the flux attribution to PSs vs Poissonian components may also be significantly influenced by the choice of priors (see also discussions in \cite{Chang:2019ars,Collin:2021ufc}). In this appendix we discuss the effect of changing selected priors from log to linear: specifically, we modify the priors that correspond to the normalization factors of the smooth GCE component $A_{\text{gce}}$ and the GCE PS population $A_{\text{GCE}}^{\text{PS}}$. We keep the prior boundaries fixed.

\begin{figure*}
    \centering
    \includegraphics[width=0.49\linewidth]{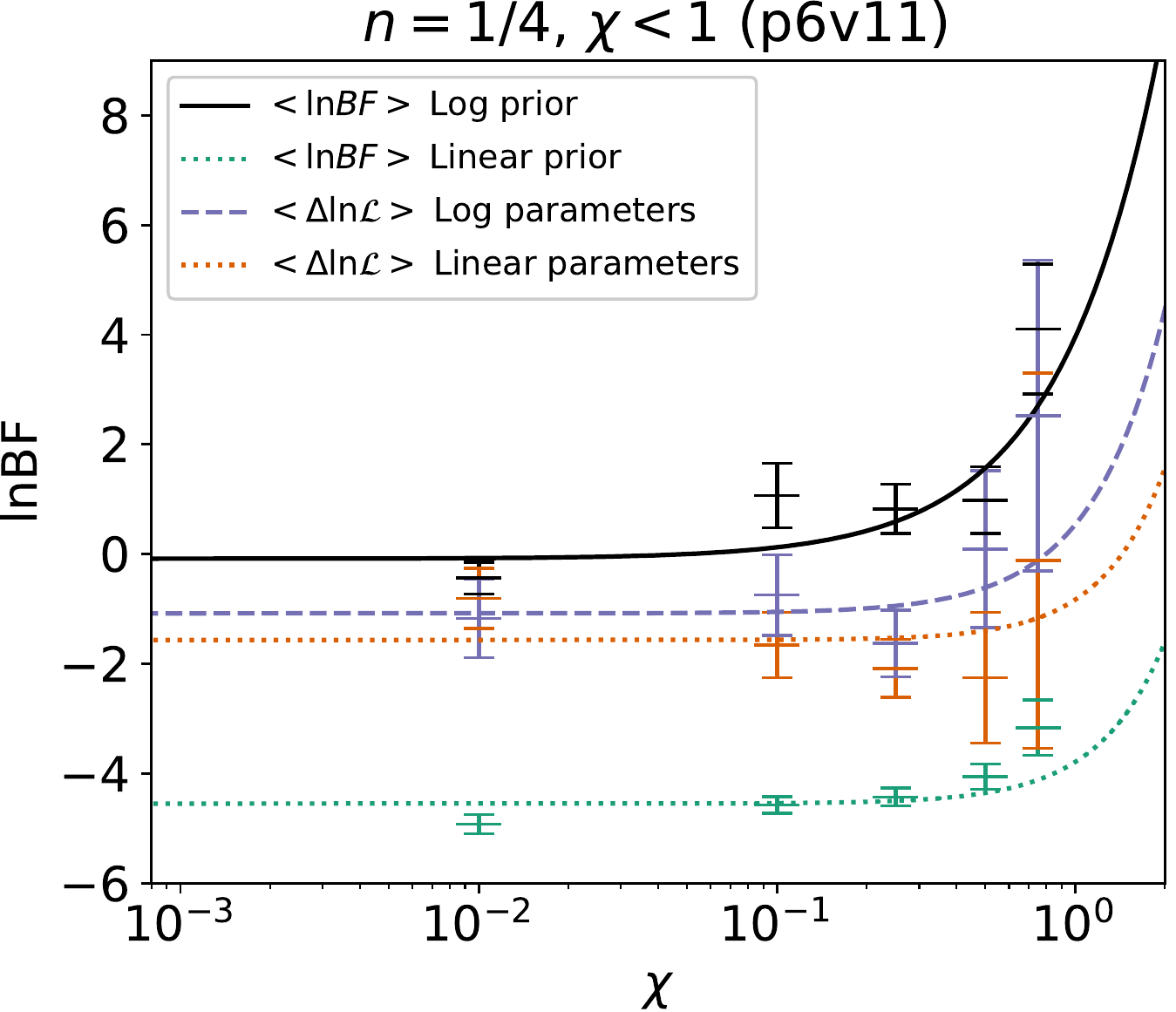}
    \includegraphics[width=0.49\linewidth]{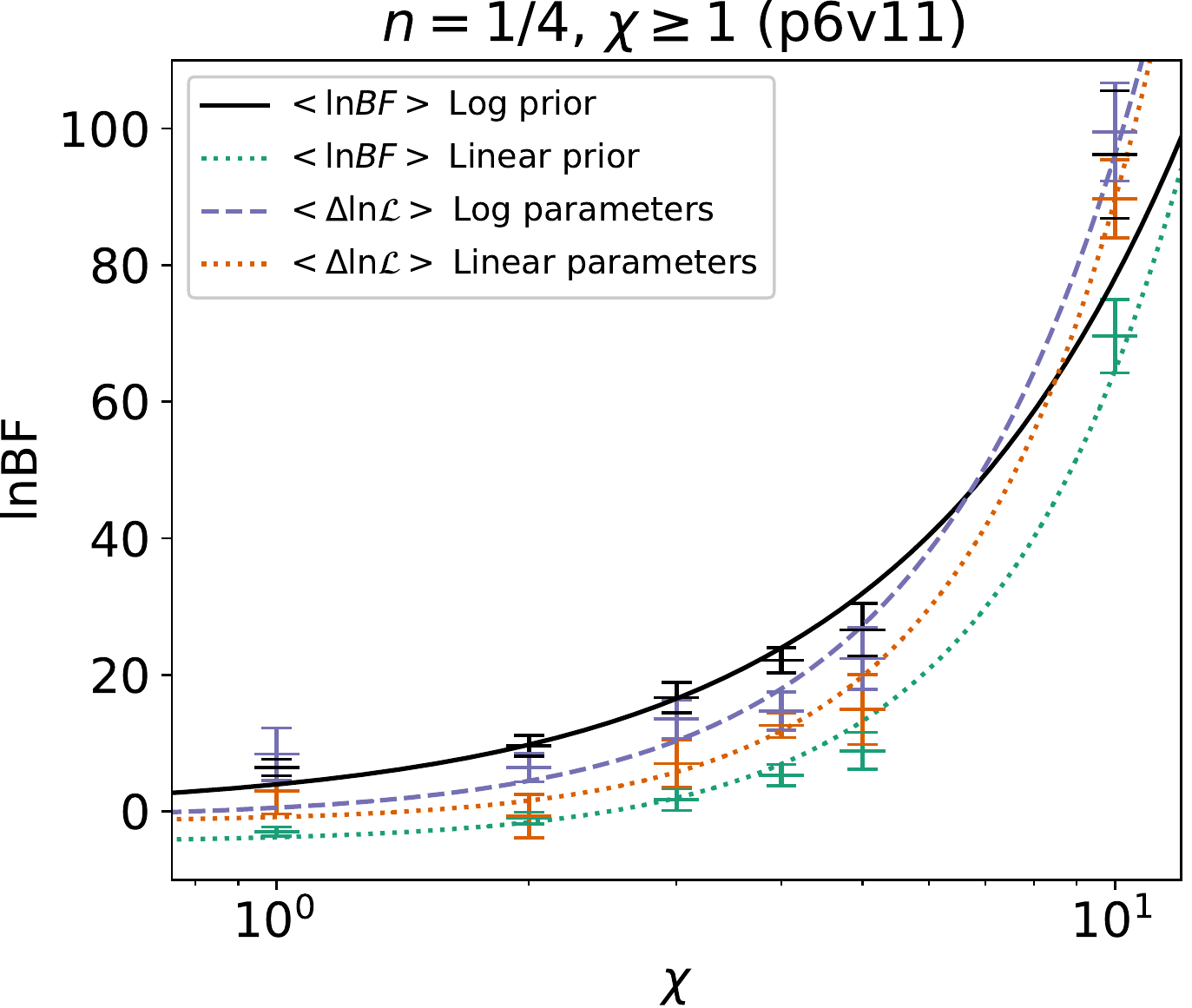}
    \caption{$\braket{\ln{\text{BF}}}$ and $\braket{\Delta \ln{ \mathcal{L}}}$, with error bars obtained from the $\sigma/\sqrt{10}$ standard error of the mean across 10 realizations, varying the $\chi$/exposure factor.  We show results for log uniform and linearly uniform priors on the normalization of the GCE components, and likelihoods are evaluated at the posterior median parameters.} 
    \label{fig:priortest}
\end{figure*}

We compared the $\ln{\text{BF}}$ in favor of PSs for a set of simulations of the full \textit{Fermi} skymap, across exposure factors $\chi =0.01$ to $\chi =10$, for both the case of the best-fit $S_b$ from the real data, and with $S_b$ reduced to $1/4$ of its baseline value (corresponding to fainter point sources). Fig.~\ref{fig:priortest} shows the effect of changing the priors for the faint ($1/4$ baseline brightness) sources, where the impact is most pronounced. In this figure we also plot the results of evaluating $\Delta\ln\mathcal{L}$ between the posterior median parameters obtained from the runs with/without GCE PSs; the likelihood itself does not depend on the priors, but since the posterior distribution is prior-dependent, we observe a small residual difference at the level of the likelihoods as well. (Note that the posterior median parameters do not generally correspond to the maximum likelihood point; evaluating the likelihood ratio between the maximum likelihood points should of course give a prior-independent result.)

We observe that in general using linearly uniform priors on $A$ leads to a smaller sensitivity figure of merit compared to the use of log uniform priors (this also holds for point sources with the baseline brightness level); the linearly uniform priors also give rise to a modestly steeper power-law slope in $\langle\ln{\text{BF}}\rangle$ with increasing $\chi$, as the difference between log and linear priors becomes less pronounced at high exposure. For sufficiently faint sources and/or low exposures, the use of linear priors can lead to an apparent preference for the model with no GCE PSs even when they are present in the simulation (this is as expected; in the absence of data favoring PSs, the Bayesian approach should prefer the simpler model).

\section{Additional analyses for the simplified isotropic scenario} 
\label{app:isotropic}

\begin{figure*}
    \centering
    \includegraphics[width=0.49\textwidth]{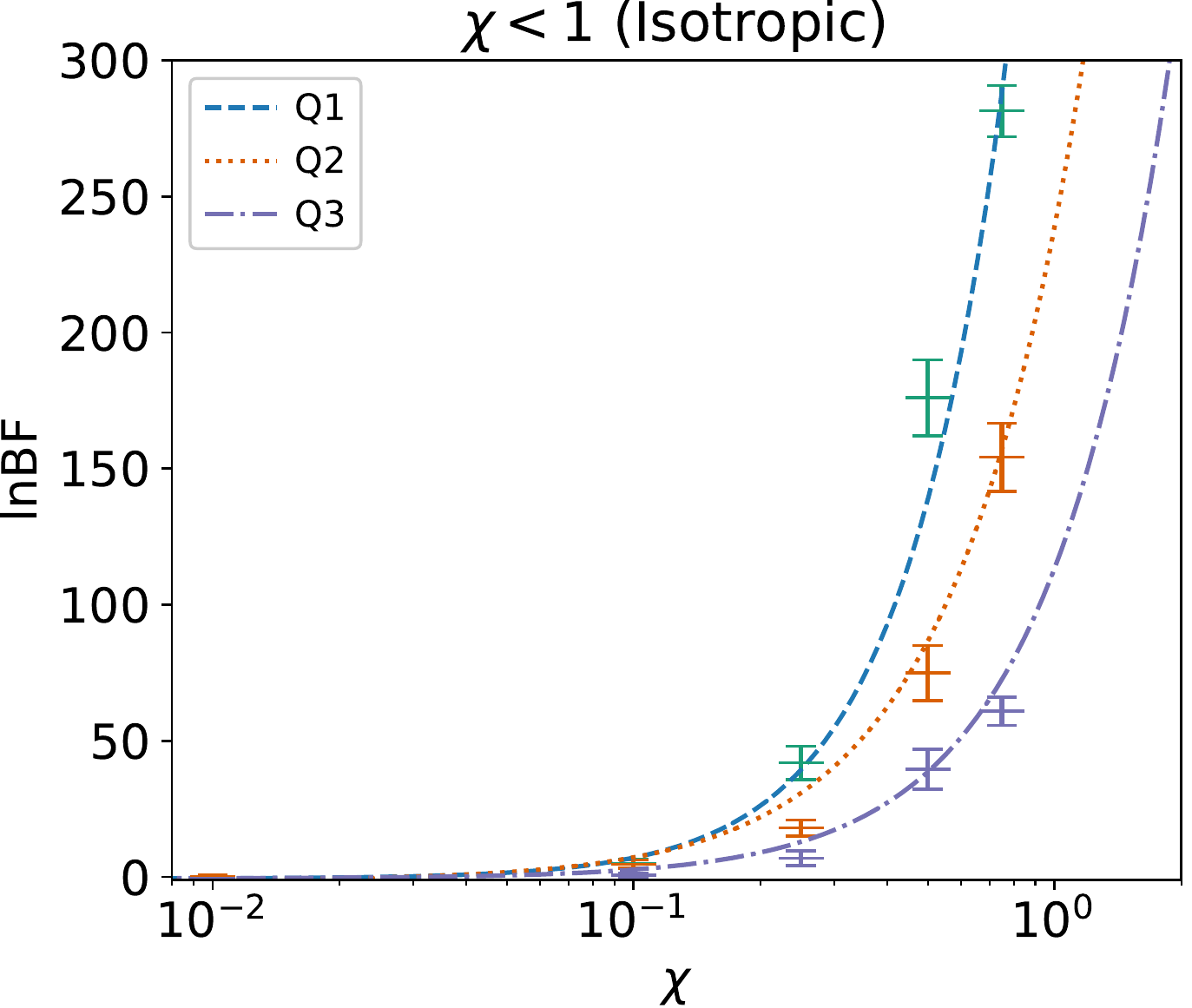}
    \hspace{0.4em}
    \includegraphics[width=0.49\textwidth]{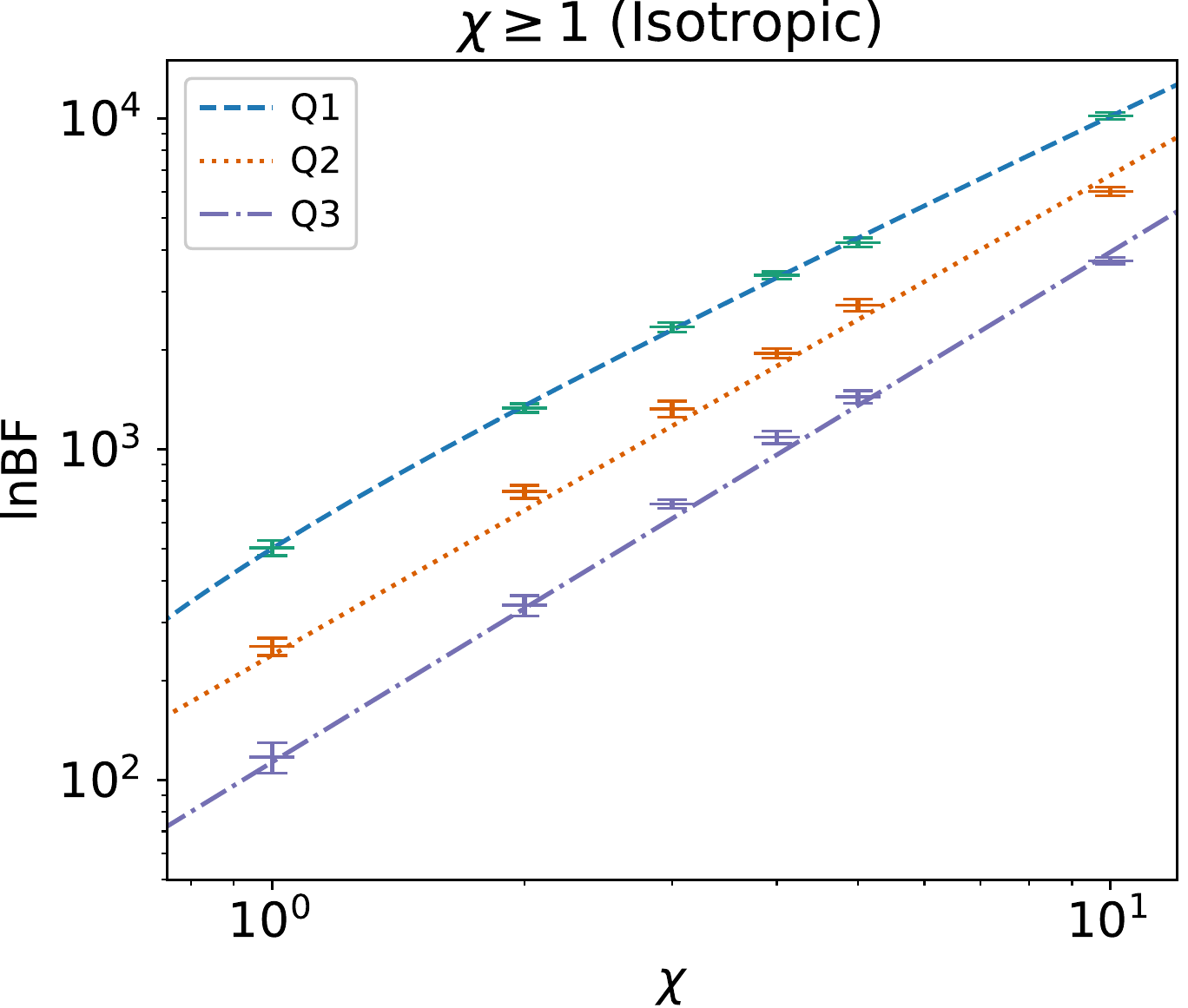}
    \caption{$\langle \ln{\text{BF}}\rangle$ and the $\sigma/\sqrt{10}$ standard error of the mean across 10 realizations sampled at each quartile across varying levels of $\chi$. \textit{Left}: realizations with $\chi<1$. \textit{Right}: realizations with $\chi \geq 1$. The parameters that describe the best-fit lines (dashed, dotted, and dot-dashed) are located in Table~\ref{tab:isotropicdetectionlimit}.}
    \label{fig:isotropicangexp1sb}
\end{figure*}

In this appendix we discuss several additional analyses of the simplified isotropic case discussed in Sec.~\ref{sec:isotropic}, to test the degree to which results from the full inner Galaxy analysis (Sec.~\ref{sec:results}) carry over to this simpler scenario.

\subsection{Tradeoffs between PSF and exposure} \label{appendix:tradeoffspsf}
We apply the same procedure as described in the main text (Sec.~\ref{sec:results}) to test the response of \texttt{NPTFit} to a simultaneous variation of angular resolution and exposure level, in the presence of a population of PSs. Note that for these simulations, the number of live points we used for the scans is set to nlive=500.

Fig.~\ref{fig:isotropicangexp1sb} shows the results of the simulations and analysis. We fitted the exposure dependence for the three separate quartiles using a power law with a constant additive shift; results are given in Table~\ref{tab:isotropicdetectionlimit} (the $n=1$ column). We find similar patterns as those described in Sec.~\ref{subsec:tradeoffs}, although the overall sensitivity is higher (presumably because the signal-to-background ratio is much higher) and the sensitivity increases modestly faster than linearly with exposure in this case, with a slope of $\sim 1.4-1.5$. We observe that moving from Quartile 1 to Quartile 2, and Quartile 2 to Quartile 3, produces a fairly exposure-independent degradation in sensitivity of roughly a factor of two.

\subsection{Detection limit for faint sources} \label{appendix:detectionlimit}
\begin{figure*}
    \centering
    \begin{subfigure}
        \centering
        \includegraphics[width=0.318\linewidth]{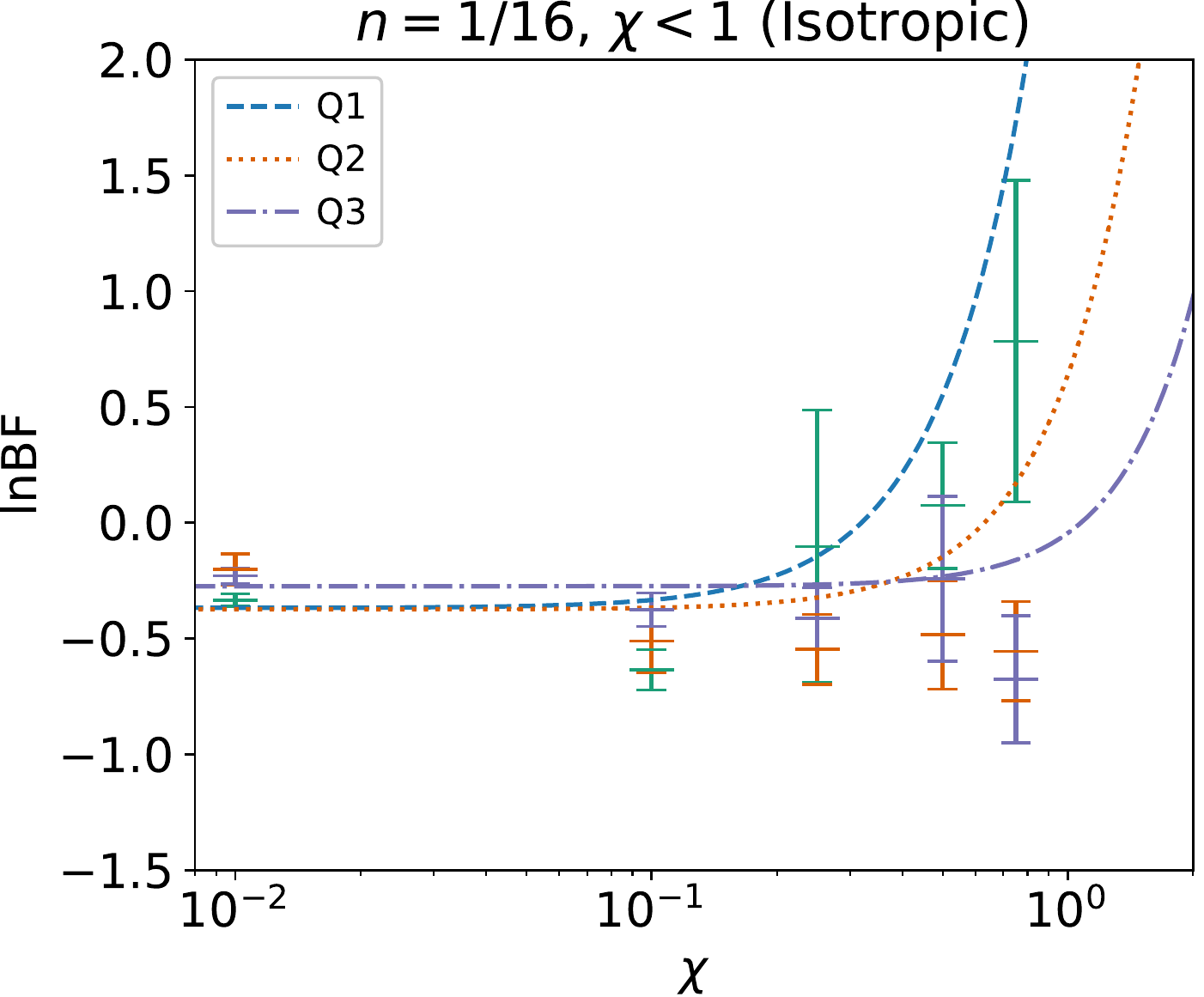}
    \end{subfigure}
    \begin{subfigure}
        \centering
        \includegraphics[width=0.318\linewidth]{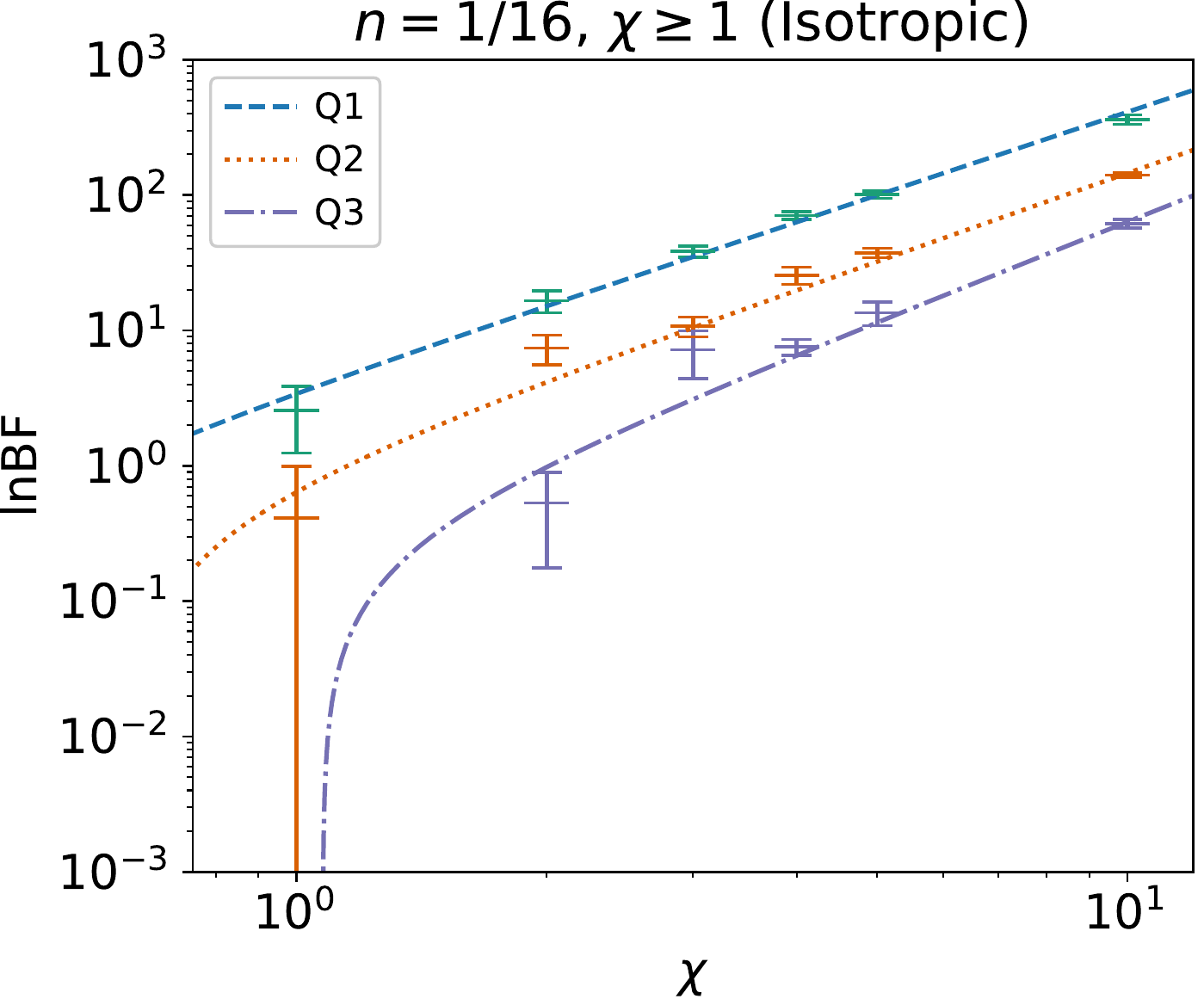}
    \end{subfigure}
    \begin{subfigure}
        \centering
        \includegraphics[width=0.318\linewidth]{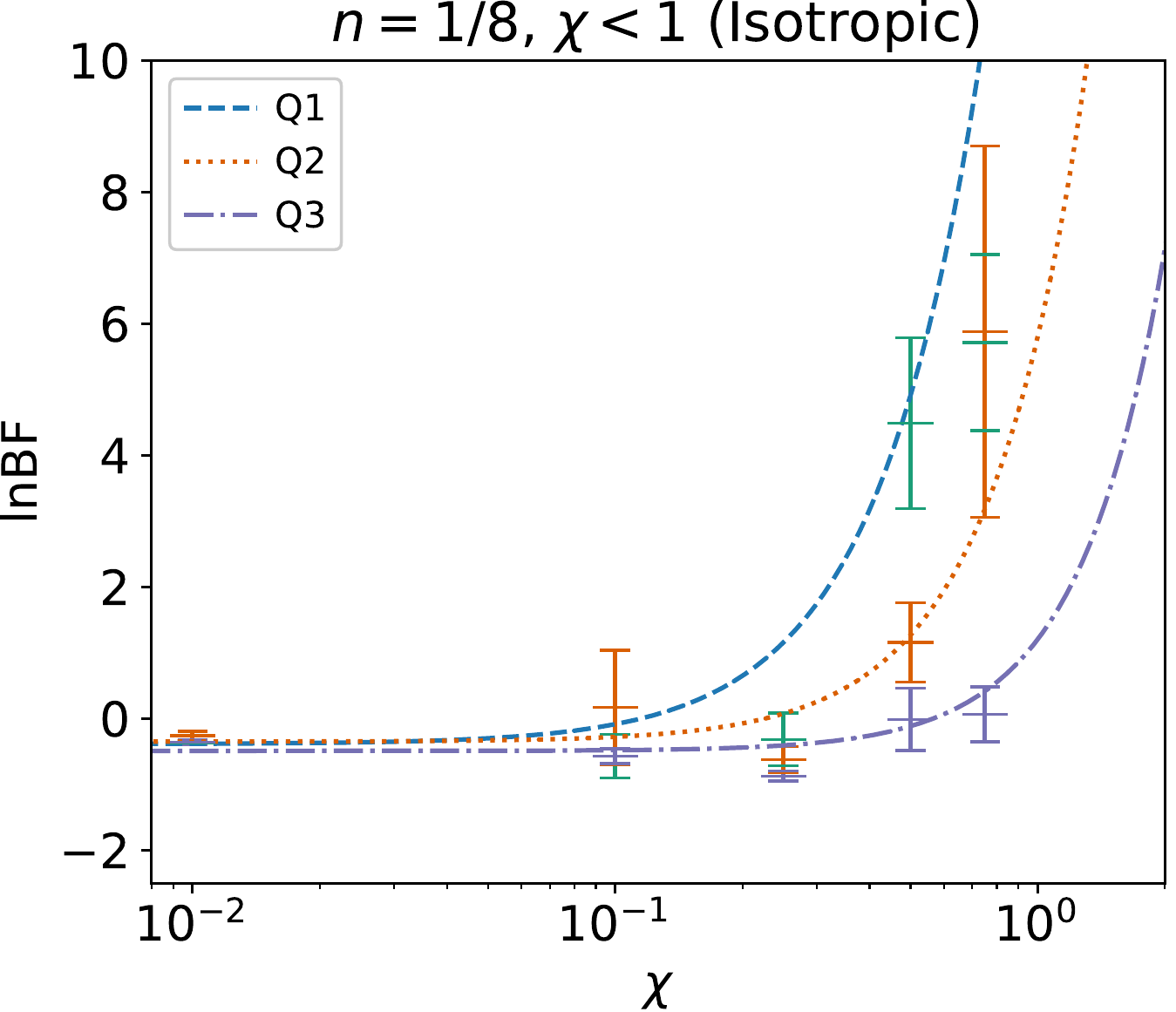}
    \end{subfigure}
    \begin{subfigure}
        \centering
        \includegraphics[width=0.318\linewidth]{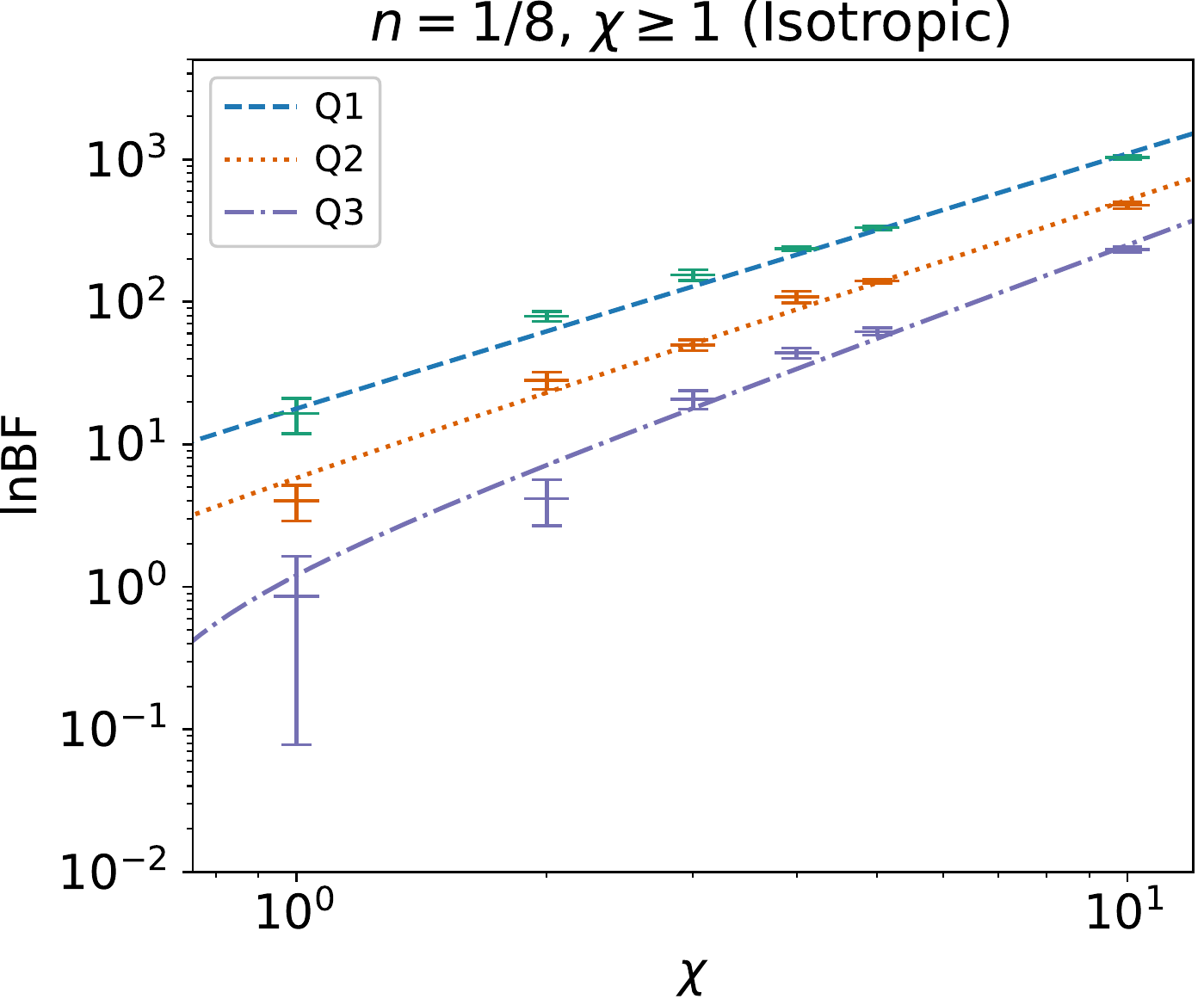}
    \end{subfigure}\begin{subfigure}
        \centering
        \includegraphics[width=0.318\linewidth]{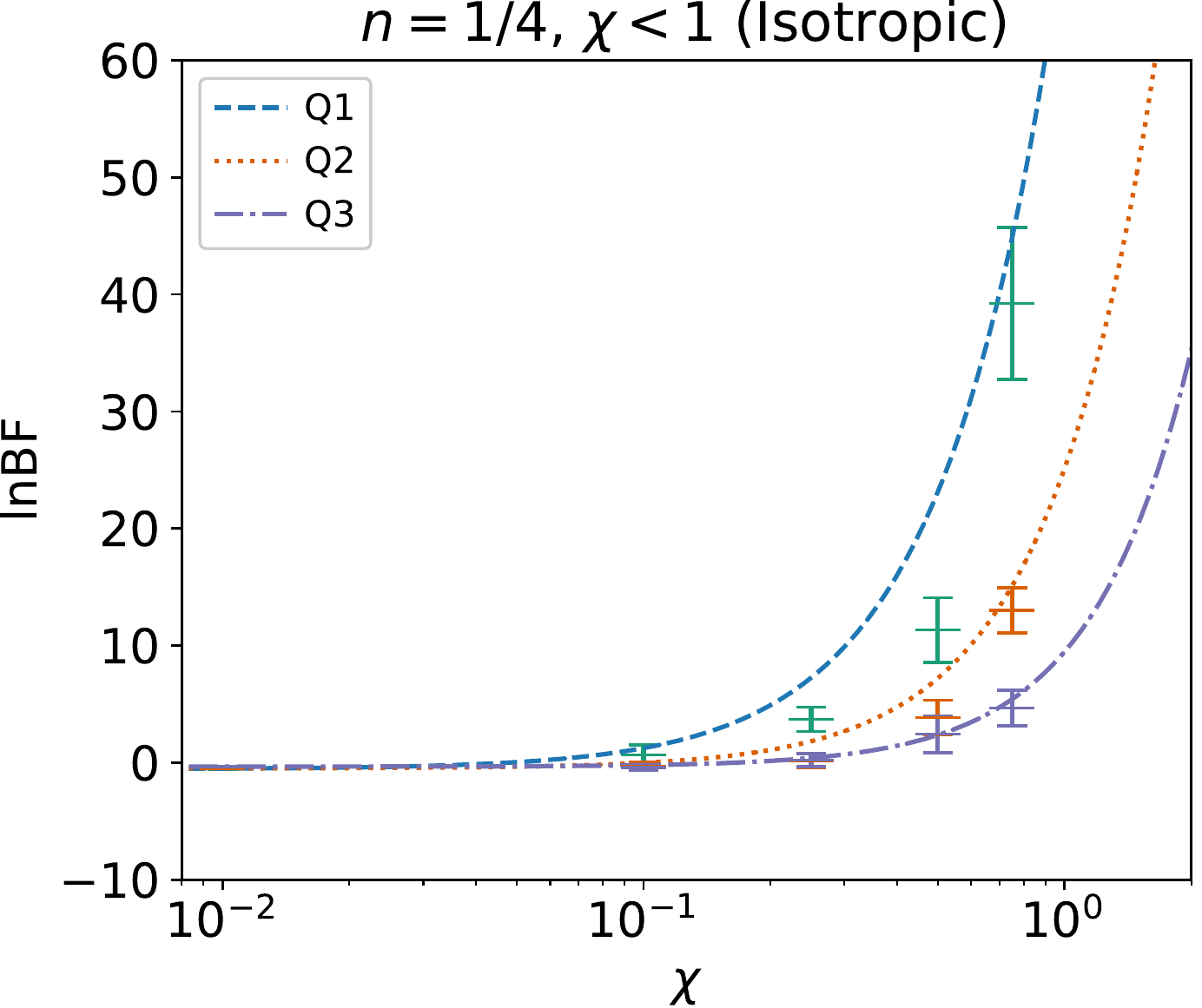}
    \end{subfigure}
    \begin{subfigure}
        \centering
        \includegraphics[width=0.318\linewidth]{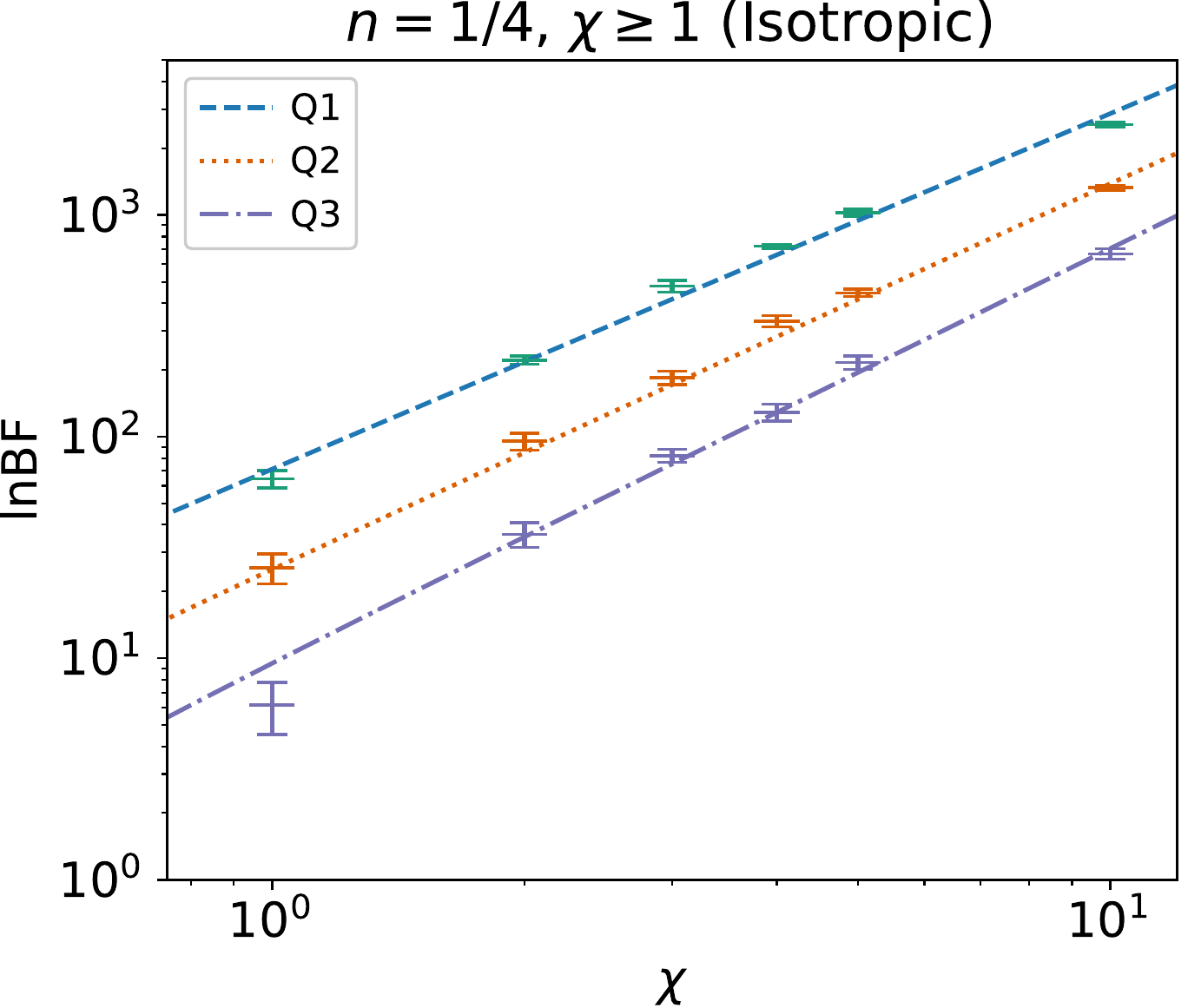}
    \end{subfigure}
    \begin{subfigure}
        \centering
        \includegraphics[width=0.318\linewidth]{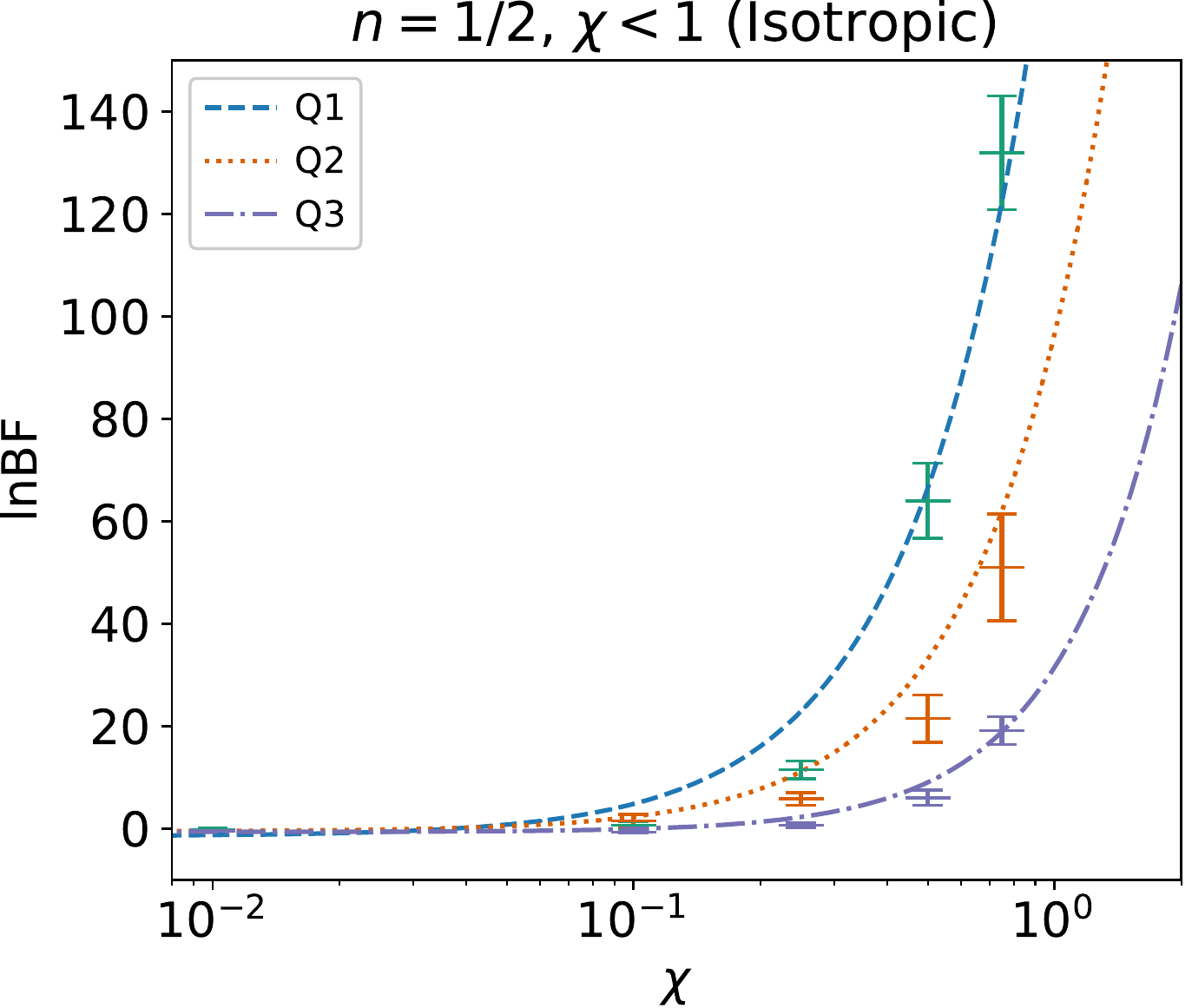}
    \end{subfigure}
    \begin{subfigure}
        \centering
        \includegraphics[width=0.318\linewidth]{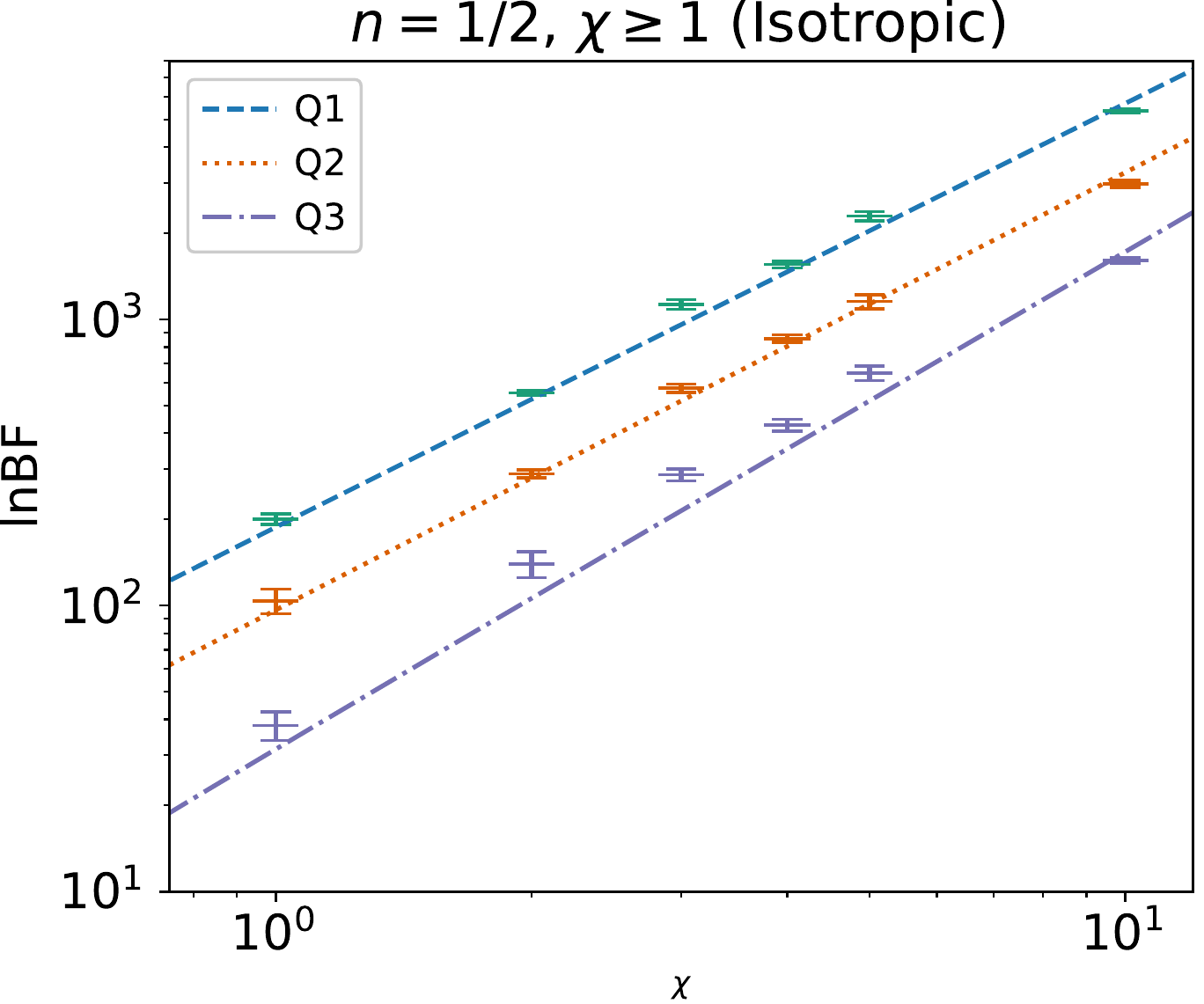}
    \end{subfigure}
    \begin{subfigure}
        \centering
        \includegraphics[width=0.318\linewidth]{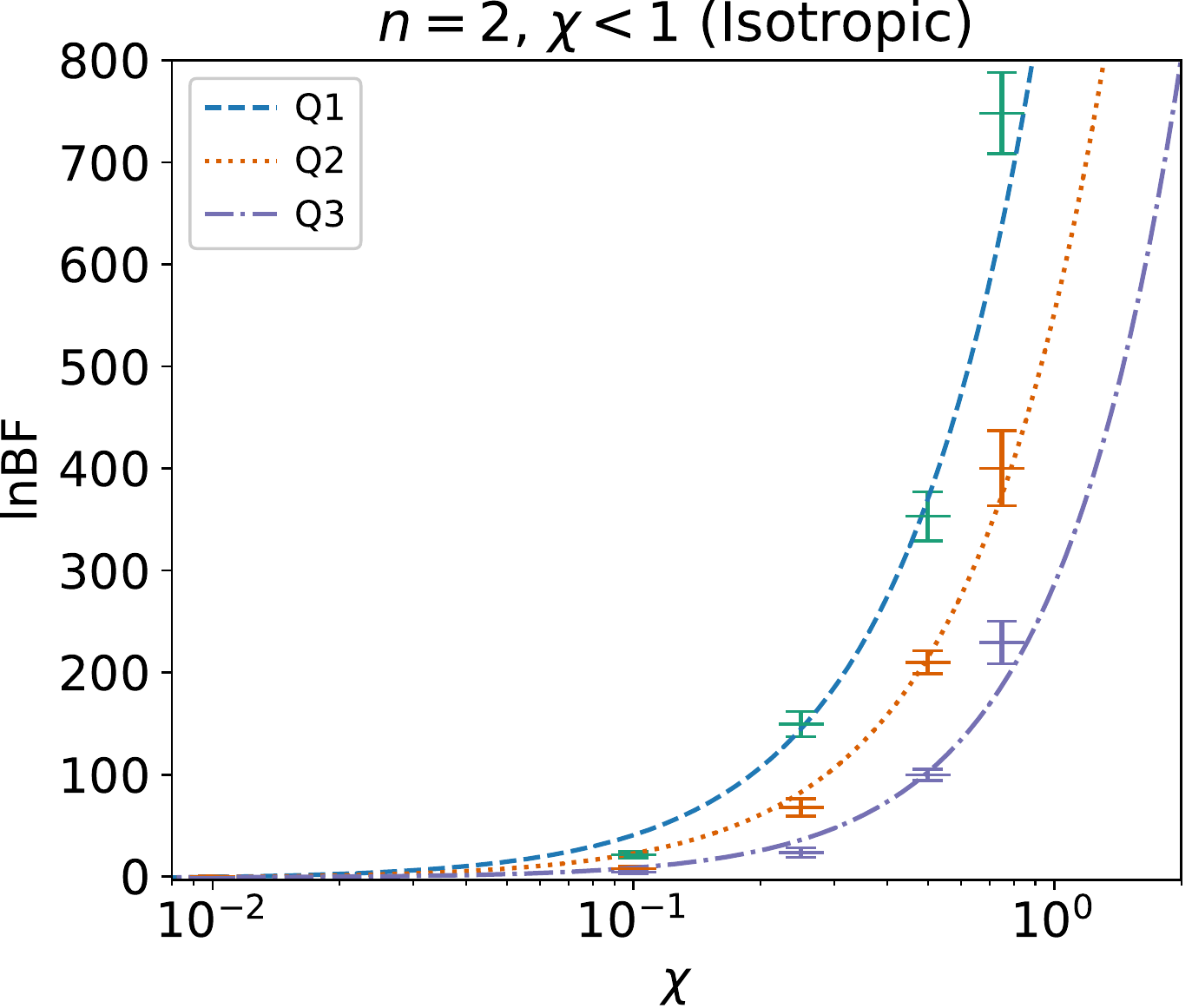}
    \end{subfigure}
    \begin{subfigure}
        \centering
        \includegraphics[width=0.318\linewidth]{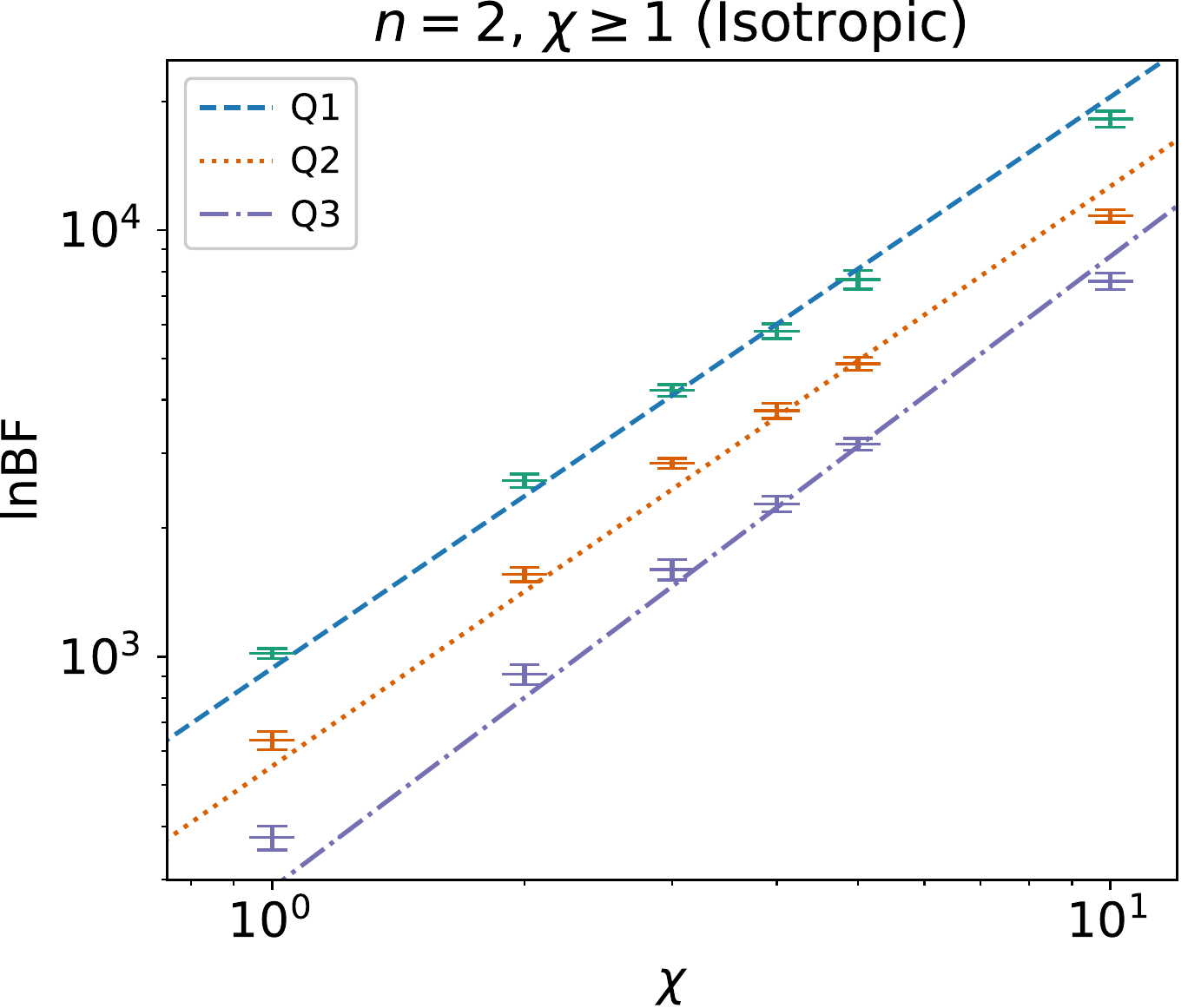}
    \end{subfigure}

    \caption{$\langle \ln{\text{BF}} \rangle$ and the $\sigma/\sqrt{10}$ standard error of the mean across 10 realizations sampled across different $\chi$ values for each of the top three quartiles graded by angular resolution, in the simplified isotropic analysis. Each plot is labeled with the $\chi$ (to denote which exposure rescaling factor range is covered) and $n$ values, where $n$ corresponds to the source brightness as described in Eq.~\ref{eqn:countsbrightness}.}
    \label{fig:isotropicangexpvarsb}
\end{figure*}

Similarly to the inner Galaxy analysis in the main text, we also tested the faintest isotropic PSs \texttt{NPTFit} is able to detect. The following graphs show the results of this analysis. In the panels of Fig.~\ref{fig:isotropicangexpvarsb}, $n$ is used in the same way as in Eq.~\ref{eqn:countsbrightness}. Aside from testing fainter PSs, we also tested brighter PSs. The flux remained constant by reducing the number of sources present. Similar to the approach described in the main text, we model the dependence of $\langle \ln \text{BF}\rangle$ on the exposure as a power law with a constant additive shift, separately for $n$ varying from $1/16$ to $2$.

Note that at lower values of $\chi$, especially for faint sources, the data is not well-described by the best-fit curve. As in the main text, this is most likely just a signal that the power-law fit is inappropriate when the sensitivity to PSs is low.

\begin{table*}
\centering
\begin{tabular}{l r r r r r r r }
\hline
 & $n=1/16$ & $n=1/8$ & $n=1/4$ & $n = 1/2$ & $n=1$ & $n=2$\\
\hline
\textit{Quartile} 1\\
\hline
$\alpha$ (coefficient) & $3.77 \pm 0.69$ & $18.21\pm 2.47$ & $71.90 \pm 7.07$ & $189.36 \pm 14.38$ & $438.79 \pm 30.71$ & $941.70 \pm 47.79$\\
$\beta$ (power) & $2.04 \pm 0.11$  & $1.78\pm0.07$ & $1.60 \pm 0.06$ & $1.48 \pm 0.04$ & $1.41 \pm 0.04$ & $1.34 \pm 0.40$\\
$\gamma$ (shift) & $-0.37 \pm 0.05$  & $-0.39\pm 0.07$& $-0.56 \pm 0.28$ & $-1.50 \pm 1.22$ & $-1.23 \pm 0.44$ & $-2.49 \pm 1.21$\\
\hline
\textit{Quartile} 2\\
\hline
$\alpha$ (coefficient) & $1.01 \pm 0.34$ & $6.16 \pm 0.96$& $25.63 \pm 2.54$ & $97.05 \pm 7.25$  & $240.09 \pm 19.31$ & $554.90 \pm 43.34$\\
$\beta$ (power) & $2.16 \pm 0.16$  & $1.93 \pm 0.09$& $1.73 \pm 0.05$ & $1.53 \pm 0.06$ & $1.45 \pm 0.05$ & $1.36 \pm 0.05$\\
$\gamma$ (shift) & $-0.37 \pm 0.11$  & $-0.35\pm 0.12$& $-0.50 \pm 0.06$ & $-0.58 \pm 0.23$ & $-1.40 \pm 1.75$ & $-1.67 \pm 1.87$\\
\hline
\textit{Quartile} 3\\
\hline
$\alpha$ (coefficient) & $0.23 \pm 0.08$ & $1.70 \pm 0.64$ & $9.83 \pm 0.87$ & $32.16 \pm 5.70$ & $113.91 \pm 9.10$ &$287.81 \pm 19.51$\\
$\beta$ (power) & $2.44 \pm 0.17$  & $ 2.17 \pm 0.19$& $1.86 \pm 0.05$ & $1.73 \pm 0.09$ & $1.45 \pm 0.04$ &$1.48 \pm 0.04$\\
$\gamma$ (shift) & $-0.28 \pm 0.04$  & $-0.49 \pm 0.10 $& $-0.35 \pm 0.05$ & $-0.70 \pm 0.26$ & $-0.58 \pm 0.29$ &$-1.02 \pm 0.66$\\
\hline
\botrule
\end{tabular}
\caption{Best-fit parameters and the $1\sigma$ errors obtained using least-squares regression method for the power law fit to the $\ln{\text{BF}}$ values of the isotropic scenario where we varied the source brightness while keeping the overall flux and the flux ratio between the PS and smooth component constant.}
\label{tab:isotropicdetectionlimit}
\end{table*}

\begin{figure}
    \centering
    \includegraphics[width=0.98\linewidth]{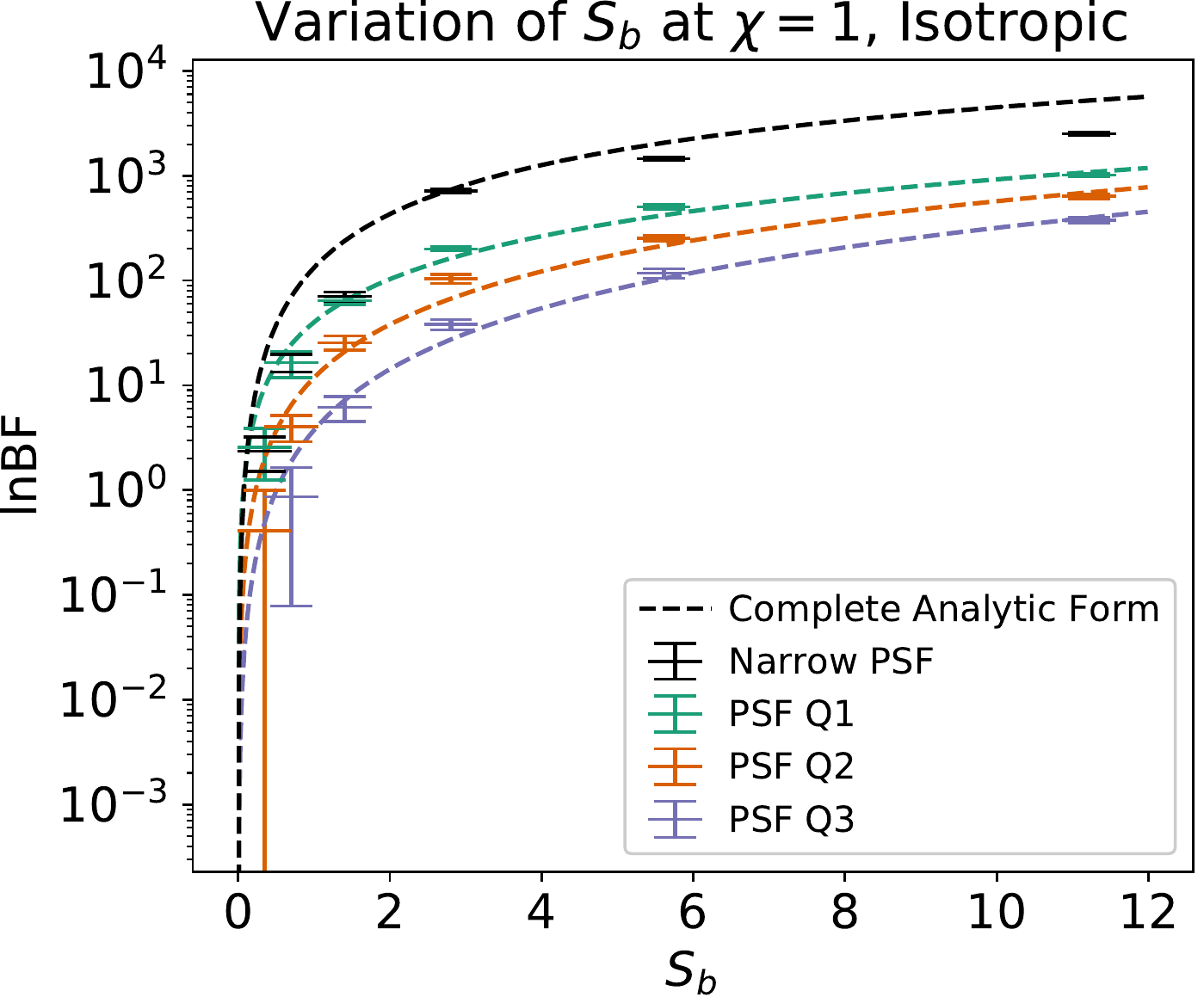}
    \caption{$\langle \ln{\text{BF}}\rangle$ with error bars obtained from the $\sigma/\sqrt{10}$ standard error of the mean across 10 realizations at each of the top three quartiles and narrow PSF at varying brightness level for the isotropic case, with the baseline exposure. The best-fit $S_b$ in the real data is 5.61; we test the effect of reducing and increasing $S_b$. All cases were scanned at nlive=500. The dashed lines represent the results of a power-law fit to $\langle \ln{\text{BF}} \rangle$ as a function of $S_{b}$ (holding exposure fixed at $\chi=1$). The explicit form of the equation is Eq.~\ref{eqn:alphapowerlaw}. For comparison, we also plot the analytic description of Eq.~\ref{eqn:lnBFcase2} as a function of $s$.
    }
    \label{fig:isotropicsensitivitychi1}
\end{figure}

We also plotted $\langle \ln{\text{BF}} \rangle(\chi=1)$ across 10 realizations for varying levels of $S_{b}$ for the isotropic scenario as shown in Fig.~\ref{fig:isotropicsensitivitychi1}. The baseline case corresponds to $S_{b}=5.61$. Similar to the full \textit{Fermi} case the sensitivity increases as the population of PSs brightens. In the isotropic case, the increase in sensitivity is more stable and consistent, likely due to a lack of cross-talk with other templates that can occur (and impact the sensitivity especially for faint signals) in the baseline case. For comparison, we also plotted the analytic solution of Eq.~\ref{eqn:lnBFcase2} and we find that the power law scaling behavior of the simulated data is well-captured by the estimate.

Comparing the full baseline case with the isotropic case, we find qualitatively similar patterns within the coefficients. At higher source brightness (increased $n$), the parameter $\alpha$ (corresponding approximately to the sensitivity at $\chi=1$) increases. More quantitatively, we fitted:
\begin{equation}
   \langle \ln{\text{BF}}\rangle|_{\chi=1}=a S_{b}^{p},
    \label{eqn:alphapowerlaw}
\end{equation}
where $a,p$ are numerical constants. We find that $\langle \ln{\text{BF}}\rangle$ scales as $O(S_{b}^{1.4}),O(S_{b}^{1.7}),O(S_{b}^{1.9})$, for quartiles 1, 2, and 3 respectively. We plotted the results of this fit in Fig.~\ref{fig:isotropicsensitivitychi1} and find that the scaling behavior is approximately the same as the analytic form Eq.~\ref{eqn:lnBFcase2}, if we allowed $s$, the number of photons per source to vary.  The $\beta$ parameter, describing the rate at which sensitivity increases with respect to exposure, is enhanced at lower values of $n$ for the range of exposure values we test, rising to $\sim 2$ (compared to $\sim 1.3$ for the brightest sources). This is consistent with our analytic estimates, where we found that we expected the significance to scale with the square of the exposure in the faint-source limit.

Similar to the analysis performed in Sec.~\ref{subsec:nptfitdetectionlimit}, if we assume that Eq.~\ref{eqn:powerlawshift} is a reasonable description of the underlying relation of sensitivity to exposure, we can use this fit to determine $\chi$ values at each brightness level (for PSFQ1) that correspond to $\ln{\text{BF}} \gtrsim  1$, indicating the first hint of evidence for a detection. Using the best-fit parameters in Table~\ref{tab:isotropicdetectionlimit}, we find that for $n=1/16$, $n=1/8$, $n=1/4$, $n=1/2$, $n=1$ and $n=2$ these ``exposure thresholds'' correspond respectively to $\chi \geq 0.61, 0.24, 0.091, 0.054, 0.024, 0.015$, scaling roughly inversely with $n$.

\subsection{Pixel size variation} \label{appendix:isotropicpixelvar}
\begin{figure*}
    \centering
    \includegraphics[width=0.318\linewidth]{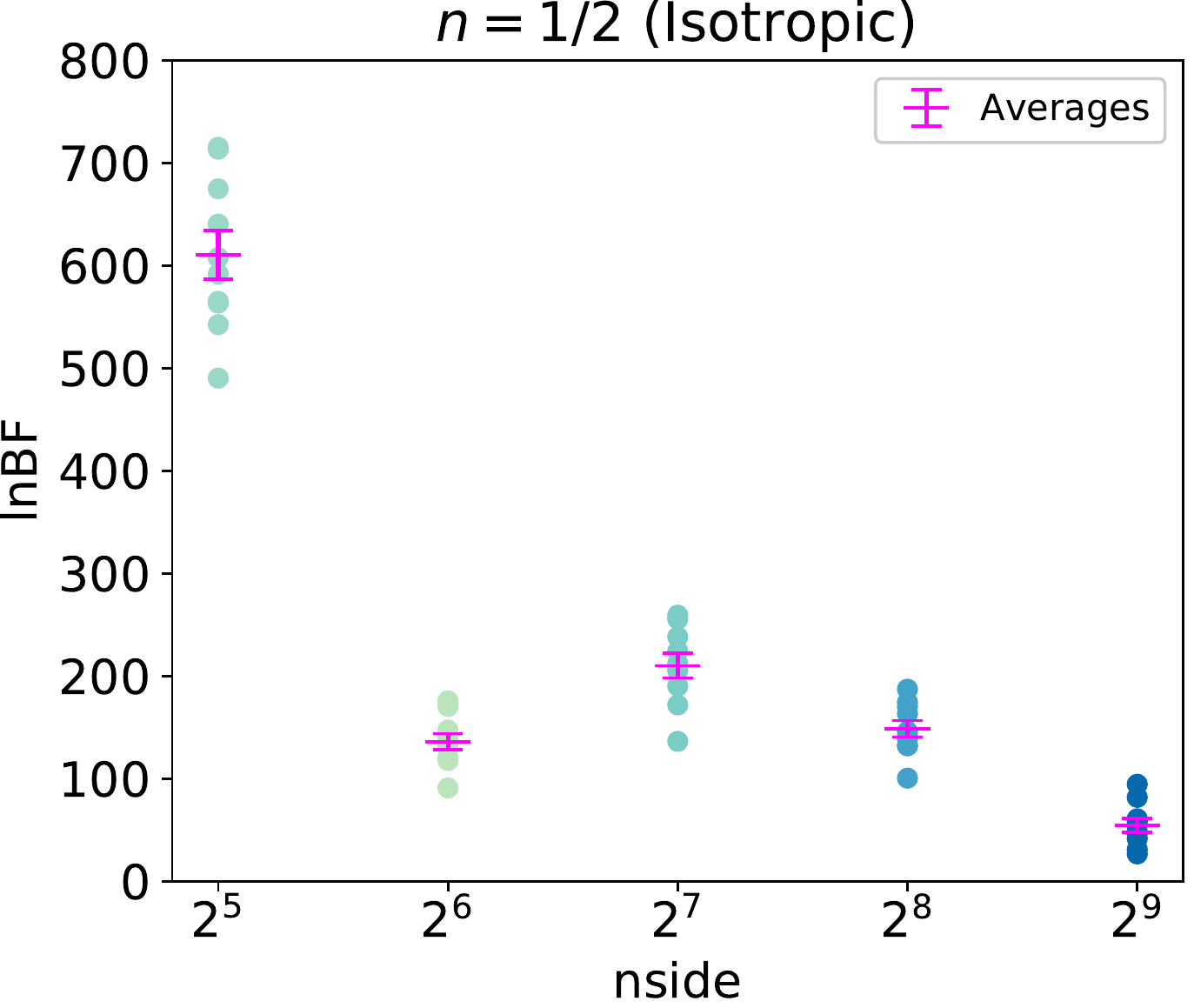}
    \hspace{0.05em}
    \includegraphics[width=0.318\linewidth]{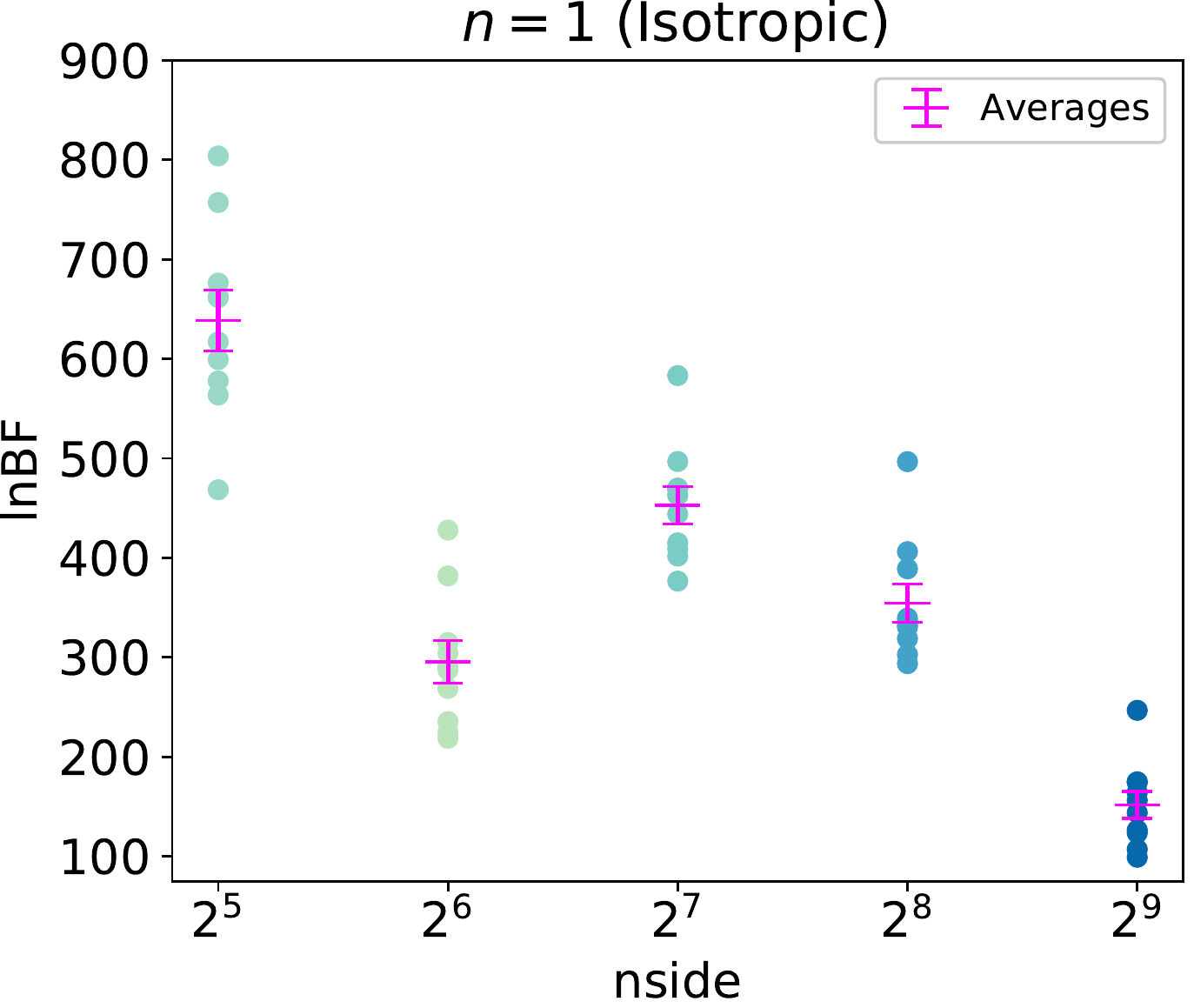}
    \hspace{0.05em}
    \includegraphics[width=0.318\linewidth]{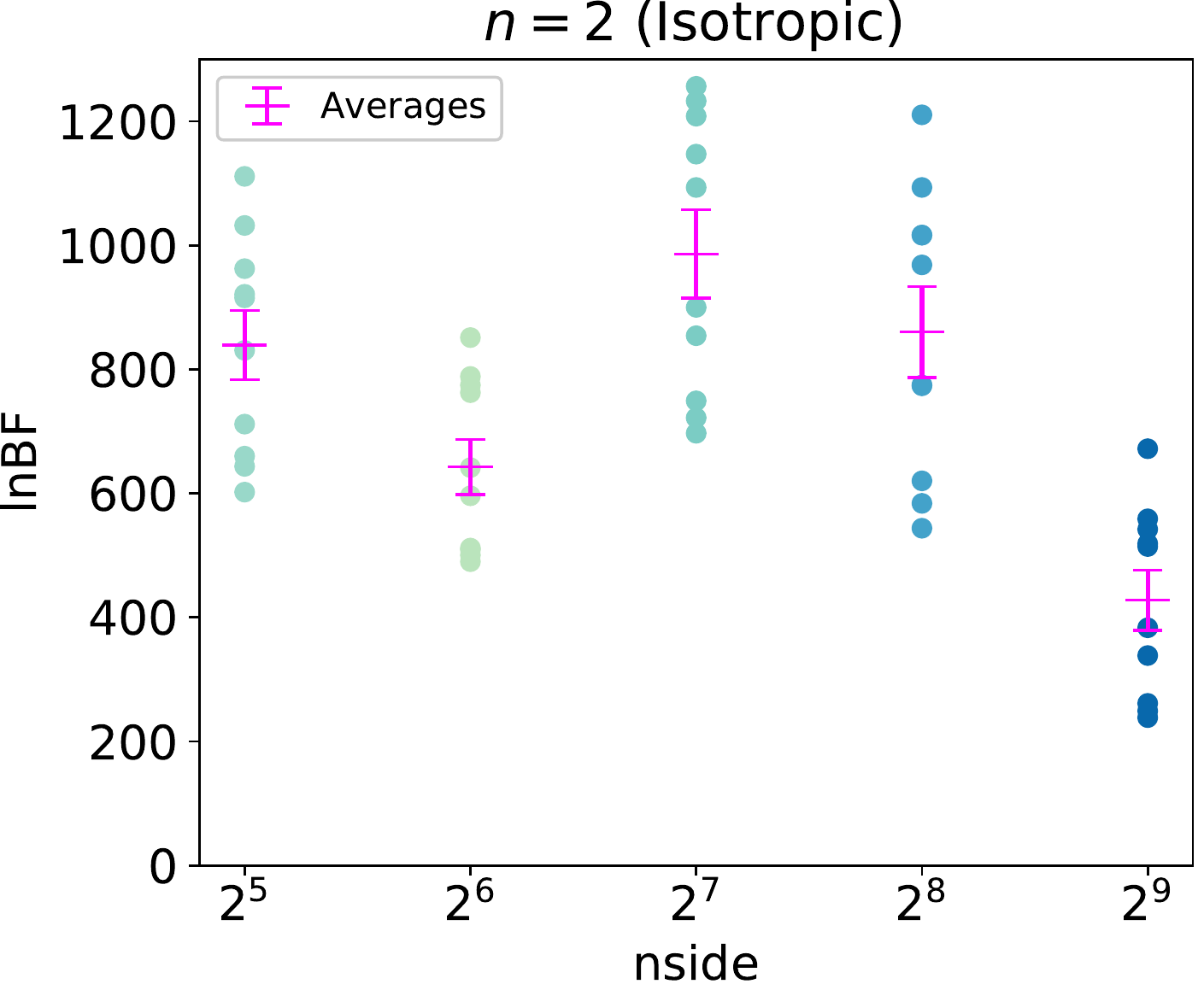}
    \caption{
$\ln{\text{BF}}$ across 10 realizations (circle markers), and $\langle \ln{\text{BF}} \rangle $ with error bars obtained from the $\sigma/\sqrt{10}$ standard error of the mean (magenta), sampling different pixel sizes (nside) at various brightness levels of an isotropic PS population. Note that the anomalously high significance values attached to the largest pixel size (nside = 32) are associated with a failure to correctly reconstruct the input parameters.}
    \label{fig:isotorpicpixelvarsb}
\end{figure*}

As in the main text, we tested the effect of varying the pixel size.  In this simplified (isotropic) scenario, we also considered the interplay between choosing smaller or larger pixel sizes and having fainter or brighter PSs -- for example, we might ask whether different pixel sizes are more or less optimal depending on the source brightness. Fig.~\ref{fig:isotorpicpixelvarsb} shows the results of our analysis; similarly to our previous results, we found that the ideal pixel size for optimizing sensitivity to PSs lies in the intermediate range around nside=128. The relative enhancement at the intermediate range, however, decreases with fainter sources.

An interesting feature present across the plots in Fig.~\ref{fig:isotorpicpixelvarsb} is the sudden spike in sensitivity at large pixel sizes (low nside). Upon closer inspection of the returned flux fraction, \texttt{NPTFit} incorrectly recovers the flux fraction for both smooth and PS isotropic emission (similar to the misattribution at low nside in the main text). Specifically, the fit attributes most of the emission to the isotropic PS population instead of distributing the flux equally to both sources. Examining the flux fraction plots for smaller pixel sizes, on the other hand, the correct flux fraction ratio is recovered.

\subsection{Tests with an overly conservative PSF}

The convention in the literature (which we have generally followed in this work, e.g. in Fig.~\ref{fig:multiquartilesuperposed}) has been to use the PSF associated with the worst-angular-resolution photons in the dataset, for the purpose of \texttt{NPTFit} scans. In this section we test the impact of using an overly conservative (i.e.~too large) PSF on data with a better angular resolution, in the simplified isotropic scenario. 

We simulated narrow-PSF isotropic data as previously discussed. We then performed three sets of \texttt{NPTFit} scans of the same data set: (1) using the narrow PSF, (2) using the realistic \textit{Fermi} PSF appropriate to the top PSF quartile, Q1 and (3) using the realistic \textit{Fermi} PSF appropriate to the third PSF quartile, Q3.

\begin{figure*}
    \centering
    \includegraphics[width=0.49\linewidth]{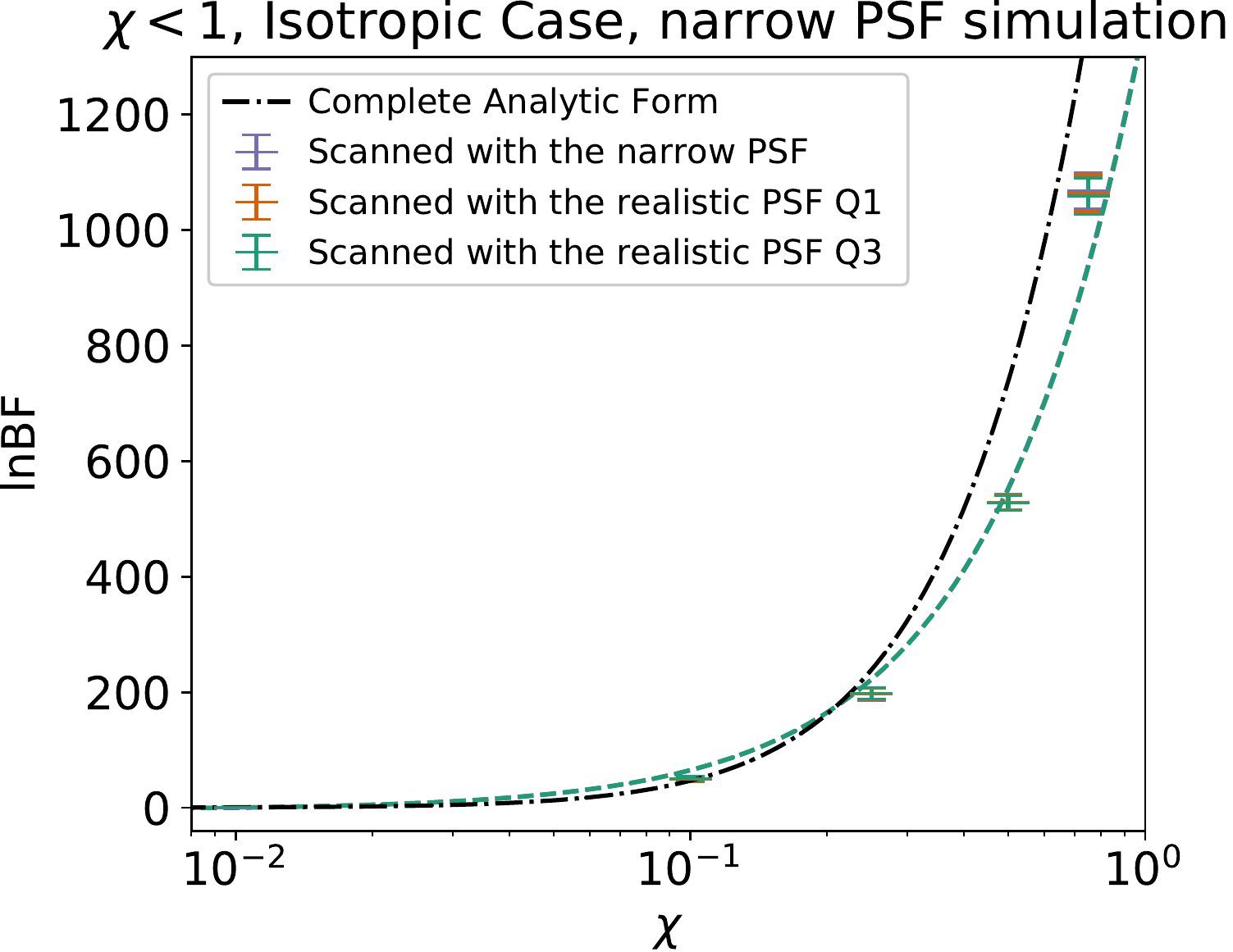}
    \includegraphics[width=0.49\linewidth]{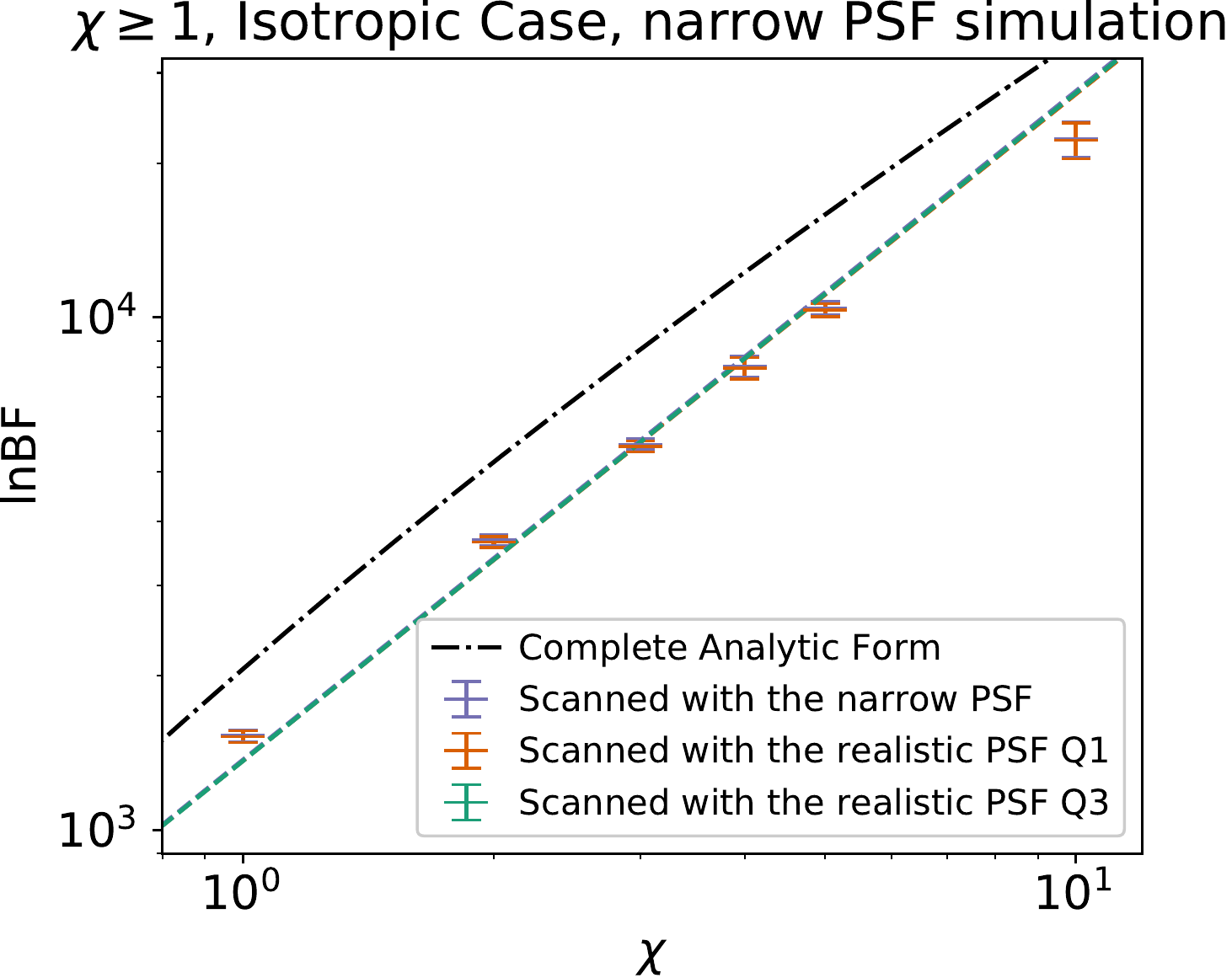}
    \caption{$\braket{\ln{\text{BF}}}$ with errors obtained from the $\sigma/\sqrt{10}$ standard error of the mean across 10 realizations across different levels of exposure $\chi$. The data was simulated using a narrow PSF and scanned in three separate ways: (1) using the narrow PSF that matches the simulation, (2) using the realistic \textit{Fermi} PSF at Q1, and (3) using the realistic \textit{Fermi} PSF at Q3.} 
    \label{fig:tests_with_overly_conservative_PSF}
\end{figure*}

Fig.~\ref{fig:tests_with_overly_conservative_PSF} displays the results of this test. We find that the three cases are generally consistent with each other. Thus, utilizing an overly conservative PSF within the \texttt{NPTFit} scan does not significantly impact the resultant sensitivity to PSs, at least in the cases we have tested. However, it is possible that assuming an isotropic background and signal reduces the effect of PSF mismodeling; the sensitivity to PSF mismodeling may also depend on the pixel size.

\section{Tests with an Alternative Diffuse Model}
\label{app:modela}

In this appendix we repeat selected analyses with a different model for the Galactic diffuse emission. \texttt{Model A} \cite{Calore:2014xka} splits the diffuse emission amongst two different components: $\pi^{0}$ decay from proton-proton collisions plus bremsstrahlung (``Pibrem") and inverse Compton scattering (``ICS"). Separating the diffuse emission into two different components grants an extra degree of freedom during the fitting process, and allows for a better fit. We note that we scanned all realizations for this subsection at nlive=100 for computational efficiency.

\subsection{Variation of exposure} \label{sec:appendixexposure}

\begin{figure*}
    \centering
    \includegraphics[width=0.49\linewidth]{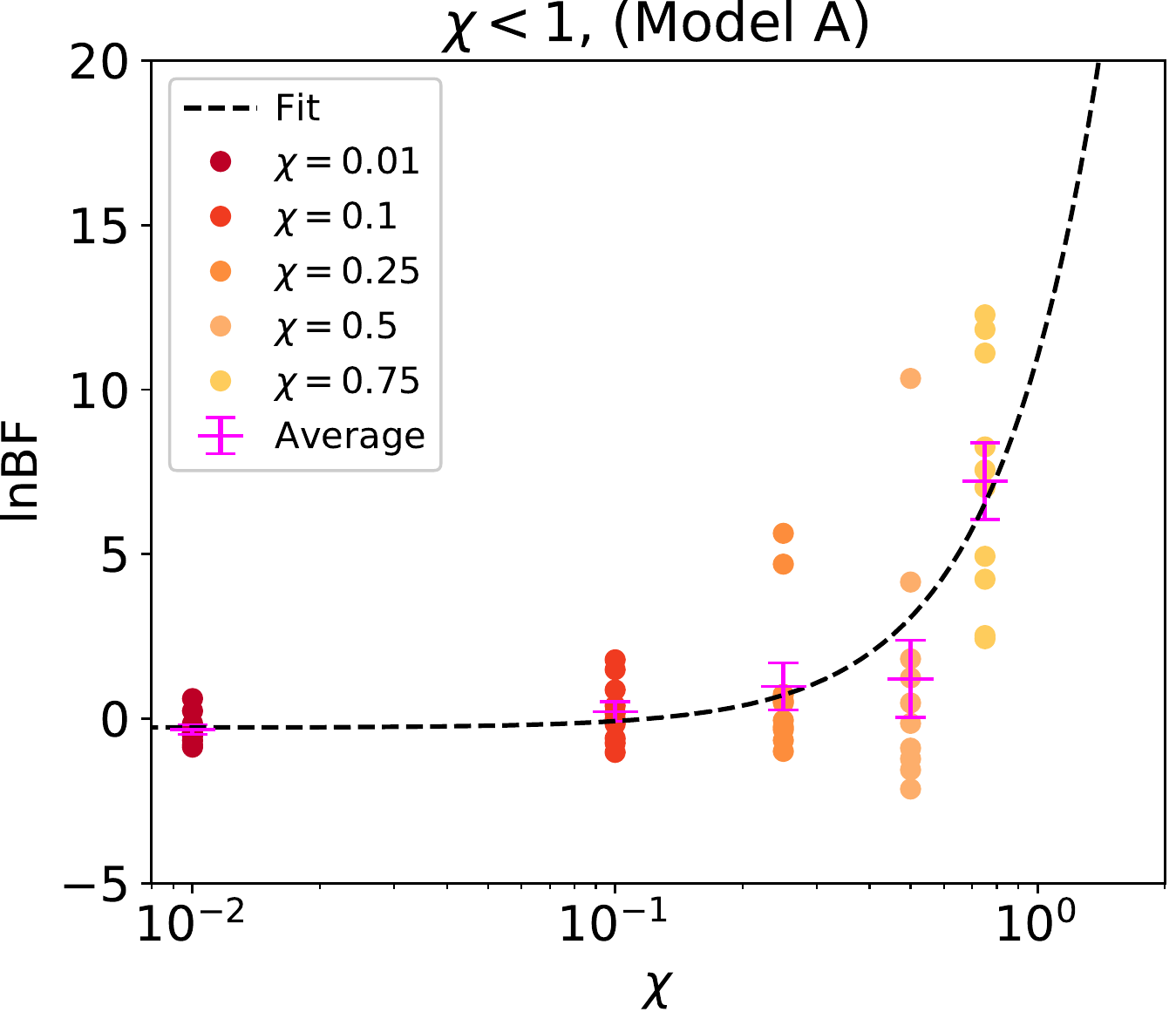}
    \hspace{0.4em}
    \includegraphics[width=0.49\linewidth]{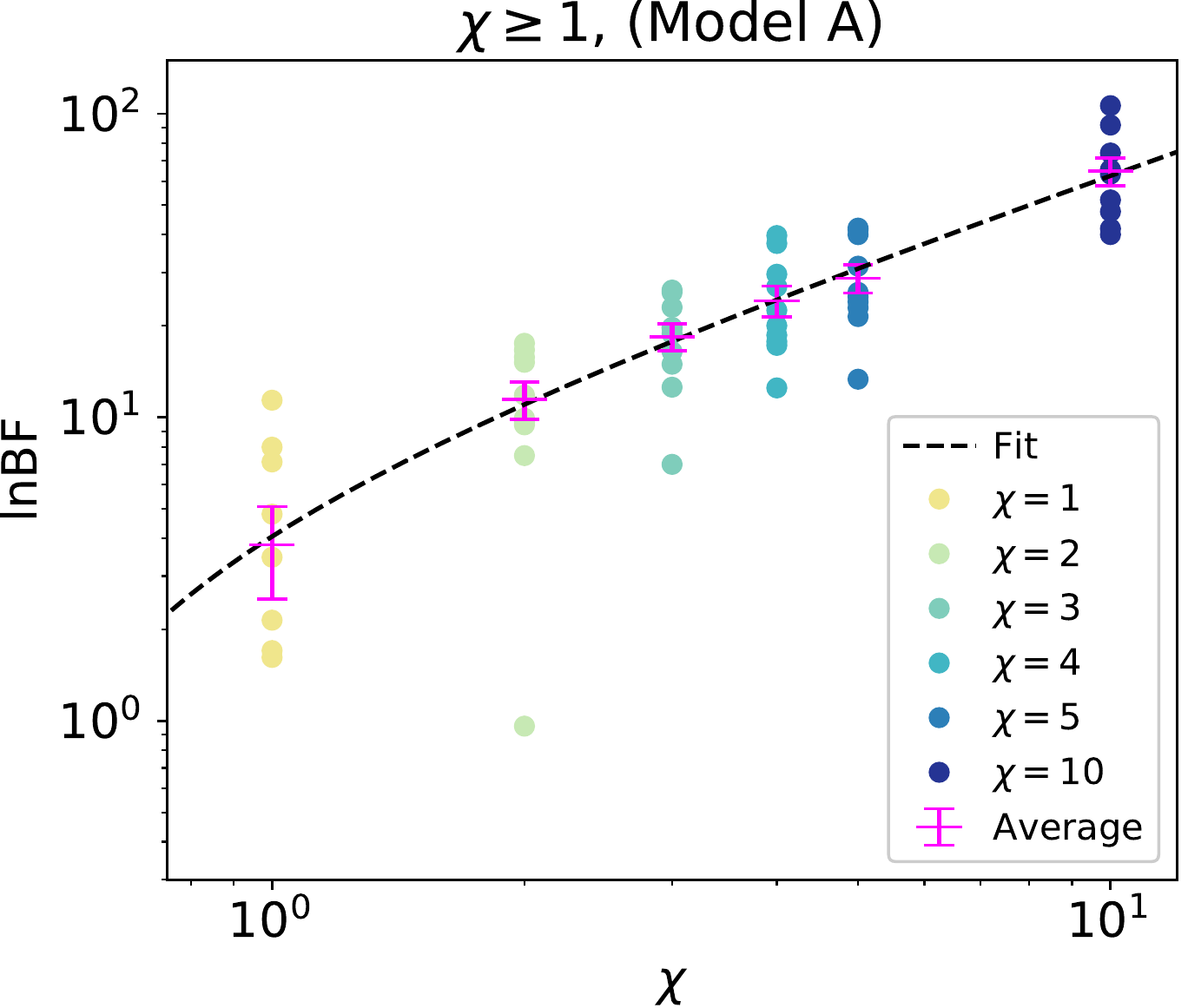}
    \caption{$\ln{\text{BF}}$ across 10 realizations (circle markers), and $\langle \ln{\text{BF}} \rangle$ with error bars obtained from the $\sigma/\sqrt{10}$ standard error of the mean (magenta), sampling across various levels of exposure $\chi$ with \texttt{Model A} as the diffuse template. \textit{Left}: realizations with $\chi<1$. \textit{Right}: realizations with $\chi \geq 1$. The best-fit line is a standard power law with an additive shift, defined in Eq.~\ref{eqn:powerlawshift}.}
    \label{fig:exposuremodela}
\end{figure*}

We began by repeating the exposure-variation analysis of Sec.~\ref{sec:exposuretest}. Fig.~\ref{fig:exposuremodela} shows the sensitivity as a function of the exposure rescaling factor $\chi$. As previously, we fit this curve with a power law with a constant additive offset, and plot the resulting best-fit parameters in Tab.~\ref{tab:exposurefitmodela}.

The behavior of $\langle \ln{\text{BF}} \rangle$ as a function of the rescaling factor $\chi$ is quite similar between \texttt{p6v11} and \texttt{Model A} analyses, but the power-law slope in the current analysis is 1.0, corresponding to a slightly steeper scaling than the slope of 0.76 found in the main text. Both are qualitatively similar to the expectation of linear scaling from the analytic estimates for the behavior of the delta log likelihood.

\begin{table}
\centering
\begin{tabular}{r r   }
\hline

Parameter & \texttt{Model A} \\
\hline
$\alpha$ (coefficient) & $6.0 \pm 0.7$ \\
$\beta$ (power) & $1.0 \pm 0.1$   \\
$\gamma$ (shift) & $-0.4 \pm 0.2$   \\
\botrule
\end{tabular}
\caption{Best-fit parameters obtained using least-squares regression method for the power law fit (\texttt{Model A}). Parameter uncertainties obtained from the standard error of the mean across 10 realizations are also displayed.}
\vspace{-10pt}
\label{tab:exposurefitmodela}
\end{table}

\subsection{Variation of PSF quartile}
\label{appendix:angularresmodela}

\begin{figure}[H]
    \centering
    \includegraphics[width=0.85 \linewidth]{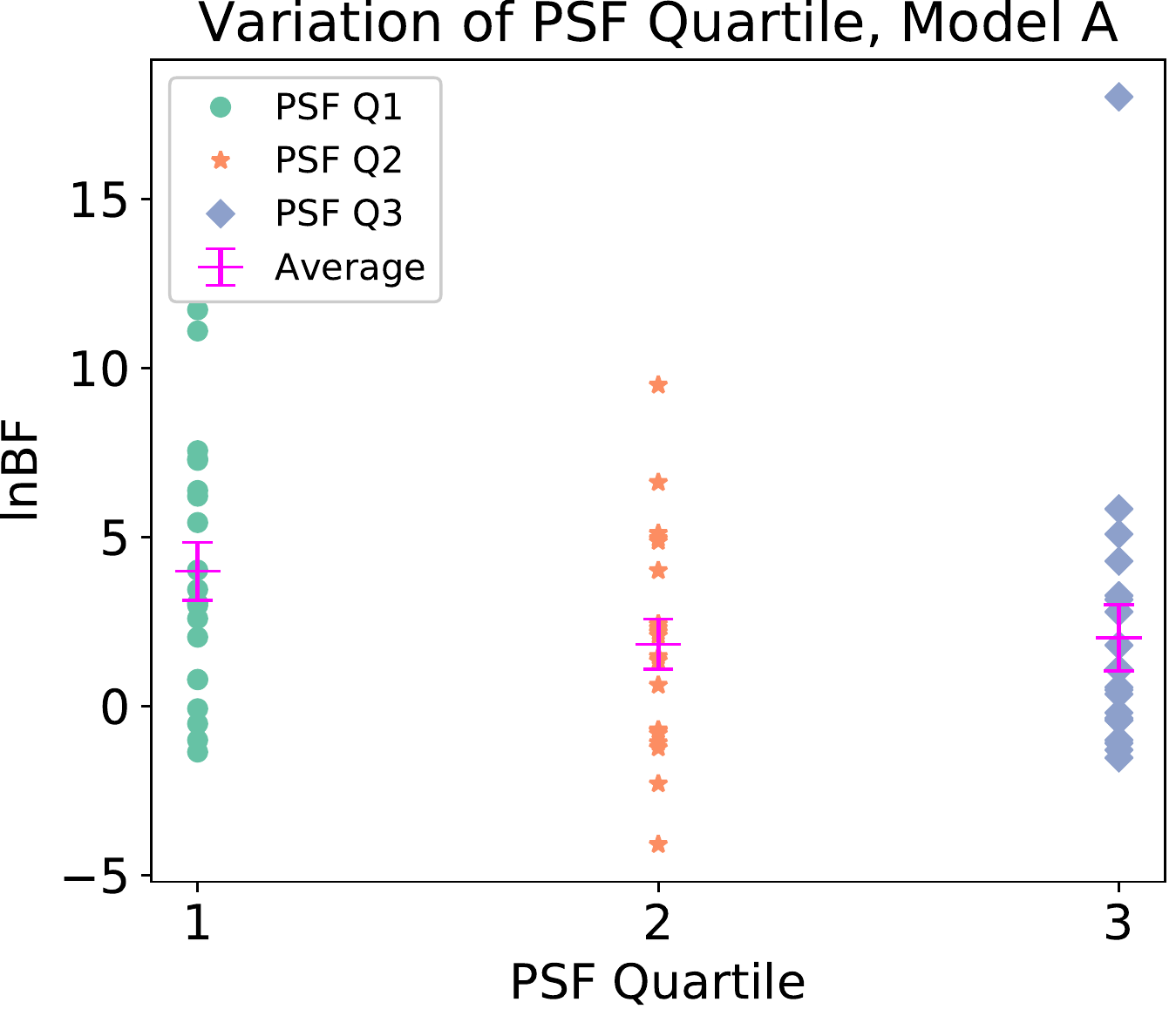}
    \caption{$\ln{\text{BF}}$ across 20 realizations (circle markers), and $\langle \ln{\text{BF}} \rangle $ with error bars obtained from the $\sigma/\sqrt{20}$ standard error of the mean (magenta), at each PSF quartile, using \texttt{Model A}.}
    \label{fig:psfvarmodela}
\end{figure}

We performed the same PSF variation test described in \ref{subsec:varyingpsf} using the alternative \textbf{(\texttt{Model A})} diffuse templates. The results are shown in Fig.~\ref{fig:psfvarmodela}; we find the sensitivity is somewhat degraded in the other PSF quartiles relative to the best quartile. For \texttt{Model A}, even in the top quartile the sensitivity is rather modest, and so an inverse scaling with the containment angle would be expected to lead to low sensitivity in the other quartiles, which is observed (but with substantial uncertainties). 

\subsection{Trade-offs in sensitivity between exposure and angular resolution} \label{appendix:expangrestest}

\begin{figure*}[ht]
    \centering
    \includegraphics[width=0.49\linewidth]{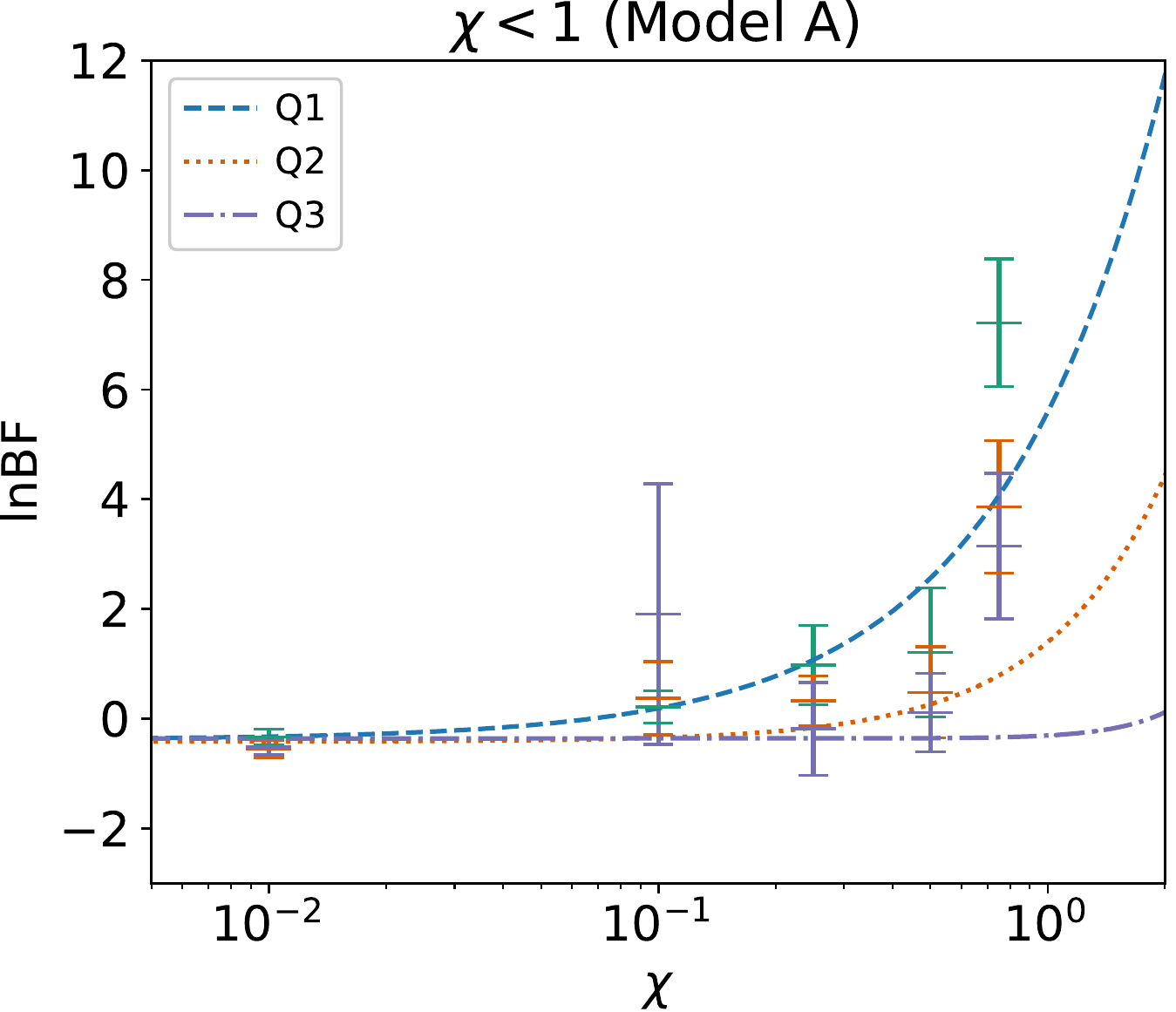}
    \hspace{0.4em}
    \includegraphics[width=0.49\linewidth]{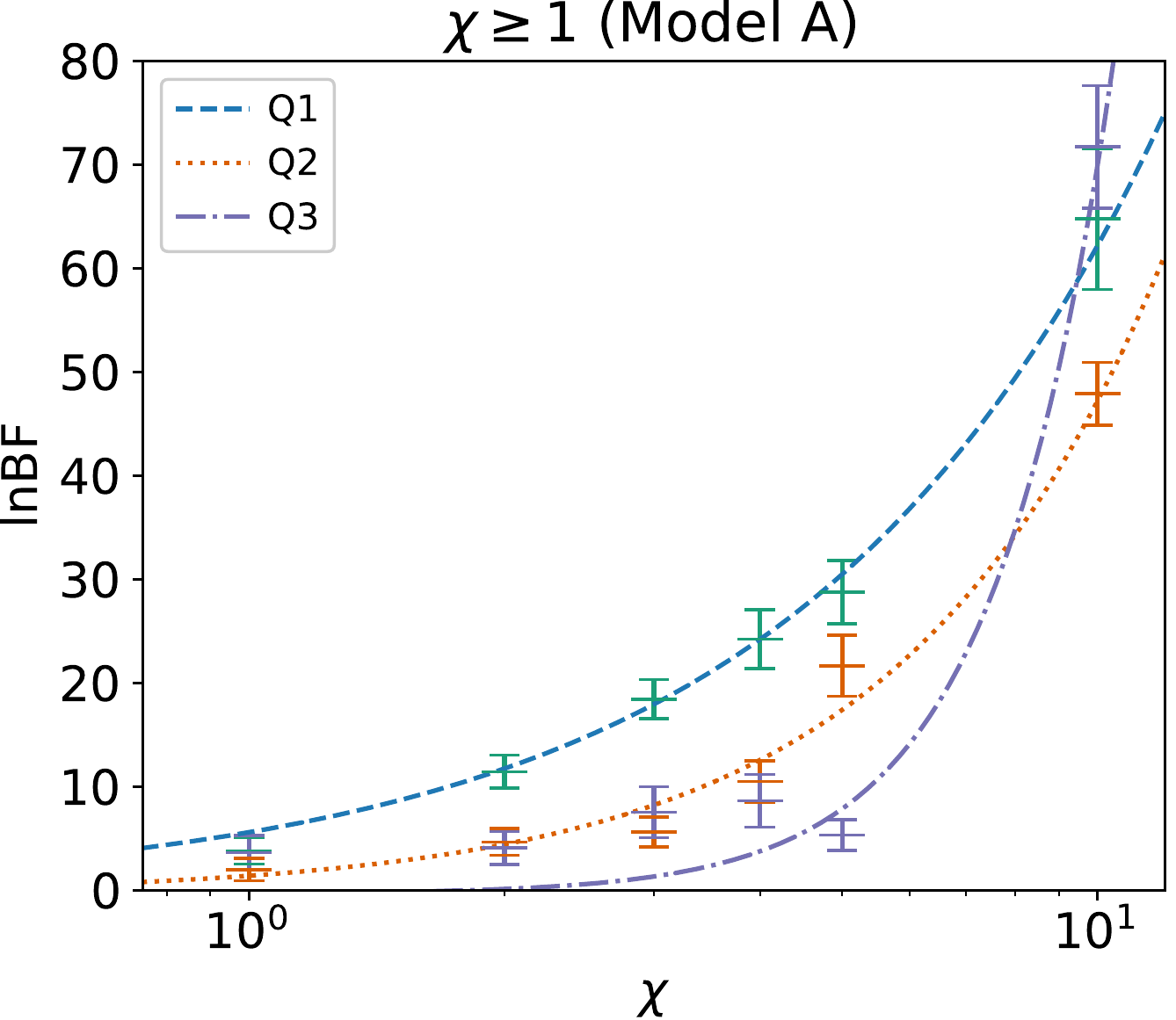}
    \caption{$\langle \ln{\text{BF}}\rangle$ and the $1
    \sigma$ standard error of the mean across 10 realizations for top three quartiles graded in angular resolution sampling different values of $\chi$ using an alternative diffuse template (\texttt{Model A}). \textit{Left}: realizations with $\chi<1$. \textit{Right}: realizations with $\chi \geq 1$. The best-fit lines show the Eqn~\ref{eqn:powerlawshift} fit to the data.}
    \label{fig:angexp1sbmodela}
\end{figure*}

We repeated the exposure variation test for each of the PSF quartiles, to test the tradeoff between increased exposure and improved angular resolution. Fig.~\ref{fig:angexp1sbmodela} shows the results. Best-fit parameters for the power-law fits to each set of results are given in Tab.~\ref{tab:tradeoffmodela}. The power-law slope with respect to exposure steepens slightly in the second quartile relative to the first. In Q3, the apparent best-fit slope is substantially steeper than in the other two quartiles, but the quality of the power-law fit is quite poor; the steep slope is driven by a high sensitivity at $\chi=10$ combined with a relatively low and stable sensitivity at $\chi=2-5$.

\begin{table}
\centering
\begin{tabular}{r r r r r }
\hline
Parameter & Q1 & Q2 & Q3 \\
\hline
$\alpha$ (coefficient) & $ 6.0\pm 0.7$ & $ 1.8\pm 0.5$ & $ 0.1\pm 0.1$ \\
$\beta$ (power) & $ 1.0\pm 0.1$  & $1.4 \pm 0.1$ & $3.1\pm 0.5$\\
$\gamma$ (shift) & $ -0.4 \pm 0.2 $  & $ -0.4 \pm 0.2$ &$ -0.4\pm 0.3$ \\
\botrule
\end{tabular}
\caption{\textbf{Model A} power law parameters across the top three quartiles.}
\vspace{-10pt}
\label{tab:tradeoffmodela}
\end{table}

\begin{figure}
    \centering
    \includegraphics[width=0.85\linewidth]{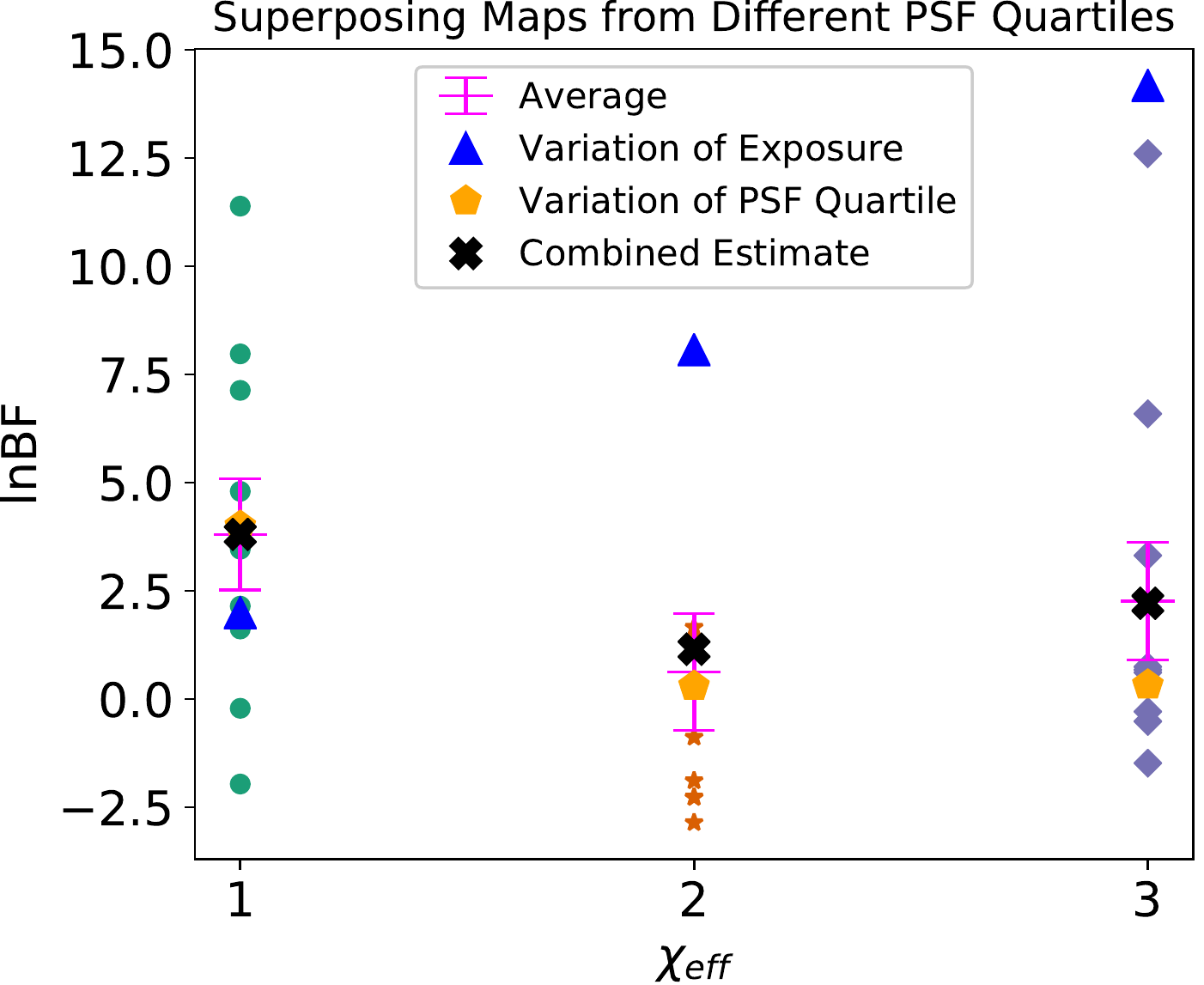}
    \caption{$\ln{\text{BF}}$ and $\langle \ln{\text{BF}}\rangle$ with error bars obtained from the $\sigma/\sqrt{10}$ standard error of the mean (magenta) across 10 realizations that stacked skymaps generated with different angular resolutions (PSF quartiles) for skymaps simulated using the \texttt{Model A} diffuse template. The scans assumed the worst angular resolution. The blue triangles indicate the increased sensitivity as predicted by varying the exposure, while the orange pentagons indicate the worsening of sensitivity due to angular resolution degradation. The black filled ``X" display the combined estimate from varying the two parameters.}
    \label{fig:multiquartilemodela}
\end{figure}

We examine the effect of ``stacking" skymaps generated at different PSF quartiles to increase the effective exposure. We perform the same comparison described in Sec.~\ref{subsec:tradeoffs} to obtain the combined estimate of varying exposure and angular resolution. Similar to the results obtained from \texttt{p6v11}, the $\langle \ln{\text{BF}}\rangle$ across the three levels of $\chi_{\text{eff}}$ are consistent with each other to within $1\sigma$. Hence, we find that the increase in sensitivity from a larger dataset is offset by the worsening of angular resolution. Fig.~\ref{fig:multiquartilemodela} displays the results along with the combined estimate.

\subsection{Pixel size variation} \label{appendix:pixelvar}

\begin{figure}
    \centering
    \includegraphics[width=0.85\linewidth]{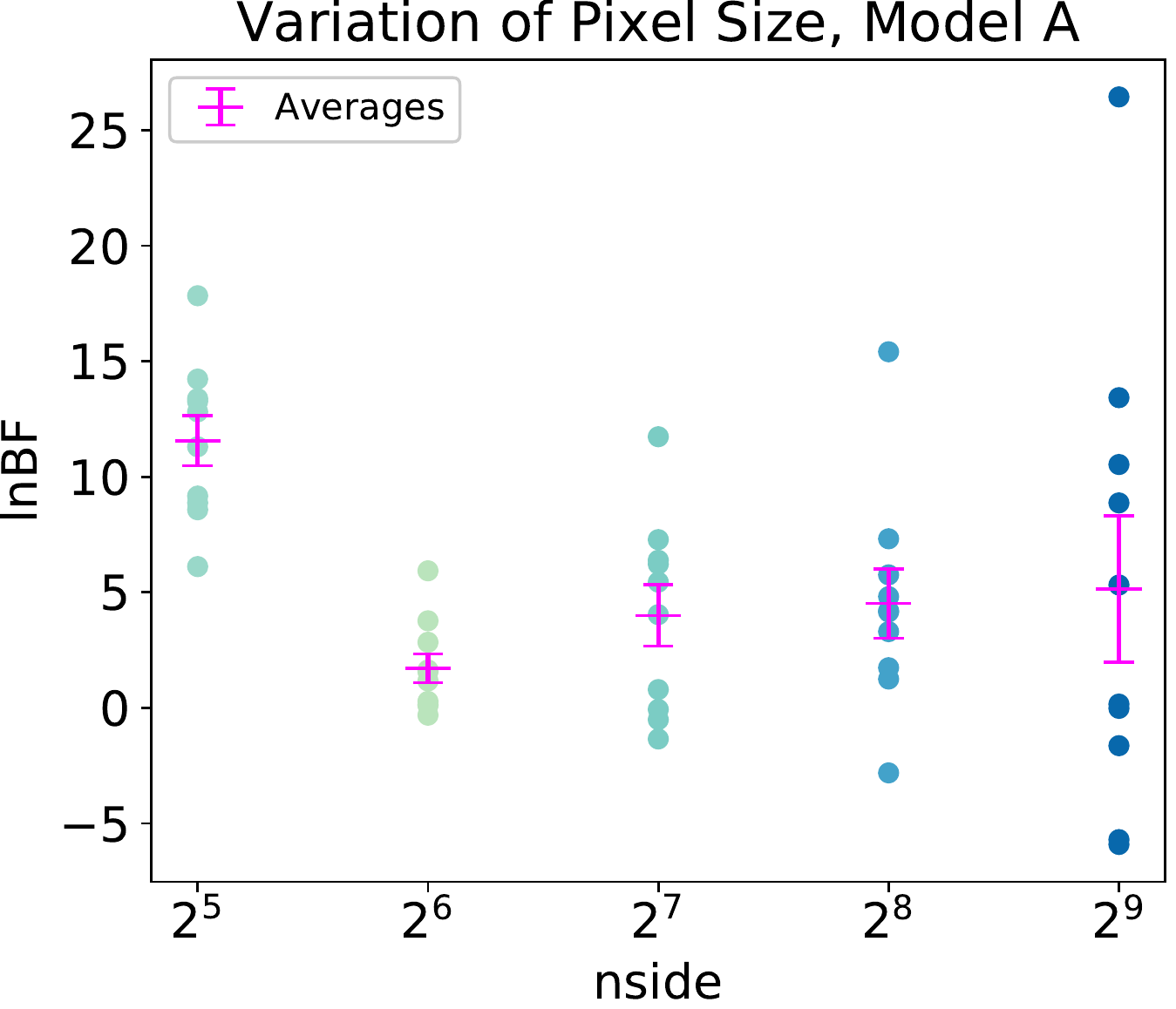}
    \caption{$\ln{\text{BF}}$ across 10 realizations (circle markers), and $\langle \ln{\text{BF}} \rangle $ with error bars obtained from the $\sigma/\sqrt{10}$ standard error of the mean (magenta), using an alternative diffuse model (\texttt{Model A}) at five different pixel sizes. The pixel size decreases as nside increases.}
    \label{fig:pixelvarmodela}
\end{figure}

Finally, we tested the effects of varying the pixel size when using \texttt{Model A} for the Galactic diffuse emission. We applied the same procedure as described in Sec.~\ref{sec:pixelsizevar}. Fig.~\ref{fig:pixelvarmodela} shows the sensitivity level as a function of pixel size.

\begin{figure}
    \centering
    \includegraphics[width=0.8\linewidth]{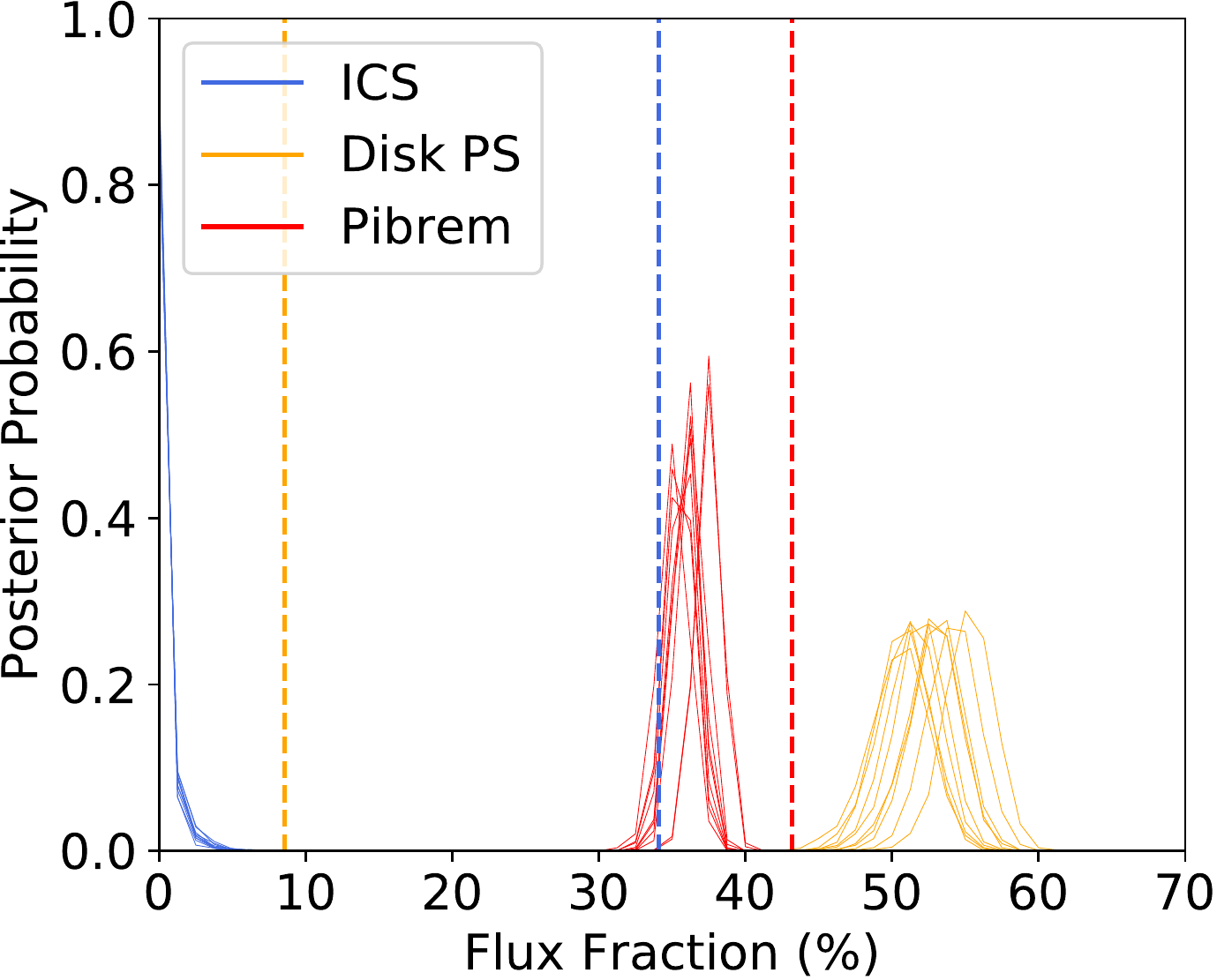}
    
    \caption{Flux fraction plot demonstrating the mis-reconstruction due to a low nside value (nside = 32). The vertical dashed lines denote the injected values obtained from the parameter posterior medians from the real \textit{Fermi} data. The flux fraction posterior probability distributions across 10 realizations (fainter lines) are displayed for the components of \texttt{Model A} diffuse emission and emission from Disk PSs.}
    \label{fig:pixelvarintensitymodela}
\end{figure}

A first look at this figure seems to indicate that the optimal pixel size is the smallest value tested, nside=32. However, this high apparent sensitivity turns out to be accompanied by a failure to correctly reconstruct the various template fluxes, similar to the effect discussed for isotropic emission in Appendix \ref{app:isotropic}. In particular, the diffuse component of \texttt{Model A} which is comprised of the inverse Compton scattering (ICS) and the pion plus bremsstrahlung (Pibrem) component appears to be degenerate with the disk PS component, and the flux associated with these components is frequently mis-reconstructed, as shown in Fig.~\ref{fig:pixelvarintensitymodela}. Taking this into account, it appears that the  ideal pixel size remains at intermediate values, nside=128-256.

\section{Derivation of the PDF for the test statistic in the Gaussian and high-detectability limit}
\label{app:PDFderivation}

To derive Eq.~\ref{eq:analyticPDF}, we first evaluate the log likelihood between  two Gaussian models, parameterized by $(X,\sigma^2)$ and $(Y,\tau^2)$, as a function of the observed number of counts $N$:
\begin{align}\Delta \ln \mathcal{L} & = P(N|\{X,\sigma^2\}) - P(N|\{Y,\tau^2\}) \nonumber \\
&= -\frac{(N-X)^2}{2\sigma^2} - \frac{1}{2}\ln(2\pi \sigma^2) + \frac{(N-Y)^2}{2\tau^2} + \frac{1}{2}\ln(2\pi \tau^2) \nonumber \\
& = \frac{-(N-X)^2}{2\sigma^2} - \frac{-(N-Y)^2}{2\tau^2} - \frac{1}{2}\ln(\sigma^2/\tau^2).  \end{align}

In the high-detectability limit, we can make the approximation that the log term is small and can be ignored. Furthermore, we are interested in the comparison of two distributions that have the same expectation value but different variances (parameterizing the degree to which the distribution is non-Poissonian), so we can set $X=Y$. Writing $\delta=(\sigma^2/\tau^2)-1$ as in the main text, we then have $\tau^2 = \sigma^2/(1 + \delta)$, and so:

\begin{align}\Delta \ln \mathcal{L} & \approx - \frac{(N-X)^2}{2} \left(\frac{1}{\sigma^2} - \frac{1}{\tau^2}\right)
\nonumber \\
& = \delta \frac{(N-X)^2}{2 \sigma^2}. \end{align}

Note that under these approximations, $\Delta \ln \mathcal{L}$ is always non-negative for $\delta > 0$. Now we can evaluate the probability that this expression takes some value $x > 0$ under the true distribution of $N$, which we take to be given by $P(N|\{X,\sigma^2\})$.

\begin{align}P(\Delta \ln \mathcal{L} = x) & = \int \frac{1}{\sqrt{2\pi\sigma^2}} e^{-(N-X)^2/2\sigma^2} \delta\left(x - \delta \frac{(N-X)^2}{2\sigma^2}\right) dN \nonumber \\
& = \frac{1}{\sqrt{2\pi\sigma^2}} e^{- x/\delta} \int  \delta\left(x - \delta \frac{(N-X)^2}{2\sigma^2}\right) dN \nonumber \\
& =\frac{1}{\sqrt{2\pi\sigma^2}} e^{- x/\delta} \sqrt{\frac{\sigma^2}{2 x \delta}} \nonumber \times \int dN \\
& \left[\delta\left(N - X + \sqrt{2\sigma^2 x/\delta}\right) + \delta\left(N - X - \sqrt{2\sigma^2 x/\delta}\right) \right] \nonumber \\
 &= \frac{1}{\sqrt{\pi x \delta}} e^{- x/\delta} \label{eq:PDFproof} \end{align}
where in the second-last line we have evaluated:
\begin{align} \left|\frac{d}{dN} (x - \delta(N-X)^2/(2\sigma^2)\right| & = \left|(N-X)\right|\delta/\sigma^2 \nonumber \\
& = \sqrt{\frac{2 x \sigma^2}{\delta}} \delta/\sigma^2 \nonumber \\
& = \sqrt{2 x \delta/\sigma^2}\end{align}
on the support of the delta function. Eq.~\ref{eq:PDFproof} matches Eq.~\ref{eq:analyticPDF} and is the desired result.

\clearpage

\bibliography{nptfurop}

\end{document}